\documentclass[a4paper, 11pt]{article}
\pdfoutput=1

\usepackage[english]{babel}
\usepackage{booktabs}
\usepackage{amsmath, amsthm}
\usepackage[left=2.5cm,right=2.5cm,top=2.5cm,bottom=2.5cm]{geometry}

\usepackage[charter]{mathdesign}
\usepackage[usenames, dvipsnames, svgnames, table]{xcolor}
\usepackage{multirow}
\usepackage{threeparttable}
\usepackage{longtable}
\usepackage{setspace}
\usepackage{comment}
\usepackage{rotating}
\usepackage{pdflscape}
\usepackage[font=small,skip=5pt]{caption}
\usepackage{subcaption}
\usepackage{listings}
\usepackage[stable]{footmisc}
\usepackage{hyperref}
\hypersetup{colorlinks=true,pdfborder={0 0 0},breaklinks=true}
\usepackage{bbm}
\usepackage{mdframed}
\usepackage{cleveref}
\usepackage{afterpage}
\usepackage{blindtext} % for titles
\usepackage[section]{placeins}
\usepackage{threeparttablex}
\usepackage{tabularx}
\usepackage{graphicx}
\usepackage{subcaption}
\usepackage[title]{appendix}
\usepackage{censor}

\usepackage{adjustbox}
\usepackage{array}
\newcolumntype{R}[2]{%
    >{\adjustbox{angle=#1,lap=\width-(#2)}\bgroup}%
    l%
    <{\egroup}%
}
\newcommand*\rot{\multicolumn{1}{R{45}{1em}}}% no optional argument here, please!

\definecolor{darkblue}{rgb}{0.0,0.0,0.3}
\hypersetup{colorlinks,breaklinks,linkcolor=darkblue,urlcolor=darkblue,anchorcolor=darkblue,citecolor=darkblue}

\newcommand{\indep}{\rotatebox[origin=c]{90}{$\models$}} % independence sign

\usepackage[mathscr]{euscript}

\usepackage[round]{natbib}

\usepackage{floatrow}
\usepackage{amsthm}    %pour mypred
   \newtheoremstyle{definition} % name
    {1em}         % Space above
    {0em}         % Space below
    {\bfseries \itshape}       % Body font
    {0em}           % Indent amount
    {\bfseries}  % Theorem head font
    {:}                          % Punctuation after theorem head
    {.5em}                       % Space after theorem head
    {}  % Theorem head spec (can be left empty, meaning ‘normal’)

  \theoremstyle{definition}
  
  \crefname{assumption}{Assumption}{P}

\usepackage{titlesec}
\titlespacing{\chapter}{0pt}{20pt}{10pt}
\titlespacing{\section}{0pt}{1.25cm}{0pt}
\titlespacing{\subsection}{0pt}{0.75cm}{0pt}
\titlespacing*{\subsubsection}{0pt}{0.25cm}{-0.25cm} %the star removes the indentation after it

\title{Estimating returns to special education: combining machine learning and text analysis to address confounding\thanks{\setlength{\baselineskip}{11pt} I am grateful to my supervisor Beatrix Eugster, as well as Elliott Ash, Simone Balestra, Brianna Ballis, Uschi Backes-Gellner, Caroline Chuard, Michael Knaus, Edward Lazear, Michael Lechner, Helge Liebert, Bryan S. Graham, Fanny Puljic, participants of the Law, Economics, and Data Science Group seminar at the ETH, the Causal Data Science Meeting at Maastricht University \&Copenhagen Business School, the Machine Learning in Labor, Education, and Health Economics Workshop at the IAB, the EffEE Workshop on Causal Analyses of School Reforms at the WZB in Berlin, the Brown Bag seminar at the University of St.\ Gallen, the Young Swiss Economists Meeting 2021, the Spring Meeting of Young Economists 2021, the International Conference on Econometrics and Business Analytics (iCEBA) 2021 for their constructive comments and suggestions on early drafts of this paper. I acknowledge financing from the Swiss National Science Foundation (grant no.\ 176381). This paper reflects the views of the author alone. The usual disclaimer applies.}\vspace{0.5cm}} 

\author{Aurélien Sallin\thanks{\setlength{\baselineskip}{11pt} University of St.\ Gallen, Switzerland. aurelien.sallin@unisg.ch.}\vspace{0.5cm}}

\date{This version: February 2022 }%\today}

\begin{document}

\maketitle

%\begin{center}
%\textcolor{red}{Check updated version \href{https://drive.google.com/file/d/107Nj2l3LgDpGxCkFDXnll7oCCeZ-Yz4c/view?usp=sharing}{\color{red}{\textbf{HERE}}}}
%\end{center}

\begin{singlespace}
\begin{abstract}
\noindent
Leveraging unique insights into the special education placement process through written individual psychological records, I present results from the first ever study to examine short- and long-term returns to special education programs with causal machine learning and computational text analysis methods. I find that special education programs in inclusive settings have positive returns in terms of academic performance as well as labor-market integration. Moreover, I uncover a positive effect of inclusive special education programs in comparison to segregated programs. This effect is heterogenous: segregation has least negative effects for students with emotional or behavioral problems, and for nonnative students with special needs. Finally, I deliver optimal program placement rules that would maximize aggregated school performance and labor market integration for students with special needs at lower program costs. These placement rules would reallocate most students with special needs from segregation to inclusion.
%While the number of students with identified special needs is increasing in developed countries, there is little empirical evidence on academic and labor market returns to special education. By leveraging unique insights into the special education placement process through written individual psychological records, I present results from the first ever study to examine short- and long-term returns to special education programs with causal machine learning and computational text analysis methods. I find that special education programs in inclusive settings have positive returns in terms of academic performance as well as labor-market integration. Moreover, I uncover a positive effect of inclusive special education programs in comparison to segregated programs. This effect is heterogenous: segregation has least negative effects for students with emotional or behavioral problems, and for nonnative students with special needs. Finally, I deliver optimal program placement rules that would maximize aggregated school performance and labor market integration for students with special needs at lower program costs. These placement rules would reallocate most students with special needs from segregation to inclusion.
\end{abstract}
\end{singlespace}

\vspace{1cm}


\setcounter{page}{0}
\thispagestyle{empty}
\setlength{\parskip}{0.5em}

\def\figuresintext{0} 
% !TeX root = Submission_ASallin_v9.tex

\section{Introduction}
A growing number of students in OECD countries are identified with special needs (SEN)\footnote{Following ICD-10 diagnosis guidelines, students with ``special-needs'' (SEN) are students suffering from learning impairments, behavioral, emotional or social disorders, communication disorders, physical or developmental disabilities.}. Taking the US as an example, 14.1 percent of US public school students received Special Education services in 2018--2019, compared to 13.3 percent in 2000-2001, and 10.1 percent in 1980-1981 (\href{https://nces.ed.gov/programs/digest/d20/tables/dt20_204.30.asp}{NCES, 2020}). At the same time, the inclusion of students with SEN in mainstream education has been set as an educational objective by developed countries since the late 1990's.\footnote{According to the \href{https://www.un.org/development/desa/disabilities/convention-on-the-rights-of-persons-with-disabilities.html}{United Nations Convention on the Rights of Persons with Disabilities (2006)}, ``States Parties recognize the right of persons with disabilities to education. With a view to realizing this right without discrimination and on the basis of equal opportunity, States Parties shall ensure an inclusive education system at all levels''.} To this end, most OECD countries have reduced segregation of students with SEN and developed a variety of services such as alternative teaching methods and curricula, Individualized Education Programs (IEPS), and increased staff to accommodate individualized support within mainstream education.\footnote{See \citet{Schwab2020}, and the following \href{https://gpseducation.oecd.org/revieweducationpolicies/}{OECD reports}: ``Students with Disabilities, Learning Difficulties and Disadvantages'' (OECD Publishing, Paris, 2005), and ``TALIS 2018 Results (Volume I): Teachers and School Leaders as Lifelong Learners'' (OECD Publishing, Paris, 2019).} 

%and increased the involvement of mainstream public education by offering a variety of services such as alternative teaching methods and curricula, Individualized Education Programs (IEPS), and increased staff to accommodate individualized support within the main classroom.\footnote{See \citet{Schwab2020}, and the following \href{https://gpseducation.oecd.org/revieweducationpolicies/}{OECD reports}: ``Students with Disabilities, Learning Difficulties and Disadvantages'' (OECD Publishing, Paris, 2005), and ``TALIS 2018 Results (Volume I): Teachers and School Leaders as Lifelong Learners'' (OECD Publishing, Paris, 2019).} %\href{https://nces.ed.gov/programs/coe/indicator_cgg.asp}

Despite the growing number of students with SEN and the increasing implementation of inclusive education programs, empirical evidence regarding how special education (SpEd) placements affect academic and labor market outcomes for students with SEN remains scarce. Existing research shows inconclusive effects of SpEd on academic performance \citep{Hanushek2002,Lavy2005,Keslair2012,Schwartz2021} and on educational attainment \citep{Ballis2019}. Since (early) interventions in children's school curricula have a profound impact on children's academic and lifelong prospects \citep[see among others][]{Cappelen2020,Heckman2013,Duncan2013,ChettyEtal2011}, it is crucial to provide teachers, parents, and policy makers with insights into which SpEd programs are effective. These insights should also help them allocate the most efficient interventions to the students who would benefit the most. This is important in light of the considerable additional financial costs SpEd programs generate for public schools in comparison to standard education \citep{Duncombe2005,Figlio2021}.\footnote{For reference, an annual total of \$40 billion was spent exclusively on SpEd in the USA for the 2015 academic year (\citealt{Figlio2021}; NCES, 2015), and educating a student in SpEd can cost twice to three times as much as educating a mainstreamed student. As an example, \href{https://lao.ca.gov/Publications/Report/4110}{the State of California} estimates that a SpEd student each year costs \$26,000, compared to \$9,000 for a mainstreamed student (Overview of Special Education in California report, LAO, 2019).}

In this study, I set out to investigate returns to SpEd on academic performance, labor participation, use of disability insurance, and wages. I analyze returns to SpEd in a comprehensive way, and assess returns of six different SpEd programs. The first four programs are offered in inclusive academic settings, and are comprised of counseling, academic support (or tutoring), individual therapies (such as speech therapy), inclusion (students with SEN are mainstreamed but with additional support by a SpEd teacher). Two programs are offered in segregated settings, i.e., semi-segregation (same school but special classrooms), and full segregation (separate schools). In addition, I assess whether inclusive programs are more efficient than segregated programs in generating positive academic and labor market outcomes. I use student-level administrative data on school performance on a compulsory standardized test and social security administrative records. These data are combined with detailed information and psychological written records on each individual student, uniquely linking students’ school performance, labor market integration and written psychological assessments for ten consecutive cohorts of students with SEN enrolled in SpEd in the Swiss State of St.\ Gallen. 

I conduct these analyses in the context of the Swiss education system, an academic context which is similar to most OECD countries and the US in terms of inclusive SpEd structures \citep{DeBruin2019}. Moreover, the Swiss academic setting offers ideal conditions for the investigation of returns to SpEd. A first ideal feature is that the diagnosis of special needs and the SpEd placement decision are conducted by the School Psychological Service (SPS), an external and independent administrative entity. This ensures that treatment is given by professional psychologists, rather than by parents, teachers, or schools. In addition, each school is free to implement the SpEd programs of its choice from a catalogue of measures provided by the Education Ministry. In practice, this means that there is variation in program assignment across schools which is not explained by student or by school characteristics. This is reinforced by the fact that schools in Switzerland were strongly encouraged to implement inclusive programs instead of segregated programs after the Swiss Equality Act for People with Disabilities was passed in 2004. However, not all schools started replacing segregated programs with inclusive programs at the same time. All in all, these features allow me to observe students with similar characteristics and similar SEN but assigned to different SpEd programs. %and that schools cannot strategically assign programs that would not best align with the interests of students. 

Special education programs are difficult to evaluate because SpEd placement is based on students’ characteristics that are usually unobservable to the econometrician. To tackle the problem of potential selection into SpEd programs, I leverage data that offer unique and unprecendented insights into the placement process: for each student with SEN, I observe all the psychological records and session transcripts written by the caseworkers in charge of both diagnosing the student's special needs and assigning the student to treatment. These records allow me to gain a deep understanding of the students' background, and the nature and complexity of their special needs. To make use of the information contained in text, I implement newly developed techniques in computational text analysis and natural language processing (NLP) adapted to a causal framework \citep{Gentzkow2019,Mozer2020,Roberts2020,Egami2018,Keith2020}. Leveraging text information with methods from the small but steadily growing literature on Double Machine Learning for flexible program evaluation (see, for instance, \citet{Chernozhukoveetal2018, Athey2019b,Davis2017}, \citealt{Knaus2020b}), I am able to account for confounding in unprecedented detail and to plausibly assume unconfoundedness for identification of returns to programs. This study is among the first studies to take advantage of computational text analysis, causal inference, and Double Machine Learning methods to evaluate education programs and returns to education.

I compare students assigned to various SpEd interventions in a pairwise manner (comparing programs that are the most similar, from the most to the least inclusive), as well as students assigned to SpEd interventions with students that were referred to the school psychological service for assessment but not treated. I find that, among all SpEd programs, inclusive programs pay off: first, returns to SpEd programs provided in mainstream education are mostly positive or null in comparison to being referred but receiving no SpEd. I present evidence that targeted individual therapies (such as speech therapy, dyslexia therapies, etc.) are effective at treating preexisting learning disabilities. Moreover, returns to inclusive education in comparison to segregated programs are strongly positive: students with SEN who remain in the mainstream classroom perform better at school, are more likely to participate in the labor market and earn a 15 percentage points higher salary on average than students with SEN segregated into small classes. By conditioning on all the information psychologists report when assigning treatment through written records, I compare students that are similar in all their observed characteristics. On average, I find that estimates based on both covariates and text information are 29\% smaller in magnitude than estimates that do not leverage the text information. Moreover, my study suggests that students with SEN who exhibit ``disruptive'' tendencies \citep[e.g.,][]{Lazear2001,CarrellEtal2018}, i.e., students with social and emotional problems, psychological problems, and nonnative speakers with SEN, are the students who would benefit the most from semi-segregation in comparison to inclusion. %My empirical conclusions however do not extend to students with SEN who are fully segregated (in special schools), as these students differ substantially in their characteristics from other students. 
Finally, my results highlight that the magnitude of returns vary greatly with the type of program, and thus that sound evaluation of SpEd should account for program specificities.

Getting insights from the literature on statistical treatment rules \citep[e.g.,][]{Kitagawa2018,Manski2004}, I further explore optimal policy allocations to inclusive and segregated SpEd programs and make placement recommendations to reach higher aggregate school performance and improve on labor market integration. I propose a set of optimal policies using machine learning algorithms \citep{Athey2020,Zhou2018} and compare implemented policies with optimal policies in terms of costs and outcomes. By implementing my proposed optimal policies, a policy maker could significantly increase average school performance and, to a lesser extent, labor market integration at lower overall costs. Easily implementable policies would send all segregated students to inclusion, while more refined policies suggest to keep students with social and emotional problems as well as nonnative students with SEN in semi-segregated settings. I further conduct welfare computations to see whether including students who were previously segregated in the classroom would harm mainstreamed students. I integrate the findings of the quasi-experimental study from \citet{BalestraEtal2020} using the same dataset to my analysis, and I find that my optimal policies would generate negligible negative effects on mainstreamed students while significantly increasing average school performance of the reallocated students with SEN. 

The present paper contributes to the understanding of returns to SpEd programs. Most studies investigate SpEd as a single, all-encompassing treatment intervention, and compare students in SpEd with students outside of SpEd. Given that SpEd is usually a multifaceted intervention with programs that differ in quality and intensity, these studies fail to provide insights into the effectiveness of different types of programs.\footnote{As exceptions, \citet{Lavy2005} and \citet{Lovett2017} focus on targeted remedial education only, and \citet{Blachman2014} look at reading remediation.} Studies have shown moderate effectiveness of SpEd considered as a single program on the academic performance of SEN students \citep{Schwartz2021, Keslair2012, Harrison2013,Lavy2005}\footnote{\citet{Scruggs2010} conduct a meta-analysis and find overall positive effects of remediation interventions for students with disabilities. SpEd has been shown to have negative or no effects on reading skills, mathematics skills and behavior of SEN students in comparison to non-SEN students in the US \citep{Morgan2010, Dempsey2016}. Similar results are documented for Norway \citep{Nilsen2018b,Lekhal2018}, but with positive impact on math skills development. Early preschool SpEd has also been shown to have little to no effects on reading and mathematics skills \citep{Sullivan2013,Kohli2015,Judge2011,Morgan2009}.} and positive returns for SEN students with learning and/or emotional disabilities \citep{Hanushek2002}. In addition, evidence on the effects of SpEd  on high-school graduation rates are found to be both positive \citep{Ballis2019} and negative.\footnote{\citet{McGee2011} for the US and \citet{KIRJAVAINEN201633} for Finland report that SEN students have a higher high-school graduation rate than their cognitively equivalent non-SEN peers due to more lenient graduation rules, but lower college enrollment, lower employment rates, and lower wages. \Citet{Blachman2014} document that the effects of a randomized reading intervention fade out 10 years after completion of the program.} Studies on the effects of SpEd on labor market integration are quasi nonexistent \citep[to the exception of][]{McGee2011,KIRJAVAINEN201633}. To my knowledge, this is the first study that assesses short- and long-term returns to SpEd programs at a granular level by ordering SpEd interventions according to their scope and intensity.

Moreover, this study expands on insights from the literature about the factors influencing the emergence of special needs and leading to referrals to SpEd interventions. Many studies highlight the fact that assignment to programs depends heavily on confounders that together influence identification of SEN, assignment to treatment, and the investigated outcomes. For instance, students from non-resilient, low-SES family backgrounds are more likely to develop SEN and to be referred to SpEd \citep{Case2002, Currie2003, Smith2009, Nilsen2018b}. Other factors that influence referrals include starting school earlier \citep{Balestra2020a,Elder2010}, racial or ethnic background \citep{Figlio2021} or suspicion of intellectual giftedness \citep{BalestraEtal2021}. In this paper, I am able to explore many of these confounders by leveraging individual written psychological records and background information about each student with SEN. Furthermore, I use all this information not only to investigate heterogeneities in returns to programs, but also to devise placement rules that are welfare increasing for all students, and cost-reducing for school officials. 

Lastly, this study contributes to investigations of the effects of inclusion in comparison to segregation. On the one hand, existing research offers inconclusive results on the short-term and long-term impacts of inclusion for SEN students \citep{Freeman2000,Cole2004,Sermier2011,Daniel1997,Peetsma2001,Eckhart2011}\footnote{In comparison to segregated SEN students, SEN students in inclusive education perform as well in mathematics and even better in literacy \citep{Sermier2011}, exhibit lower motivation but better math performance \citep{Peetsma2001}. However, \citet{Daniel1997} find that mainstreamed students with SEN generate more behavioral disruptions, exhibit lower self-esteem, and marginally improve in academic performance. \citet{Eckhart2011} reports that segregated students are less likely to be integrated in the job market and have smaller social networks than students in inclusive environments. The attitude of teachers towards inclusion is also a major influential factor of success for inclusive schooling \citep{Avramidis2002,DeBoer2011}.} On the other hand, inclusion is reported to have negative effects on peers without SEN in the mainstream classroom \citep{BalestraEtal2020,Rangvid2019,Fletcher2009}. This study bridges the gap between these two strands of literature by investigating in more detail the short-term and long-term impacts of inclusion from the perspective of SEN students, and by investigating optimal inclusive policy rules.

%The remainder of this study is organized as follows. \Cref{section:background} provides an overall picture of the institutional settings, describes the data, and discusses the text analysis. \Cref{section:identification} presents the identification strategy, and briefly describes the use of causal machine learning for the estimation of treatment effects. \Cref{section:results} shows the main results, and \Cref{section:policy} discusses reallocation policies. Finally, \Cref{section:conclusion} concludes.

\section{Background and Data\label{section:background}}

\subsection{Institutional background: special education programs}
The implementation of SpEd policies in Switzerland is conducted independently by each Swiss federal state (``canton''). To foster inclusion, the Swiss Equality Act for People with Disabilities (2004) made the equality of access to education for SEN students a priority, and emphasized the promotion of inclusion of SEN students in the main classroom rather than segregation. Thus, inclusion is promoted as the main SpEd intervention tool \citep{SKBF2006}, and as a direct substitute for segregation in small special needs classrooms (semi-segregation) \citep{Hafeli2005}. As a result, the share of students sent to segregated schooling has decreased since the Equality Act, while the share of students sent to inclusive schooling has increased. According to the European Agency Statistics on Inclusive Education \citep{EASIE2014,EASIE2018}, the enrollment rate in mainstream education in Switzerland is similar to other European countries. However, the share of segregated Swiss students with SEN varies substantially across Swiss cantons. The Canton of St.\ Gallen ranked 5th as the canton with the most segregated SEN students (3.33\% of the overall student population vs. 1.85\% in Switzerland) in 2010.\footnote{Canton St.\ Gallen, \emph{Nachtrag zum Volkschulgesetz 2013}, p.38.}  %relatively high in international comparison \citep{Sermier2011,SKBF2014}, and this share

This study focuses on students enrolled in SpEd during their mandatory schooling in the Swiss Canton of St.\ Gallen (around 6\% of the Swiss population). The St.\ Gallen Ministry of Education defines a catalogue of SpEd measures and programs for SEN children.\footnote{As elaborated in the official document ``Kantonales Konzept fördernde Massnahmen'' in 2006 by the Canton of St.\ Gallen, the basic offer includes ``SE, speech therapy, rhythm therapy, psychomotor therapy, therapy for dyslexia and dyscalculia, tutoring, special classes''. I describe all the therapies given in the canton of St.\ Gallen in \Cref{table:list_therapies}.} These measures are counseling, academic support, individual therapies, inclusion, semi-segregation, and full segregation. Counseling refers to traditional visits to a therapist or a counselor in which the student's difficulties in school or at home are discussed. It is mostly offered by therapists outside of the School Psychological Service (SPS). Academic support refers to tutoring for children needing additional support for their homework or for learning. Individual therapies are one-to-one or small group sessions; they typically take place during class time, and they target particular learning disabilities for which a particular treatment is required (such as speech therapy, dyslexia or dyscalculia therapy). The inclusion measure refers to all students who received individual inclusive SpEd. These students are provided adapted and goal-oriented complementary teaching by a SpEd teacher who works in the main classroom alongside the main teacher. Semi-segregation refers to small classes (with 10 to 15 students) within the main school. Both inclusive SpEd and semi-segregation are targeted at students with learning and social disabilities, special diagnoses (such as autism, dyslexia, etc.) as well as students who fall behind the class schedule. Full segregation refers to schooling in special schools and targets students for whom mainstream schooling is too challenging (e.g., students with severe disabilities or students suffering from physical impairments such as deafness). Finally, I also observe students who were referred to the SPS for diagnosis but who were not assigned to any treatment (``No placement'').

\if\figuresintext0
  	\begin{center}
  	[Insert \Cref{fig:kk.isf.year} here]
	\end{center}
\else
  \begin{figure}[t!]
	\centering
  	\caption{Share of Special Education placements among students referred to the School Psychological Service over the years \label{fig:kk.isf.year}}
  	\begin{minipage}{\linewidth}
  	    \includegraphics[width=0.9\textwidth]{D: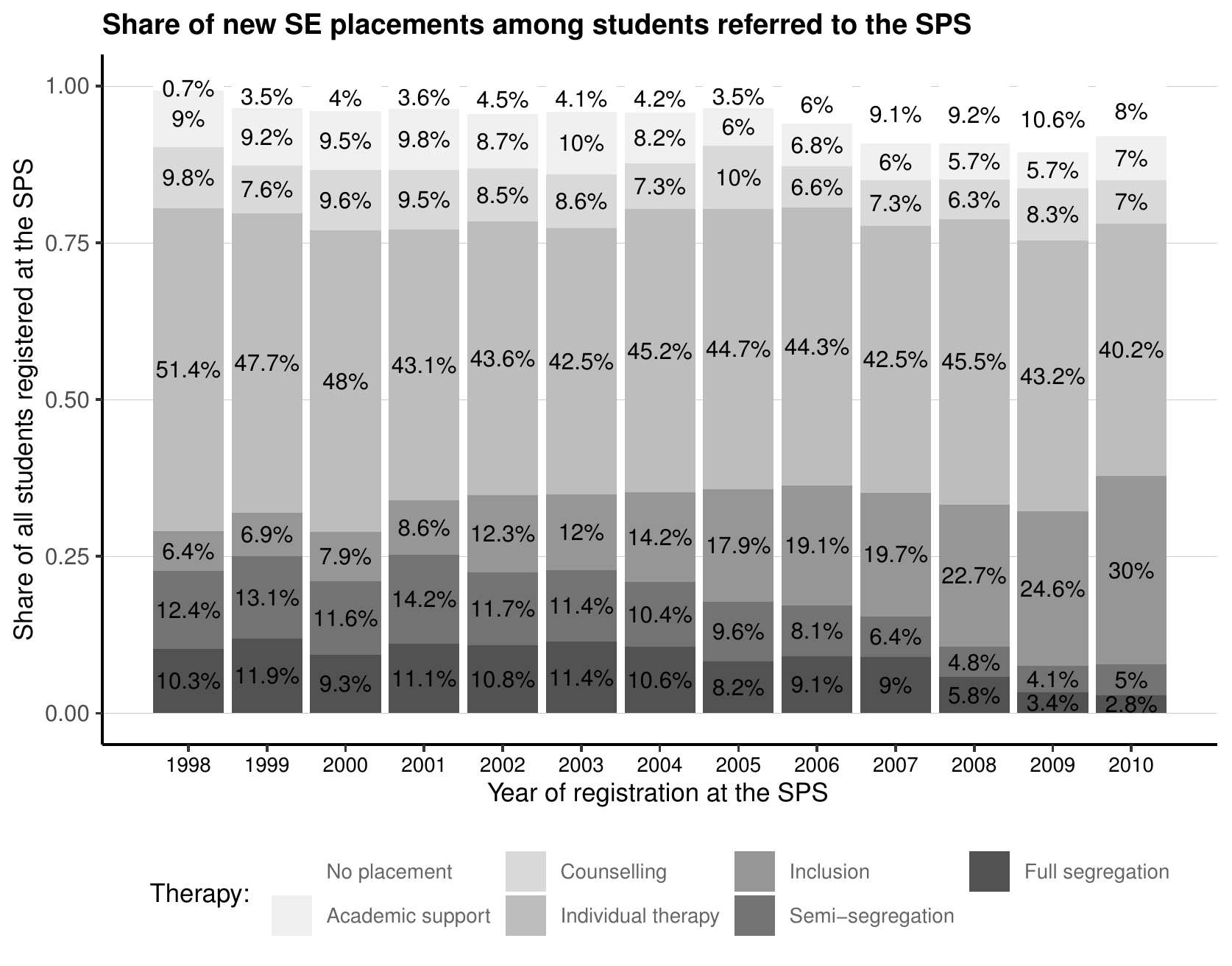}
  	    
  	    \footnotesize \emph{Notes:} This figure displays the newly assigned special education interventions per year. It gives the share of students assigned to a particular SpEd program among all students referred to the School Psychological Service over the years. \emph{Source: SPS}.
  	\end{minipage}
	\end{figure} \fi

\Cref{fig:kk.isf.year} displays the newly assigned SpEd interventions in St.\ Gallen per year. The most frequently assigned therapies are individual therapies. The number of students newly assigned to inclusive SpEd increased from around 6\% of students referred to the SPS in 1998 to around 30\% in 2010, whereas the number of students assigned to small classes steadily decreased (from 12.4\% to around 5\%). These figures reflect the actual number of students with SEN being taught in a semi-segregated setting at the primary level (``stocks''), which dropped from 9.17\% in 1999 to 6.4\% of all students with SEN in 2009, as documented by the official placement register data. 

The St.\ Gallen setting offers many advantages to estimate returns to SpEd programs. First, the diagnosis and SpEd placement decision of students with SEN is conducted by the School Psychological Service, which is an external and independent administrative entity. Therefore, diagnoses and placement decisions are made by SPS psychologists, rather than by parents, teachers, or school administrators. The SPS is organized in eight regional offices. The main task of the SPS is to independently provide diagnoses of learning disabilities, behavioral difficulties, and developmental deficiencies. It assigns therapies and treatments, and offers counseling to students, parents and teachers. As part of the diagnoses, an intelligence test (IQ test) is often administered. After the first consultation, the caseworker, in agreement with parents and teachers, assigns the student to the necessary program. For most students (about nine out of ten), services of the SPS are requested directly by the teacher and/or school official, but some requests are also filed by the parents or the child's medical doctor. Most of the requests to the SPS are made when the student is in Kindergarten/Preschool (see \autoref{fig:registration}).\footnote{The end of Kindergarten is the moment when teachers decide whether the student is ready for primary school or whether the student needs to take a bridge year. This is in line with \citet{Greminger2005}, who report that most segregation decisions happen in Kindergarten in Switzerland.}

\if\figuresintext0
\begin{center}
  [Insert \Cref{fig:IV_variation_school_year} here]
  %[Insert \Cref{fig:KK.ISF.share} here]
\end{center}
\else
	%\begin{figure}[t!]
	%\includegraphics[width=0.7\textwidth]{D:/HSG_research/SpecialneedsHSG/Text/Graphs/KK_ISF_share.png}
	%\caption{Inclusion and semi segregation across school years \label{fig:KK.ISF.share}}
	%\floatfoot{\emph{Notes:} This figure depicts the fraction of school-by-year units offering semi segregation programs (KK), inclusion programs (ISF), or both/none. One school-by-year unit is one school during one school year, and there are approximately 1326 units (102 schools over the years 1998-2010). Many schools have changed their Special Education strategies over the years. \emph{Source: %Pensenpool}.}
	%\end{figure}
	\begin{figure}[p]
\centering
\caption{Distribution of school-years deviations in assignment to SpEd from mean year inclusion assignment rate \label{fig:IV_variation_school_year}}
\begin{minipage}{\linewidth}
    \includegraphics[width=0.9\textwidth]{D:/HSG_research/SpecialneedsHSG/Text/Graphs/IV_variation_school_within_year.png}

    \footnotesize \emph{Notes:} This figure shows the distribution of deviations (residuals) in the assignment rate of students assigned to all SpEd programs per school-year from the mean year assignment rate for the population of students with SEN. Both the raw deviation and the regression-adjusted deviation are displayed. The adjusted deviation shows deviation adjusted for student-level and school-level covariates. Student-level covariates include gender, nonnative status, IQ, reason for referral, and who sent the student for referral (Panel A of \Cref{table:summary_stats}). School-level covariates are shown in Panel B of \Cref{table:summary_stats} and include share of nonnative speakers, share of students with SEN, school size, school socio-economic composition, and urban status. Full segregation is not represented. \emph{Source: Pensenpool, SPS}.
\end{minipage}
	\end{figure} \fi

Second, each school is in charge of setting up their own SpEd policies. This offers valuable variation in program assignment within years across schools which is not explained by the students' characteristics. Schools choose, on a yearly basis, which programs to offer among the programs in the catalogue of interventions provided by the Canton. Schools vary substantially in the therapies they offer, as well as in the extent to which they implement inclusive schooling. \Cref{fig:IV_variation_school_year} shows the distribution of deviations in the assignment rate of students assigned to each SpEd program per school-year from the mean year inclusion assignment rate for the population of students with SEN. The figure shows that there is substantial variation in assignment to each program across schools within the same year, and that this variation in program assignment is not fully explained by students' and schools' characteristics such as socio-economic score and per-student expenditure (regression-adjusted mean assignment). For instance, some schools have a probability to assign students to inclusion which is more than 50 percentage points higher than the mean assignment to inclusion in the same year. This is valuable information, especially given that students in St.\ Gallen are assigned to schools on the sole basis of their location of residence. Parents and students must comply with the assignment to the treatment offered by the school.\footnote{This strict assignment procedure is thoroughly implemented, such that parents have no say about their child's school other than moving permanently to a different municipality or enrolling their students in a private school. Private schooling remains uncommon in Switzerland: in 2014, around 95\% of students attend public-funded schools of their community of residence \citep{SKBF2014}. } 

%Second, each school is in charge of setting up their SpEd policies. This offers valuable variation in program assignment within schools across cohorts and within cohorts across schools which is not explained by the students' characteristics. Schools choose, on a yearly basis, which programs to offer among the programs in the catalogue of interventions provided by the Canton. Schools vary substantially in the therapies they offer, as well as in the extent to which they implement inclusive schooling. For instance, \Cref{fig:KK.ISF.share} shows the fraction of schools-by-years units offering semi-segregation programs, inclusion programs, or both/none: forty-two percent of all school-years offered only inclusive programs and 17\% offered only semi-segregated classes. Moroever, students in St.\ Gallen are assigned to schools on the sole basis of their location of residence. Thus, parents and students must comply with the assignment to the treatment offered by the school.\footnote{This strict assignment procedure is thoroughly implemented, such that parents have no say about their child's school other than moving permanently to a different municipality or enrolling their students in a private school. Private schooling remains uncommon in Switzerland: in 2014, around 95\% of students attend public-funded schools of their community of residence \citep{SKBF2014}. } 

Third, due to the centralized administration and monitoring of SpEd interventions, programs are similar across schools and use comparable educational technology. Moreover, schools have no real budget constraints when it comes to SpEd programs. This prevents strategic program assignment against additional budget, as documented for some US States \citep[e.g.,][]{Cullen2003}. Schools receive a target amount of therapy-hours from the cantonal central administration, which is calculated on the basis of their ``socio-economic score''.\footnote{The school socio-economic score is based on the following four indicators: ratio of foreigners with citizenship of non-German-speaking countries in the population group of 5-14-year-olds, share of unemployed in the 15-64-year-old permanent resident population, ratio of 5-14-year-olds dependent on social assistance to the 5-14-year-old population, quota of low-income households with 0-13-year-old children. It is provided by Competence Center for Statistics within the Department of Economic Affairs of the Canton of St. Gallen.} Within this given amount of therapy-hours, schools are free to allocate SpEd programs according to their preferred strategy. Schools are obliged to satisfy demand, and often offer more hours than the number of allocated hours. Schools also have a duty to report yearly statistics on the number of SpEd hours offered.

\subsection{Data: main variables and summary statistics\label{section:data}}

The main data source on students in SpEd are the administrative records from the SPS, academic test scores, data on labor market integration provided by the Swiss Social Security Administration (SSA), and data on schools' statistics about SpEd (``Pensenpool''). \Cref{fig:dataflow} summarizes the dataset structure and gives an overview of the cohorts represented in the sample. In what follows, I discuss in detail each element of the figure.

\if\figuresintext0 \begin{center}
  [Insert \Cref{fig:dataflow} here]
\end{center} \else
	\begin{figure}[t!]
	\centering
	\caption{Visualization of the sample structure \label{fig:dataflow}}
	\begin{minipage}{\linewidth}
		\includegraphics[width=\textwidth]{D:/HSG_research/SpecialneedsHSG/Text/Graphs/Data_workflow.png} %made with creately.com
		\footnotesize
		\emph{Notes:} This figure presents the sample structure as a timeline. All students referred to the School Psychological Service (SPS) between years 1998 to 2012 are observed and receive a treatment. Students' academic performance is observed in the test data (``Stellwerk8'') for all students reaching the age of 14 or 15 in years 2008 to 2017. Labor market outcomes are observed in the Swiss Social Security Administration (SSA) data for students reaching the labor market in years 2007 to 2016. Because of attrition and the particular data structure, not all students are observed in both the Stellwerk8 and the SSA data (blue arrow).
	\end{minipage}
\end{figure} \fi

\paragraph{Administrative records from the SPS}
The administrative records from the SPS provide information on all students referred to the SPS for a clarification/diagnosis interview between 1998 and 2012. They contain information about the student's characteristics, the therapy assigned, the number of visits to the SPS, and the entirety of the psychological records written by the caseworker. All summary statistics are reported in \Cref{table:summary_stats}, and more detailed statistics per treatment status are given in \Cref{table:summary_stats_treatment} (columns are ordered from the most inclusive program to the least inclusive program).\footnote{\Cref{table:summary_stats_SMD} in the Appendix gives the Standardized Mean Difference across all treatment states for all covariates.}

\if\figuresintext0 
\begin{center}
  [Insert \Cref{table:summary_stats} here]
\end{center} \else
% latex table generated in R 3.6.1 by xtable 1.8-4 package
% Thu Oct 22 10:02:26 2020
\begin{table}
\caption{Summary statistics} \label{table:summary_stats}
\footnotesize

\begin{threeparttable} 
\begin{tabular}{llllll}
  \toprule
  & Mean & Sd & Min & Max & N. obs \\ 
  \midrule

  \addlinespace[0.2em]
  \multicolumn{5}{l}{\textbf{A: Individual and school characteristics}}\\
  Female                                      & 0.407     &         & 0     & 1     & 17,822 \\ 
  Foreign language                            & 0.126     &         & 0     & 1     & 17,822 \\ 
  IQ                                          & 94.92     & 11.9    & 41    & 152   & 13,021  \\ 
  IQ measured                                 & 0.730     &         & 0     & 1     & 17,822 \\ 
  Birth year                                  & 1995.35   & 4.3     & 1982  & 2003  & 17,822 \\ 
  Had bridge year (intro class)               & 0.134     &         & 0     & 1     & 17,822 \\ 
  Age at first interview                      & 8.563     & 2.3     & 3     & 18    & 17,822 \\ 
  %Reasons: other                              & 0.043     &         & 0     & 1     & 17,822 \\ 
  Reasons: social and emotional problems      & 0.209     &         & 0     & 1     & 17,822 \\ 
  Reasons: performance and learning problems  & 0.886     &         & 0     & 1     & 17,822 \\ 
  Reasons: problems with teachers or school   & 0.027     &         & 0     & 1     & 17,822 \\ 
  Reasons: not specified                      & 0.011     &         & 0     & 1     & 17,822 \\ 
  Sent by Caseworker                          & 0.029     &         & 0     & 1     & 17,822 \\ 
  Sent by Others                              & 0.024     &         & 0     & 1     & 17,822 \\ 
  Sent by Parents                             & 0.052     &         & 0     & 1     & 17,822 \\ 
  Sent by Parents and teacher                 & 0.656     &         & 0     & 1     & 17,822 \\ 
  Sent by Teacher                             & 0.237     &         & 0     & 1     & 17,822 \\ 
  Total number of SPS visits                  & 10.587    & 8.6     & 1     & 152   & 17,822 \\ 
  %Regional office: G.                         & 0.099     & 0.308   & 0     & 1     & 17,822 \\ 
  %Regional office: RJ.                        & 0.138     & 0.356   & 0     & 1     & 17,822 \\ 
  %Regional office: R.                         & 0.146     & 0.356   & 0     & 1     & 17,822 \\ 
  %Regional office: Ro.                        & 0.134     & 0.338   & 0     & 1     & 17,822 \\ 
  %Regional office: S.                         & 0.181     & 0.382   & 0     & 1     & 17,822 \\ 
  %Regional office: Wa.                        & 0.136     & 0.328   & 0     & 1     & 17,822 \\ 
  %Regional office: W.                         & 0.163     & 0.371   & 0     & 1     & 17,822 \\ 

  \addlinespace[0.8em]
  \multicolumn{5}{l}{\textbf{B: School characteristics }}\\
  Schools: share of nonnative speakers          & 0.223     & 0.1     & 0     & 0.59  & 17,812 \\
  Schools: share of SEN students                & 0.180     & 0.1     & 0     & 1.00  & 17,812 \\
  Schools: total number of students             & 170.028   & 151.0   & 0     & 1063  & 17,812 \\
  Schools: urban                                & 0.453     &         & 0     & 1     & 17,812 \\
  Schools: socio-economic score                 & 0.980     & 0.1     & 0.79  & 1.20  & 17,812 \\
  Schools: per-student expenditure (2017, std.) & 0.00      & 1       & -0.94 & 5.02  & 13,052 \\

  \addlinespace[0.8em]
  \multicolumn{5}{l}{\textbf{C: Treatment assignment}}\\
  Counseling                                  & 0.081     &         & 0   & 1   & 17,822 \\ 
  Academic support                            & 0.077     &         & 0   & 1   & 17,822 \\ 
  Individual therapy                          & 0.449     &         & 0   & 1   & 17,822 \\ 
  Inclusive SE                                & 0.152     &         & 0   & 1   & 17,822 \\ 
  Semi-segregation                            & 0.095     &         & 0   & 1   & 17,822 \\ 
  Full segregation                            & 0.090     &         & 0   & 1   & 17,822 \\ 
  No therapy (but sent to SPS)                & 0.056     &         & 0   & 1   & 17,822 \\ 
  %\midrule

  \addlinespace[0.8em]
  \multicolumn{5}{l}{\textbf{D: Outcomes}}  \\
  SW8 in SW8 cohort                                   & 0.763   &           & 0       & 1       & 13,890 \\ 
  SW8 composit score (SW8 cohort)                     & 0.000   & 1         & -3.70   & 4.28    & 10,602 \\ 
  %Chose VET                                          & 0.541   & 0.498     & 0       & 1       & 14943 \\ 
  %Chose no further education                         & 0.408   & 0.492     & 0       & 1       & 14943 \\ 
  Used disability insurance (SSA cohort)              & 0.075   &           & 0       & 1       & 11,979 \\ 
  Used unemployment insurance (SSA cohort)            & 0.234   &           & 0       & 1       & 11,979 \\ 
  Monthly wage: last registered year (std., SSA cohort) & 0.000 & 1         & -1.85   & 6.85   & 11,979 \\ 
  
  \addlinespace[0.8em]
  \multicolumn{5}{l}{\textbf{E: Sample attrition}}\\
  In SW8 cohort (1992-2003)                   & 0.779   &      & 0     & 1     & 17,822 \\ 
  In SSA cohort (1982-1998)                   & 0.672   &      & 0     & 1     & 17,822 \\ 
  In both SW8 and SSA cohorts                 & 0.463   &      & 0     & 1     & 17,822 \\ 
  %\midrule

  \bottomrule
\end{tabular}
\begin{tablenotes}[para,flushleft]
\footnotesize
\emph{Notes:} Summary statistics for the population of students referred to the SPS in the Canton of St.\ Gallen. The sample is composed of SN students from the Canton of St.\ Gallen having visited the SPS between 1998 and 2012. ``SW8'' refers to students observed in the \emph{Stellwerk8} academic performance dataset, and ``SSA'' to students observed in the the Swiss Social Security dataset. Standard deviations are not reported for dummy variables. \emph{Source: SPS, SW8, SSA and Pensenpool data}.
\end{tablenotes}
\end{threeparttable} 
\end{table}
 \fi

Students' characteristics are presented in Panel A of \Cref{table:summary_stats}. Forty percent of students in the whole sample are female, and 13\% do not have German as their mother tongue. The IQ score is available for 73\% of the students, mostly for students in later years as IQ testing at the SPS has become more systematic over the years. At an average of 95, sample IQ scores for SEN students are slightly lower than the population average of 100. Students had on average 10.6 contacts with the SPS, and the number of contacts is strongly positively correlated with the intensity of the program (more contacts are needed for students in segregated programs). Age at first registration is almost 9 on average, which coincides with the start of grading for students attending second grade. The (not mutually exclusive) reasons for referral most commonly mentioned are performance and learning problems (89\%), and social or emotional problems (21\%). Sixty-six percent of all decisions for referrals are made by the teachers together with the parents of the child. Around 13\% of students were enrolled in bridge years between Kindergarten and primary school because of slow development or poor school readiness. 

The identification of returns to SpEd in this paper relies mostly on the text contained in the student-level psychological records written by caseworkers. The valuable information contained in the text records makes the assignment process observable. For each visit to the SPS, the caseworker in charge of the student documents the visit, reports the discussion, and gives a recommendation for SpEd placement. Most comments are quite detailed and offer a comprehensive picture of the problems addressed in the discussion, such as family background, psychological issues, the diagnoses of the student, and the particularities of the case. 

To be used in estimation, text records must be reduced to some usable representation. In the context of this study, psychological text records are modeled with the intention of learning about the assignment process and adjusting for confounding, while remaining as low dimensional as possible to avoid problems of support and of computational complexity. The text representations should map concepts of the students' mental health, learning/behavioral disabilities, and other background information as well as possible; they should also account for the context of words and offer enough nuance to adequately represent the situation of each student. Using text for the purpose of causal analysis to adjust for confounding is a recent enterprise and depends heavily on the empirical setting: there is so far no established standard practice \citep[see relevant discussions in][]{Mozer2020,Weld2020,Keith2020,Roberts2020,Egami2018}.

\if\figuresintext0 \begin{center}
  [Insert \Cref{table:listmeth} here]
\end{center} \else
\begin{table}[t!]
\caption{List of used methods for text information retrieval \label{table:listmeth}} 
\centering
\scriptsize

\begin{threeparttable}
\begin{tabularx}{0.9\textwidth}{XXX}%{llXc}
  \toprule
\multicolumn{2}{l}{\footnotesize\textbf{Text Representation}}     %&\textbf{Tuning parameters}   
& \textbf{\footnotesize Dimension of covariate matrix}  \\ 
\midrule
\textbf{``Bag-of-words''}       &\emph{tf}              				 &$N\times$782 tokens    \\   
                                &\emph{tf-idf}                   &$N\times$921 tokens    \\
                                &\emph{tf-tf-idf}                &$N\times$914 tokens    \\
\midrule
\textbf{Structural Topic Modelling (STM)}     &10 topics         &$N\times$10 topics     \\
                                              &80 topics         &$N\times$80 topics     \\
\midrule
\textbf{Topical Inverse Regression Matching (TIRM)}         	   &10 topics + 1 treatment projection   &$N\times$10 topics \\
%\textbf{(for propensity score only)}                             &  & \\
\midrule
\textbf{Word2Vec}               &50--dimensional                   &$N\times$50    \\            
                                &100--dimensional                  &$N\times$100   \\    
\midrule
\textbf{Professional diagnosis} &Dictionary/Keyword approach  %&Predict dictionary from independent dataset with diagnoses made by health professionals. Per diagnosis, a set of features unique to each diagnosis are selected.  
								&$N\times$16 diagnoses \\       
\bottomrule
\end{tabularx}
\begin{tablenotes}[para,flushleft]
\footnotesize
\emph{Notes:} This table describes the different Natural Language Processing (NLP) methods for text information retrieval used in this paper. A discussion of these methods, examples and summary statistics can be found in Appendix \Cref{appendix:appendix_text}.
\end{tablenotes}
\end{threeparttable}
\end{table}
 \fi

\Cref{table:listmeth} summarizes the computational apparatus used to extract information from text. To avoid making estimates too dependent on the choice of text information retrieval method, I extract information from the text using five different state-of-the-art NLP methods and nine different specifications: the term-document matrix (TDM) representation, or ``bag-of-words'' \citep[see, for instance,][]{Mozer2020}; structural topic modeling and topical inverse regression matching, which learn topics and context of words in a semi-supervised manner \citep{Roberts2016,Roberts2020,Blei2003}; neural network embeddings such as Word2Vec in which words are embedded in a lower-dimensional space \citep{Mikolov2013}; dictionary representations that map professional diagnoses. For each method, the final dimension of the text representation matrix is presented. The features contained in the representation matrix are subsequently used as controls for estimation of treatment effects. I discuss how I implement each of these methods and provide descriptive statistics for each method in Appendix \Cref{appendix:appendix_text}. 

\paragraph{Program assignment} 
SpEd programs of interest are defined as the programs figuring in the cantonal catalogue of measures mentioned in \Cref{table:list_therapies}. Around 37\% of the students were given individual, one-to-one therapy only, such as speech therapy, dyslexia therapy, or dyscalculia therapy. Thirteen percent of all students are placed in inclusive settings, around 16\% in segregated settings (8\% in semi segregation and 8\% in full segregation). %Although introductory classes (or ``bridge year'') between kindergarten and primary school are SpEd interventions, they are given before primary school and might be followed by further interventions.\footnote{I consider assignment to introductory classes as a covariate instead of a treatment. Children assigned to introductory classes are reevaluated upon entry in primary school by psychologists and assigned to the programs if needed. Note that many students having had bridge years do not need further interventions.}
Some students (around 5\%) were referred by their teachers to the SPS but did not receive any SpEd intervention. These students form an interesting comparison group, since they are students who raised strong suspicion for SpEd referral but who do not receive SpEd placement after all. From the notes, I know that most of these students have been received and assessed by a caseworker who in turn decided that no further intervention was needed. %The comparison between students who received a treatment and students who were referred but received no treatment is thus an interesting and meaningful comparison. 

Even though SpEd interventions are defined at the central level, program effectiveness might vary with schools' characteristics. To account for this, I bring in statistics about schools obtained from the \emph{Pensenpool} data of the Ministry of Education. I use four measures about the school population (share of students with SEN, share of foreign students, total school population, urban or rural school). %As my sample covers students with SEN in fully segregated settings, the school composition in terms of students with SEN can be as high as 100\%. 
Moreover, I use two measures of educational inputs previously used in the literature: standardized per-student spending \citep[e.g.,][]{JacksonEtal2016}\footnote{The data on spending per primary school students comes from the official accounts published by municipalities at the end of the fiscal year. According to the data, municipalities spend on average 10,160 Swiss Francs (approximately 11,140 USD) per primary school student. This figure is higher than the OECD average (8,733 USD) but comparable to the corresponding figure in the U.S.\ (11,319 USD), as the OECD documents \citep{oecd2017}.}, and the ``socio-economic score'', introduced above, which measures the school's socio-economic composition \citep[e.g.,][]{AngristLang2004}. These school-level statistics are measured in the year in which students with SEN are assigned to treatment (with the exception of per-student spending, which is only measured in 2017). \Cref{table:summary_stats_treatment} shows that school characteristics are rather well balanced across all treatments, with the exception of inclusion and semi-segregation. Schools implementing semi-segregation tend to be more urban, larger, and with a higher share of foreign students. However, schools offering inclusion and schools offering semi-segregation overlap in their characteristics, as can be seen in \Cref{fig:KKvsISF_SCHOOLS}. 

\paragraph{Outcomes: test scores and labor-market integration} I measure different outcomes to capture school achievement as well as labor market integration. Outcomes are reported in Panel D of \Cref{table:summary_stats}. For academic performance, I use test scores from the ``Stellwerk8'' standardized test (SW8) taken in grade 8, which give the individual academic achievement for the entire population of students enrolled in 8th grade during the years 2008 to 2017. This test is mandatory for all students with SEN (except for students in fully segregated settings) and is the same in all schools. It is computer-based, and automatically adapts the difficulty of questions to the ability and knowledge revealed by the student in the previous questions. It tests core knowledge of mathematics, language (German), and, depending on the track, other subjects. I focus on the composite score in German and Math, which are compulsory subjects for all students. Test scores range between 0 and 1,000 (1,000 being the best), and are standardized by school-year for easier interpretation and comparison. The performance on the test is important both for students, who will use the test scores when choosing their post-compulsory education, and for teachers, whose relative performance can be reflected in the rate of success of their students. As students with SEN in fully segregated settings are not required to take the test and can choose to opt out, I create a test-taking indicator variable to account for attrition. %Around 76\% of SEN students subject to the test actually take it, which is significantly less than the coverage for non-SEN students. 

Data on labor market integration are provided by the Swiss Social Security Administration (SSA) for the years 2007 to 2016, and contain the individual history of wages, whether the individual has benefited from a disability insurance status (DI), and whether the individual has requested unemployment insurance. I compute the income as the last income recorded standardized over birth years. This gives the average relative position of individual income per cohort and per year, which accounts for cohort as well as year effects.\footnote{I also checked other income definitions, such as last monthly income recorded standardized over birth years. Results are robust across these alternative income definitions.} Income is defined as income from one's own labor, namely net of DI and unemployment benefits. Around 8\% of the sample have claimed disability insurance, and 23\% have claimed unemployment insurance.

\paragraph{Sample restrictions}
Some restrictions are imposed on the data (details can be found in \Cref{table:attrition}). I discard students who received therapies or measures that are not offered by the schools (for instance, private tutoring). Moreover, I conservatively discard students who received so-called secondary ``supportive measures'' only\footnote{These measures include tutoring, language classes for students with an immigration background, and gifted education. For details, see the ``Sonderpädagogik-Konzept'' of the Canton of St.\ Gallen, available on \href{https://www.sg.ch/bildung-sport/volksschule/rahmenbedingungen/rechtliche-grundlagen/konzepte.html}{the website of the St.\ Gallen schools}. Students receiving supportive measures in addition to the main measures are, however, kept in the dataset.}, and students who received more than one treatment. This ensures that multiple influences of different treatments are not confounding the main treatment. 

Cohorts registered in the school data and cohorts registered in the SSA data do not perfectly overlap (see the red arrows in \Cref{fig:dataflow}). Since the SW8 test was given in years 2008 to 2017, and given that some cohorts were not exposed to the test, I investigate subsamples for each outcome separately. Subsample sizes are reported in Panel E of \Cref{table:summary_stats}. While 78\% of the sample were in cohorts subject to the SW8 test, 67\% are from cohorts with no test but with recorded labor market outcomes. Finally, 46\% of observations are observed in both subsamples. Attrition is only due to cohort variation, and I conduct attrition analyses in my robustness checks to show that attrition is not a problem for my main results.

\section{Empirical strategy \label{section:identification}}
A plausible causal estimation of returns to SpEd programs requires comparing the academic and labor outcomes of students who are similar in all the characteristics which jointly influence their outcomes and their assignment to SpEd programs. In the absence of a randomized experiment in which students are randomly assigned to programs, I leverage the information contained in the psychological reports, and I model the assignment process with a unusually exclusive and detailed perspective. Furthermore, I make implicit use of the exogenous variation in treatment assignment within years across schools to identify effects of SpEd programs.

\subsection{Definition}
I compare the outcomes of students assigned to various SpEd interventions in a pairwise manner (from most to least inclusive interventions), as well as students assigned to SpEd interventions with students that were referred to the SPS but who were not treated. I follow a multivalued treatment framework in observational studies \citep{Imbens2000,Lechner2010}, in which I compare program $d$ with program $d'$ for student $i$. More precisely, I denote by $d$ the received treatment by student $i$ among the set of mutually exclusive seven programs $\mathscr{D}$. The observed outcome given $i$'s assigned therapy is $Y_i = \sum_{d=1}^{D}\underline{1}(D_i = d)Y_i^d$, and the potential outcome for each individual is $Y_i^d$ for all $d \in \mathscr{D}$. I further denote as $\mathscr{X}$ a set of pre-treatment variables, and as $\mathscr{Z}$ the subset of $\mathscr{X}$ that contains the variables used to conduct heterogeneity analysis. The generalized propensity score is defined as $p_d(x) = P(D_i = d | X_i = x)$, namely the conditional probability of receiving each treatment. 

I am interested in the following estimands. The first is the average potential outcome (APO) under each treatment $d$, $\text{APO}_{d} = E[Y_i^d]$. It is the average outcome for the whole population as if it was assigned to program $d$. This corresponds to the ``value'' of each program. The second is the pairwise Average Treatment Effect $\text{ATE}_{d,d'}  = E[Y_i^d - Y_i^{d'}]$, which represents the effect of treatment $d$ vs treatment $d'$ as if everyone in the population was observed under both treatment states. Since some treatments might not be available for the whole population (for instance, full segregation is not a feasible intervention for all SEN students), the ATE is not interesting for all treatment pairs. I thus compare treatment effects for the subpopulation actually observed in a given program using the Average Treatment Effect on the Treated $\text{ATET}_{d,d'} = E[Y_i^d - Y_i^{d'}|D=d]$. Comparing the ATE and the ATET gives valuable insights about the program assignment process: a large difference between the two estimates might underline effect heterogeneity or nonrandom assignment into programs. Finally, I look at Conditional Average Treatment Effects (CATEs). I consider two different cases of CATEs: first, Group Average Treatment Effects (GATEs) give the ATEs for predefined and policy relevant groups of students, i.e. $\text{GATE}_{d,d'}(z)  = E[Y_i^d - Y_i^{d'}|Z_i=z]$ where $Z_i \in \mathscr{Z}$. For instance, I investigate whether treatment effects are heterogeneous for students with and without behavioral problems. Second, I look at Individual Average Treatment Effects (IATEs) for ATEs at the most granular, individual level. Instead of focusing on groups, IATEs include all observed confounders as heterogeneity variables. This is expressed as $\text{IATE}_{d,d'}(z)  = E[Y_i^d - Y_i^{d'}|X_i=x]$, where $X_i \in \mathscr{X}$ (i.e. a vector of observed pre-treatment variables).

\subsection{Identification}
The previous section presented the estimands of interest as potential outcomes. As each student is only observed in one program, only one potential outcome per student is observable and the other potential outcomes are latent. Therefore, the estimands of interest are not identified unless the following standard assumptions hold \citep{Imbens2015}. The first key identifying assumption is unconfoundedness, i.e. that the vector of observed pre-treatment covariates $X_i$ contains all the features that jointly influence treatment and potential outcomes $Y_i^d \indep D_i|X_i=x, \forall x \in \chi,  \forall d \in \mathscr{D}$. The plausibility of this assumption is justified by the use of text information: the information extracted from the text delivers a unique and detailed overview of both pre-treatment information relevant for treatment assignment and details on the treatment assignment itself. To support this assumption, I show that text brings additional information which is richer than the information contained only in covariates not extracted from text. \Cref{appendix:appendix_text} in the Appendix, and more precisely \Cref{fig:tf}, \Cref{fig:tfidf}, \Cref{fig:topic.distr.10}, and \Cref{fig:diagnosisdistr}, provide evidence that the text delivers valuable additional information which is not contained in the covariates not extracted from text. These figures also show how text information is related to treatment assignment. In addition to the use of text, unconfoundedness is particularly plausible in my setting given the unexplained variation in treatment assignment within years across schools. 
%In addition, I show that the information retrieved from the text works well to predict nontext covariates, but is not fully predictive of nontext covariates (i.e., the information extracted from the text is not redundant).\footnote{To support this assumption, I train different classifiers with machine learning methods on text data to predict students' nontext covariates. The fraction of missclassified covariates (error rate) from the text is depicted in \Cref{fig:identification_prediction_covariates} of the Appendix for each text retrieval method presented in \Cref{table:listmeth}. I find that my text methods are able to capture students' main traits in a reliable way. For instance, my text accurately predicts learning difficulties as a reason for referral for 88\% of students, and nonnative status for 90\% of students. Gender seems to be the most difficult covariate to predict from the text, as my best text methods capture gender accurately in 70\% of cases (which remains a good accuracy score). I also show that my text variables can predict students' IQ score with a very low Mean Absolute Error. } 

The second identifying assumption states that confounders are exogenous, i.e. confounding variables in $\mathscr{X}$ (and in $\mathscr{Z}$) are not influenced by the treatment in a way which is related to the outcomes. This assumption would be violated if covariates are measured after treatment assignment. Non-text covariates are measured before treatment assignment, but text covariates require more scrutiny. To avoid text-induced post-treatment bias, I only use records written before the treatment assignment, thereby removing therapy evaluations and reports about the progress of the student. I also strip out of the text all mentions or discussions of interventions \emph{per se}. The exogeneity assumption would also be violated if referrals to the SPS and treatment assignment would be done based on expected treatment returns, i.e. teachers would refer only the students who are expected to benefit from SpEd to the SPS, and psychologists would be perfectly able to predict the outcomes from treatment assignment. On the one hand, %referrals to SpEd based on observables (e.g., in case of race discrimination) is a known phenomenon \citep[see][]{Figlio2021,Lanfranchi2004} which cannot be totally ruled out in this case. However, 
this problem is likely mitigated given that SPS centers as well as the organization of SpEd in St.\ Gallen are centralized, and that there is no budgetary constraints for SpEd. On the other hand, psychologists are not able to perfectly predict outcomes given treatment, since not all treatments are available in all schools and in all years. %As a reminder, my estimates apply to the population of \emph{students referred to the SPS}, and must be read as such.

Third, overlap (or common support) $0 < p_d(x) <1, \forall x \in \chi,  \forall d \in \mathscr{D}$ ensures that SEN students can be compared at all values of $X$ for a given treatment effect. Because of the variation offered by the school choices of supplied interventions, I can observe students with similar characteristics who were offered different interventions. To make this assumption even more plausible, I compare only interventions that are closest in terms of intensity and in terms of the special needs they target. Moreover, to deal with potential problems of overlap, I present effects for the overlap population in the \autoref{appendix:robustness}. Finally, lack of overlap is problematic for students assigned to fully segregated SpEd programs, since this particular population of SEN students exhibits more severe mental and learning disabilities than other SEN students. This will be kept in mind in the discussion of the results.

Finally, assignment to a particular SpEd program does not generate spillover effects (SUTVA), i.e. $Y_i = Y_i(D_i)$. There are mainly two cases in which SUTVA could be violated. First, the presence of a student with SEN in a program might generate spillover effects. I estimate total effects of programs on the population of SEN students only (and not on the mainstream population). In other words, my estimates incorporate potential classroom spillovers.\footnote{Note that there is no strategic assignment of SEN students to mainstream classrooms in St.\ Gallen \citep[see][]{BalestraEtal2021, BalestraEtal2020}.} Second, if therapies are budgeted at the school level, sending one student to therapy might reduce available resources for other SEN students who also need therapy. This is not a concern in this setting. As mentioned above, schools in St.\ Gallen do not engage in strategic therapy assignment against additional budget \citep[e.g.,][]{Cullen2003}, as there are no real budget constraints when it comes to SpEd. 

Under these assumptions, the estimands of interest are identified using the ``augmented'' weighted estimator (AIPW) score $\Gamma_{d,X_i}^h$. This score  combines the conditional expectations of the outcome $Y$ specific to each potential treatment, $\mu(d,x) = E[Y_i|D_i = d, X_i=x]$ with the outcome residual reweighed by some function of the treatment probability $p_d(x)$. Following the ``balancing weights'' notation of \citet{Li2019}, the general form of this estimator is:
\begin{align}\label{equation:weighting}
\Gamma^h(d,X_i)  = \mu(d, x)h(x) + {\text{\underline{1}}(D_i = d)(Y_i-\mu(d,x))\omega_d(x)},
\end{align} The ``tilting function'' $h(x)$ defines the target population as a function of the propensity score $p_d(x)$, and $\omega_d(x) = h(x)/ p_d(x)$.\footnote{Note that $\omega$ can accommodate weights for different subpopulations as additional ``balancing weights'' schemes, such as trimming weights or matching weights \citep{Li2018,Li2019}.} When the population of interest is the whole population, as in the ATE, the tilting function $h(x)$ is 1 and the estimator is doubly robust \citep{RobinsRotnitzkyZhao1994,RobinsRotnitzkyZhao1995}.\footnote{The score is doubly robust when it is still consistent if the propensity score or the outcome equation are misspecified. For more details on the APO score and the double robustness properties, see \citet{glynn_quinn_2010} for an intuitive introduction and \citet{Knaus2020} for DML.} 

All estimands of interest mentioned above are identified as follows:
\begin{alignat}{3}
&\text{APO}_{d}         &&= E[\Gamma^{h}(d, X_i)], &&\quad h(x)= 1   \\
&\text{ATE}_{d,d'}      &&= E[\Gamma^{h}(d,X_i) - \Gamma^{h}(d',X_i)],             &&\quad h(x)= 1\\
&\text{ATET}_{d,d'}     &&= E[\Gamma^{h}(d,X_i) - \Gamma^{h}(d',X_i)|D_i = d],     &&\quad h(x)= p_{d'}(x) \label{equation:AIPW_ATET}\\
&\text{GATE}_{d,d',z}   &&= E[\Gamma^{h}(d,X_i) - \Gamma^{h}(d',X_i)|Z_i = z],     &&\quad h(x)= 1 
\end{alignat}
The APO $\Gamma^{h}(d,X_i)$ for the ATE score takes $h(x) = 1$ since it applies to the whole population. The estimand for the ATET takes $h(x) = p_{d'}(x)$ as it applies to the population of the treated. Note that, in my main specifications, I do not explicitly model the variation in assignment rate across schools within years. These variations contribute to the plausibility of the four assumptions presented above. However, I estimate an IV specification with the within year across school variation in program assignment rate as an instrument in Appendix \Cref{appendix:subsection_IV}.

\subsection{Estimation with Double Machine Learning \label{section:estimation}}
The estimation procedure is represented in the stylized workflow of \Cref{fig:workflow}. In a first step, text representations are extracted from an independent, held-out sample in order to avoid risks of overfitting, and are subsequently predicted on the main sample. This ensures that text representations are meaningful across the whole dataset. Once text representations are predicted on the main sample, $K$-fold cross-fitting \citep[see][]{Chernozhukoveetal2018} is used to estimate the two nuisance parameters of interest $\hat{\mu}(d,x)$ (estimated conditional expectation of the outcome) and $\hat{p}_d(x)$ (estimated propensity score) using the covariates $\mathscr{X}$ presented in Panel A of \Cref{table:summary_stats} as well as text covariates. Succinctly, the procedure works as follows: (i) the sample is randomly split in $K$ folds of equal size, (ii) one fold is left out, and the remaining $K-1$ folds are used to train machine learning models to estimate the nuisance parameters. These models (iii) are used to predict $\hat{p}_d(x)$ and $\hat{\mu}(d,x)$ on the left-out $K$th fold, and (iv) the procedure is repeated such that each fold is left out once. Extracting text representations from an independent sample requires the availability of large amount of data, which is not always available. When not available, text representations can be retrieved alongside the training of nuisance functions within each $K-1$ folds.\footnote{For each fold, the model used for text representation is trained on the $K-1$ training folds and is in turn used as a set of covariates to train the predictive model for the propensity score or the outcome. In this case, text representations are discovered in each $K-1$ fold and thus are fold-dependent. They cannot be compared to text representations in other folds, and cannot be used as covariates of interest in the set $\mathscr{Z}$. To reduce computing times, the vocabulary (the set of tokens) is extracted for the whole dataset before cross-fitting.} In this application, text representations are extracted from the City of St.\ Gallen sample in the case of Word2Vec and the dictionary. Topics for STM and TIRM are extracted within each $K-1$ folds.
%the sample is randomly split in $K$ folds of equal size \citep[$K$-fold cross-fitting, see][]{Chernozhukoveetal2018}. On each $K-1$ folds, I train machine learning models to predict the ``nuisance parameters'' $\hat{\mu}(d,x)$ (the conditional expectation of the outcome) and $\hat{p}_d(x)$ (propensity score) using the covariates $\mathscr{X}$ presented in Panel A of \Cref{table:summary_stats} as well as text covariates. The nuisance parameters are then predicted on the left-out $K$th fold. Extracting text representations from an independent sample requires the availability of large amount of data, which is not always available. When not available, text representations can be retrieved alongside the training of nuisance functions within each $K-1$ folds.\footnote{For each fold, the model used for text representation is trained on the $K-1$ training folds and is in turn used as a set of covariates to train the predictive model for the propensity score or the outcome. In this case, text representations are discovered in each $K-1$ fold and thus are fold-dependent. They cannot be compared to text representations in other folds, and cannot be used as covariates of interest in the set $\mathscr{Z}$. To reduce computing times, the vocabulary (the set of tokens) is extracted for the whole dataset before cross-fitting.} In this application, text representations are extracted from the City of St.\ Gallen sample in the case of Word2Vec and the dictionary. Topics for STM and TIRM are extracted within each $K-1$ folds.

\if\figuresintext0
\begin{center}
  [Insert \Cref{fig:workflow} here]
\end{center}
\else
\begin{figure}[t!]
	\centering
	\caption{Workflow of Double Machine Learning and text analysis \label{fig:workflow}}
	\begin{minipage}{\linewidth}
		\includegraphics[width=\textwidth]{D:/HSG_research/SpecialneedsHSG/Text/Graphs/ML_workflow.png}
		\footnotesize
		\emph{Notes:} This figure represents a stylized workflow of the estimation procedure. First information from text is retrieved, then used in $k-$fold cross-fitting to estimate the two nuisance parameters (estimated propensity score and estimated conditional expectation of the outcome). The doubly-robust score is computed and used to estimate the estimands of interest (APO, ATE, ATET, IATE and GATE, optimal policies).
	\end{minipage}
\end{figure} \fi

The nuisance parameters are then combined to build, on each left-out fold, the doubly-robust (DR) score as: 
\begin{align}\label{equation:AIPW_DR}
\hat{\Gamma}^{h=1}_{i,d} = \hat{\mu}(d, X_i) + \frac{\text{\underline{1}}(D_i = d)(Y_i-\hat{\mu}(d, X_i))}{\hat{p}_d(X_i)}.
\end{align} Since no observation is used to estimate its own nuisance parameter, cross-fitting reduces the risk of overfitting. I estimate the nuisance parameters with a combination of many methods through an ensemble learner \citep{Vanderlaan2007}: I predict the nuisance parameters with three ML methods (Lasso, Elastic Net and Random Forest) and with 11 different text representations on top of main covariates. This results in 33 different estimations per fold. I obtain the weights of the ensemble learner by cross-validating the out-of-sample MSE of each specification and use a weighted combination of the 5 most predictive specifications in the final score. 

From the score of the APO defined in \Cref{equation:AIPW_DR}, the ATE is constructed as the mean of the difference between the APO scores for the treatments of interest, i.e. $\widehat{\text{ATE}}_{i,d,d'} = \hat{\Gamma}^{h=1}_{i,d} - \hat{\Gamma}^{h=1}_{i,d'}$. For the ATET, the doubly-robust score for $\widehat{\text{ATET}}_{i,d,d'}$ is $ \bigg[\frac{\text{\underline{1}}(D_i = d)(Y_i - \hat{\mu}_{d'}(X_i))}{\hat{p}_{d}} - \frac{\hat{p}_{d}(X_i)}{\hat{p}_{d'}(X_i)}\frac{\text{\underline{1}}(D_i = d')(Y_i-\hat{\mu}_{d'}(X_i))}{\hat{p}_{d}}\bigg]$ where $\hat{p}_{d} = P[D=d] = N_{d}/N$ \citep{Farrell2015}. For point estimates of the APO, ATE and ATET, I take the means of the different estimands and rely on single-sample $t-$tests for statistical inference.\footnote{This is possible without taking into account the fact that nuisance parameters are estimated in the first place if the nuisance parameters estimators are consistent at a relatively fast rate, asymptotically normal and semiparametrically efficient \citep{Chernozhukoveetal2018}.} The GATEs are estimated by taking the conditional mean of the $\widehat{\text{ATE}}_{i,d,d'}$ over groups determined by pretreatment variables $Z_i$, i.e. by regressing the score $\widehat{\text{ATE}}_{i,d,d'}$ on the group variables of interest and using standard heteroscedasticity robust standard errors \citep[following][]{Semenova2017}. To assess the effect heterogeneity along a continuous variable $Z_i$, \citet{Zimmert2019} and \citet{Fan2020} propose to regress the individual score of $\widehat{\text{ATE}}_{i,d,d'}$ on $Z_i$ with a kernel regression and standard inference for nonparametric regression. I estimate second-order Gaussian kernel functions and choose the 0.9 cross-validated bandwidth, as recommended by \Citet{Zimmert2019}. Finally, I estimate IATEs by using a DR-learner, i.e. I train an ensemble learner to predict the individual ATE score $\widehat{\text{ATE}}_{i,d,d'}$ out-of-sample \citep[see][]{Kennedy2020,Knaus2020}.\footnote{I follow the following procedure: in a first step, I predict in each fold the nuisance parameters and then compute the individual score $\widehat{\text{ATE}}_{i,d,d'}$. In a second step, I train an ensemble learner in the same folds to predict $\widehat{\text{ATE}}_{i,d,d'}$ from covariates $X$. In a third step, I use the trained ensemble learner to predict $\widehat{\text{ATE}}_{i,d,d'}$ on the left-out fold. This procedure is computationally heavier than an in-sample IATE prediction but has the advantage of avoiding overfitting. It is however less computationally burdensome than the cross-fitting procedure proposed by \citet{Knaus2020}, as I do not have to re-estimate, in each fold, the text measures that need to be estimated in-sample.} 

Alongside its double-robustness property, the use of Double Machine Learning (DML) and of the AIPW score has many advantages when working with text. First, it allows for leveraging text representations both in the propensity score \citep[as in][]{Mozer2020,Roberts2020} and in the outcome equation, which reduces problems of extreme propensity score accuracy \citep{Weld2020} and overcomes difficulties of matching on both covariates and text.\footnote{Matching algorithms for text as proposed by \citet{Mozer2020} are both computationally burdensome and difficult to implement, insofar as assessing match quality of text is difficult (researchers must find the relevant text reduction, the relevant text distance metrics, and the relevant matching assessment tool, such as human coders).} Second, by not relying on one particular estimation method but combining many of them in an ensemble learner, I make full use of different ML methods and use the ones that work best with each text representation. This also mitigates potential misspecification of the text and covariate functional forms.\footnote{For instance, the \emph{Generalized Random Forest} of \citet{Athey2019} or the \emph{Modified Causal Forests} of \citet{Lechner2019} rely exclusively on random forest, which might not perform well on a ``bag-of-words'' representation of text due to the high number of sparse dummy variables.}

\section{Results: returns to special education programs \label{section:results}}
In this section, I present different sets of main results: first, I present the pairwise effects for inclusive SpEd interventions (i.e., interventions that are provided in the mainstream school environment). Second, I focus more specifically on the effect of inclusion vs. semi-segregation. Third, in order to relate to existing literature, I look at the ``extensive margin'' of SpEd interventions and assess the effect of being assigned to a program vs. being assigned to no program at all. This set of results corresponds to the effect traditionally estimated in the literature. Fourth, I conduct analyses of the heterogeneous effect of inclusion. Finally, I perform a series of further analyses and robustness checks.  

\subsection{Returns to Special Education programs in inclusive school settings}
I first present, in \Cref{fig:results.main}, returns to SpEd on academic performance for interventions that are the closest in degree of severity and inclusion, and which are either provided as supportive or remediation measures (counseling, academic support or tutoring, individual therapies, and inclusion) provided in the mainstream school environment. Results read as follows: pairwise effects give the effect for being assigned to the first program (for instance, in the first column, to counseling) instead of being assigned to the second program (e.g., to no program) on academic performance (Panel a), probability to be unemployed (Panel b), the probability to use disability insurance (Panel c), and on work income (Panel d). Test scores and wages are standardized with mean 0 and standard deviation 1. The baseline  (``No SpEd'') probability of unemployment benefit recipiency is 0.19, and 0.07 for disability insurance recipiency. Effects account for all the observed confounding from covariates (such as gender and IQ) as well as all information contained in the psychologists' records. Point estimates and 95\% confidence intervals are shown graphically, and both the effect for the whole population (ATE) and the effect for the population of the treated (ATET) are represented.\footnote{Regression tables with point estimates and exact confidence intervals are available upon request.} Pairwise effects compare interventions that are the most similar, but that incrementally differ in their severity. For instance, the pairwise comparison of academic support and individual therapy compares interventions which are very similar and which target issues that are overlapping. The exception is counseling, which I compare to no treatment, as counseling does not happen in schools but with independent psychologists. 

\if\figuresintext0
\begin{center}
  [Insert \Cref{fig:results.main} here]
\end{center}
\else
\begin{figure}[t!]
	\caption{Pairwise returns to Special Education programs according to their level of inclusion. \label{fig:results.main}}
	\begin{minipage}{\linewidth}
	\includegraphics[width=\textwidth]{D:/HSG_research/SpecialneedsHSG/Text/Graphs/ATEresults/results_main_FULL-eps-converted-to.pdf}
	\footnotesize
	\emph{Notes:} This figure depicts pairwise treatment effects for Special Education programs in St.\ Gallen. Each pair compares interventions that are the closest in degree of severity and inclusion. Each pairwise treatment effect is the effect of being assigned to the first program in comparison to the second program on one of the four outcomes presented in the panel headers. Both the treatment effect on the whole population (ATE) and on the population of the treated (ATET) are presented. ``Ind. therapy'' is the abbreviation for individual therapies, ``Acad. support'' for academic support, and ``no SpEd'' for receiving no program. Nuisance parameters are estimated using an ensemble learner that includes text representations presented in the ``data'' section. 95\% confidence intervals are represented and are based on one sample $t$-test for the ATE and the ATET. Test results and wages are standardized with mean 0 and standard deviation 1. The baseline  (``No SpEd'') probability of unemployment benefit recipiency is 0.19, and 0.07 for disability insurance recipiency. \emph{Source: SPS}.
	\end{minipage}
\end{figure} \fi

Results clearly show that returns to counseling are positive for academic performance. Students who receive counseling seem to fare better academically in comparison to those who do not receive any intervention but who exhibit similar difficulties and characteristics. This effect is four times the 0.1 standard deviation effect size criterion for successful interventions suggested by \citet{Bloom2006} and \citet{Schwartz2021}. Academic support offers no benefits but does not harm either. Results suggest that individual therapies are more effective than tutoring to improve academic performance. This is due most likely to the fact that students in individual therapies work alone with a trained therapist who can address the roots of their learning difficulties (for instance, dyscalculia or speech problems). Finally, students in inclusive intervention fare worse than students in individual therapies. The main explanation for this difference is that inclusion is designed to address clusters of learning, psychological, behavioral and social problems, whereas individual therapies tackle one particular (learning) disability. Dealing with multi-faceted issues might render inclusion less effective in terms of academic achievement. 

Long-term labor market effects are consistent with the effects on academic performance. As regards the probability of being unemployed and thus of benefiting from unemployment insurance, results show that counseling (-3 percentage points) and individual therapies (-10 percentage points) have a positive effect. Students benefiting from academic support are more likely to be unemployed than students with the same issues receiving no SpEd. A possible explanation for this negative labor integration effect is that these students needed support to succeed in school, support which is no longer provided once they enter the labor market. Individual therapies are more effective than tutoring in lowering unemployment probability (-10 p.p.), and are almost as effective as inclusion. Most pairwise effects indicate that intenser programs lead to lower probabilities of benefiting from disability insurance, with individual therapies showing the strongest reduction in disability insurance recipiency. The difference between inclusion and individual therapies in terms of disability insurance recipiency is small (0.8 percentage points for the ATE, 0 for the ATET). Finally, effects in wage returns remain small, as most pairwise effects are kept within the 0.1 standard deviation effect size criterion for successful SpEd interventions. Only individual therapies increase expected wage returns by almost 0.15 standard deviations in comparison to tutoring. 

\Cref{fig:results.SW8.nontrimmed} in the Appendix shows that more severe interventions in the inclusive setting slightly increase the probability of taking the Stellwerk8 test. Although the Stellwerk8 test does not indicate graduation \emph{per se}, these results back the idea that SpEd interventions slightly increase the probability of attending high-stake tests close to the age of graduation, which brings nuance to the findings of \citet{Schwartz2021}, \citet{McGee2011} or \citet{KIRJAVAINEN201633}. Moreover, \Cref{fig:results.SW8.nontrimmed} shows that more serious interventions in inclusive settings are as effective as more benign interventions in ensuring that students with SEN take the test.

Finally, in most pairwise treatment effects, the ATE does not differ significantly from the ATET, which suggests that effects for the population of the treated are consistent with effects for the whole population. Noticeable differences between the ATE and the ATET persist for the pairwise comparisons that involve the ``no treatment'' category in the case of disability insurance. For these comparisons, the population of students who either receive counseling or academic support are more positively affected than the whole population by the interventions.

\subsection{Returns to Special Education programs in segregated school settings}

I now pay closer attention to returns to inclusion and segregation. I first compare inclusion and semi-segregation, which are two SpEd programs that are considered as close substitutes in St.\ Gallen. Second, I compare semi-segregation with full segregation. Results presented in \Cref{fig:results.ISFKKSON} speak in favor of inclusive measures when it comes to improving academic performance: students in inclusive settings perform on average 0.6 test score standard deviations better than students sent to semi-segregation. The most commonly given explanation for the success of inclusion in the literature is that mainstreaming enhances the performance of students with SEN due to a more stimulating and demanding environment \citep[e.g.,][]{Cole2004,Daniel1997,Peetsma2001}. 

As regards labor participation, students sent to semi-segregation have a 10 percentage point higher probability to become unemployed than SEN students kept in the mainstream classroom. If students in inclusive settings have an average probability of being unemployed of around 11\%, this probability reaches around 20\% for students in semi-segregation. These findings confirm findings from \citet{Eckhart2011}, who mention the lack of social network of segregated students as a plausible reason for lower employment. Moreover, these results might be explained by the fact that semi-segregation is attached to a signaling penalty, i.e. semi-segregation results in an irregular degree that considerably reduces access to regular VET programs. This is even more striking as lower employment by semi-segregated students comes exclusively from unemployment insurance and not from disability insurance. Estimates show a significant wage gap: students placed in semi-segregated settings earn on average 0.15 standard deviations less than students placed in the main classroom.

\if\figuresintext0
\begin{center}
  [Insert \Cref{fig:results.ISFKKSON} here]
\end{center}\else
\begin{figure}[t!]
	\caption{Pairwise returns to segregated SpEd programs. \label{fig:results.ISFKKSON}}
	\begin{minipage}{\linewidth}
	\includegraphics[width=\textwidth]{D:/HSG_research/SpecialneedsHSG/Text/Graphs/ATEresults/results_ISFKKSON_FULL.png}
	\footnotesize
	\emph{Notes:} This figure depicts relevant pairwise treatment effects for Special Education programs in St.\ Gallen. Each pairwise treatment effect is the effect of being assigned to the first program in comparison to the second program on one of the four outcomes presented in the panel headers. Both the treatment effect on the whole population (ATE) and on the population of the treated (ATET) are presented. ``Semi-segr.'' is the abbreviation for semi-segregation (segregation in small classes), and ``Full segr.'' stands for full segregation (in special schools). Nuisance parameters are estimated using an ensemble learner that includes text representations presented in the ``data'' section. 95\% confidence intervals are represented and are based on one sample $t$-test for the ATE and the ATET. Test results and wages are standardized with mean 0 and standard deviation 1. The baseline  (``Inclusion'') probability of unemployment benefit recipiency is 0.19, and 0.04 for disability insurance recipiency. \emph{Source: SPS}.
	\end{minipage}
\end{figure}\fi

Returns to full segregation in comparison to semi-segregation are grim: the seemingly positive returns to full segregation in terms of academic performance are mostly due to selection into test participation (see \Cref{fig:results.SW8.nontrimmed}). Since, in the Canton of St.\ Gallen, only students in segregated schooling environments are allowed to opt out of the mandatory test, SEN students in full segregation are between 20 to 30 percentage points less likely to take the SW8 test than students in semi-segregation. Contrary to the case of semi-segregation, results for full segregation correspond to the findings of \citet{McGee2011} and \citet{KIRJAVAINEN201633}, who found a lower graduation rate for students in SpEd. This emphasizes the importance of making a distinction between semi-segregation and full segregation when assessing the effects of segregation in general.

SEN students assigned to segregated schooling have a significantly higher probability of benefiting from disability insurance, but not of becoming unemployed. Interestingly, the channels of (lack of) labor market participation are different for semi-segregated and fully segregated students: whereas students in small classes are more likely to be unemployed, students in special schools are likely to end up on disability insurance. This is not surprising in light of the information extracted from the psychological records. Records strongly suggests that students in full segregation are students with a higher probability of suffering from motor disabilities, developmental problems as well as speech and language problems (see \Cref{fig:diagnosisdistr}). Students with SEN are less likely to be sent to full segregation for learning disabilities. As mentioned in the identification section, results about full segregation should be interpreted with caution, as overlap is difficult to obtain for students in full segregation.

\subsection{Returns to Special Education interventions against no intervention}

Most of the literature evaluating SpEd programs focuses on returns to SpEd as a single intervention, without distinguishing between SpEd programs. My results clearly show that SpEd cannot be considered as a single intervention, and that each program brings different returns. In order to compare my results to results presented in the existing literature, I now compare the effect of each program with receiving no program at all. Even though all effects take into account observed confounding from covariates and text, it is clear that the ATE and ATET between receiving an intervention and not receiving one becomes more and more difficult to identify as interventions become more severe (lack of overlap). I therefore remain cautious when interpreting the effect of receiving no intervention with receiving more intensive interventions such as semi-segregation. 

\if\figuresintext0
\begin{center}
  [Insert \Cref{fig:results.NO} here]
\end{center}
\else
\begin{figure}[t!]
\caption{Pairwise returns to Special Education vs. No Special Education. \label{fig:results.NO}}
	\begin{minipage}{\linewidth}
	\includegraphics[width=\textwidth]{D:/HSG_research/SpecialneedsHSG/Text/Graphs/ATEresults/results_main_NO_FULL-eps-converted-to.pdf}
	\footnotesize
	\emph{Notes:} This figure depicts relevant pairwise treatment effects for Special Education programs in St.\ Gallen. Each pairwise treatment effect is the effect of being assigned to the first program in comparison to being assigned to no program on one of the four outcomes presented in the panel headers. Both the treatment effect on the whole population (ATE) and on the population of the treated (ATET) are presented. ``Ind. therapy'' is the abbreviation for individual therapies, ``Acad. support'' for academic support, ``no SpEd'' for receiving no program, ``Semi-segr.'' for semi-segregation (segregation in small classes), and ``Full segr.'' for full segregation (in special schools). Nuisance parameters are estimated using an ensemble learner that includes text representations presented in the ``data'' section. 95\% confidence intervals are represented and are based on one sample $t$-test for the ATE and the ATET. \emph{Source: SPS}.
	\end{minipage} \end{figure}\fi

\Cref{fig:results.NO} presents main results and shows intervention effects from the least intensive interventions (left) to the most intensive interventions (right). I represent again the first two pairwise effects for sake of comparison. Counseling and individual therapies have positive academic returns (their effects exceed the threshold of 0.1 standard deviations in test score). Academic support and inclusion bring almost zero returns (for inclusion, I measure -0.0878 for the ATE and -0.121 for the ATET). Students with SEN in these four programs catch up on the labor market: they are less likely to be unemployed (\Cref{fig:results.NO}b.), are less likely to end up on disability insurance (\Cref{fig:results.NO}c.), and earn as much or even more than students having received no SpEd (\Cref{fig:results.NO}d.). Only academic support does not improve chances of labor market integration.

Effects of segregation are generally negative. Segregated interventions have negative academic returns (the negative effects of full segregation are due to attrition into test taking), and generate negative labor market outcomes. Students with SEN in semi-segregated settings have a higher probability of becoming unemployed (around 10 percentage points), but not of receiving DI. However, returns to semi-segregation in terms of wages are similar to receiving no SpEd at all. In contrast, students with SEN assigned to segregated schooling have a significantly higher probability of benefiting from disability insurance, but not of becoming unemployed.

\subsection{Inclusion and semi-segregation: who benefits from segregation?}
Is inclusion always preferable to segregation for students with SEN? Despite the (political) decision to implement inclusive SpEd rather than segregated SpEd, a nuanced analysis about which students might still benefit from segregation is lacking. In this section, I explore whether average effects of inclusion compared to semi-segregation hides treatment effect heterogeneity, and whether semi-segregation might be beneficial for some students.

I first investigate Individual Average Treatment Effects (IATEs) for the effect of inclusion in comparison to semi-segregation. IATEs give the treatment effects at the most granular level and are useful to identify students with SEN who benefit the most from each treatment assignment. \Cref{fig:distr.IATE} reports the smoothed distribution of IATEs per outcome with the light horizontal bars depicting the first and fifth quintiles of the IATE distribution.\footnote{Note that some predicted IATEs have very large values, especially for the test scores. This is due to the fact that the DR-learner is weighted by the inverse of the propensity scores, which do not sum to one in finite samples. Although I am not concerned with extreme values in this case, I computed the normalized DR-learner of \citet{Knaus2020}, which mitigates this problem. Results are very similar, and are available upon request.} The distributions of IATEs are quite spread out, indicating large heterogeneities in responses to semi-segregation in comparison to inclusion across students with SEN. This is specially salient for IATEs in academic performance. For all outcomes, some students are shown to be indifferent between semi-segregation and inclusion, or even to benefit from semi-segregation. 

\if\figuresintext0
\begin{center}
  [Insert \Cref{fig:distr.IATE} and \Cref{table:IATE} here]
\end{center}
\else
\begin{figure}[t!]
\centering
\caption{Distributions of IATEs for segregation vs. inclusion. \label{fig:distr.IATE}}
\begin{minipage}{\linewidth}
	\includegraphics[width=0.9\textwidth]{D: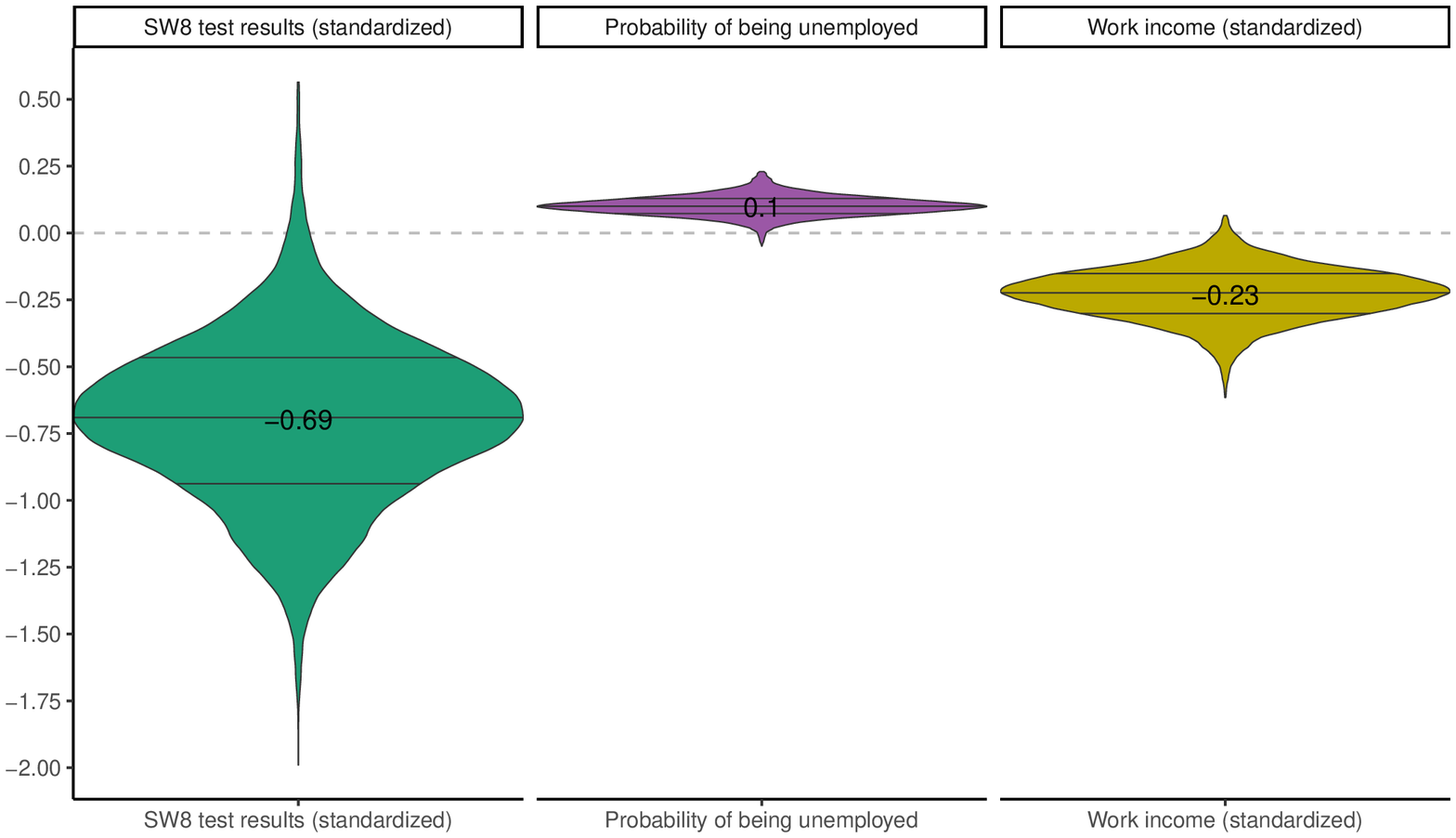}

	\footnotesize
	\emph{Notes:} This graph represents the smoothed distribution of IATEs for the four outcomes of interest. The IATEs were predicted out-of-sample using the DR-learner presented in \Cref{equation:AIPW_DR}. The average effect is reported in the figure. Light quintile bars represent the median and the 1st. and 5th. quintiles of the IATE distribution.
\end{minipage}
\end{figure}

%\newgeometry{left=1.5cm, right = 1.5cm }
\begin{table}[t]
\scriptsize
\centering
\begin{tabular}{lccccccccc}
  \toprule
                                              & \multicolumn{3}{c}{\textbf{Test scores}}         & \multicolumn{3}{c}{\textbf{Unemployment Probability}}   & \multicolumn{3}{c}{\textbf{Wage}}           \\    
                                              \cmidrule(lr){2-4}\cmidrule(lr){5-7} \cmidrule(lr){8-10}
                                              & Quint. I. & Quint. V. &SMD                    & Quint. V. & Quint. I. &SMD                    & Quint I. & Quint. V. &SMD  \\
  \midrule
\multicolumn{10}{l}{\textbf{Main covariates}} \\[1ex]
Female                                         &0.45           &0.38           &0.14             & 0.41       & 0.39     & 0.04                     & 0.48      & 0.35   & \textbf{0.28}     \\
Nonnative                                      &0.07           &0.34           &\textbf{0.73}    & 0.06       & 0.22     & \textbf{0.48}            & 0.04      & 0.21   & \textbf{0.55}     \\
IQ                                             &93.41          &97.20          &\textbf{0.34}    & 94.66      & 94.82    & 0.01                     & 95.82     & 93.66  & 0.18     \\
Age referral                                   &8.00           &8.39           &0.19             & 10.06      & 8.61     & \textbf{0.63}            & 10.06     & 8.40   & \textbf{0.75}     \\
Referral: social/emotional problems            &0.13           &0.37           &\textbf{0.56}  & 0.13            & 0.28       &\textbf{0.39}     & 0.13             & 0.27      & \textbf{0.34}     \\
Referral: performance/learning problems        &0.87           &0.90           &0.09           & 0.91            & 0.79       &\textbf{0.35}     & 0.85             & 0.84      & 0.03     \\
Referral: conflict with teacher                &0.02           &0.05           &\textbf{0.21}  & 0.01            & 0.04       & 0.141            & 0.01             & 0.03      & 0.11     \\
Need psychological treatment                   &0.13           &0.26           &\textbf{0.31}    & 0.14       & 0.24     & \textbf{0.27}            & 0.11      & 0.25   & \textbf{0.35}     \\
Nonnative $\times$ Female                      &0.04           &0.13           &\textbf{0.34}    & 0.03       & 0.08     & \textbf{0.22}            & 0.02      & 0.08   & \textbf{0.28}     \\
Nonnative $\times$ Social/emotional              &0.01           &0.11           &\textbf{0.43}    & 0.01       & 0.06     & \textbf{0.31}            & 0.00      & 0.06   & \textbf{0.33}     \\[1ex]

\bottomrule

\end{tabular}
\caption{Classification analysis of IATEs for semi-segregation vs inclusion}
\floatfoot{\emph{Notes:} This table shows the mean of each covariate and the standardized mean differences (SMD) across both SE programs between the fifth and the first quintile of the respective estimated IATE distribution. The first quintile refers to the most negative effects, whereas the fifth quintile to the most positive effects. For the probability of unemployment, the first and fifth quintiles are reversed (as students in the fifth quintiles suffer the most from semi-segregation). For two SE programs $w$ and $w'$, SMDs are computed as $\frac{\bar{x}_w - \bar{x}_{w'}}{\frac{\sqrt{s^{2}_{w} + s^{2}_{w'}}}{2}}$, where $\bar{x}_w$ is the mean of the covariate in treatment group $w$ and $s^2_w$ is the sample variance of covariate in treatment group $w$.  SMDs higher than 0.2 are depicted in bold.} 
\label{table:IATE}
\end{table}
%\restoregeometry
\fi

%\if\figuresintext1
%\newgeometry{left=1.5cm, right = 1.5cm }
%\input{"D:/HSG_research/SpecialneedsHSG/Text/Tables/IATE.quantile.tex"}
%\restoregeometry
%\fi

To have an idea of which individual characteristics are the most predictive of treatment effect size, I perform a classification analysis in the spirit of \citet{Chernozhukov2018a}. I group the predicted IATEs into quintiles and compare the standardized means difference (SMD) of covariates for students with the 20\% highest IATEs (fifth quintile) and students with the 20\% lowest IATEs (first quintile). For the probability of unemployment, the first and fifth quintiles are reversed in the table, as students in the fifth quintile suffer the most from semi-segregation. SMDs that are larger than 0.2 are considered to be large \citep{Rosenbaum1985}. I report covariates for which the standardized mean difference is higher than 0.2 in at least one of the treatment effects. Note that I do not report results for disability insurance, as average effects are almost zero. 

\Cref{table:IATE} shows the characteristics of students in the lower and higher tails of the IATE distributions according to main covariates. Students in the highest IATE quintile for academic performance are more likely to be nonnative students referred for social and emotional issues. Students in need of psychological support are also more likely to benefit from semi-segregation for academic performance. This particular population of students with SEN is also more likely to benefit from inclusion in terms of employment and wages. Finally, results clearly show that the age at referral is important for labor market outcomes: students that are referred earlier to the SPS clearly better benefit from segregation in terms of labor market integration. It is important to note that gender alone is not a predictive characteristic of different individual effects, the exception being that female students can expect worse wage outcomes as a result of semi-segregation.

\subsubsection{Group heterogeneity in the effect of semi-segregation: the disruption hypothesis}
An argument in favor of semi-segregation is that it attenuates disruption in the main classroom by removing ``disruptive'' students from the mainstream environment. However, the question whether semi-segregation is more beneficial than inclusion for ``disruptive'' peers has not yet been answered in the literature. Disruptive students are students who disturb their classmates and need additional teacher time and attention \citep[see][]{Lazear2001,CarrellEtal2018}. They might benefit from a segregated environment, which offers them increased teacher time and the right monitoring to focus on academic tasks. In this section, I explore whether returns to inclusion and semi-segregation systematically differ along pretreatment characteristics that potentially reveal disruptive behaviors: gender, nonnative speaking, whether the student has been referred for behavioral problems, and interaction between these groups. Disruptive behaviors are known to be prevalent in male students \citep{Bertrand2013,Lavy2011}, students with behavioral problems \citep{Fletcher2009}, or nonnative speakers \citep{Diette2014,Cho2012}. %In addition, I look at treatment heterogeneity along IQ scores.

\if\figuresintext0
\begin{center}
  [Insert \Cref{fig:GATE_covariates} here]
\end{center}
\else
\begin{figure}[t!]
\includegraphics[width=1\textwidth]{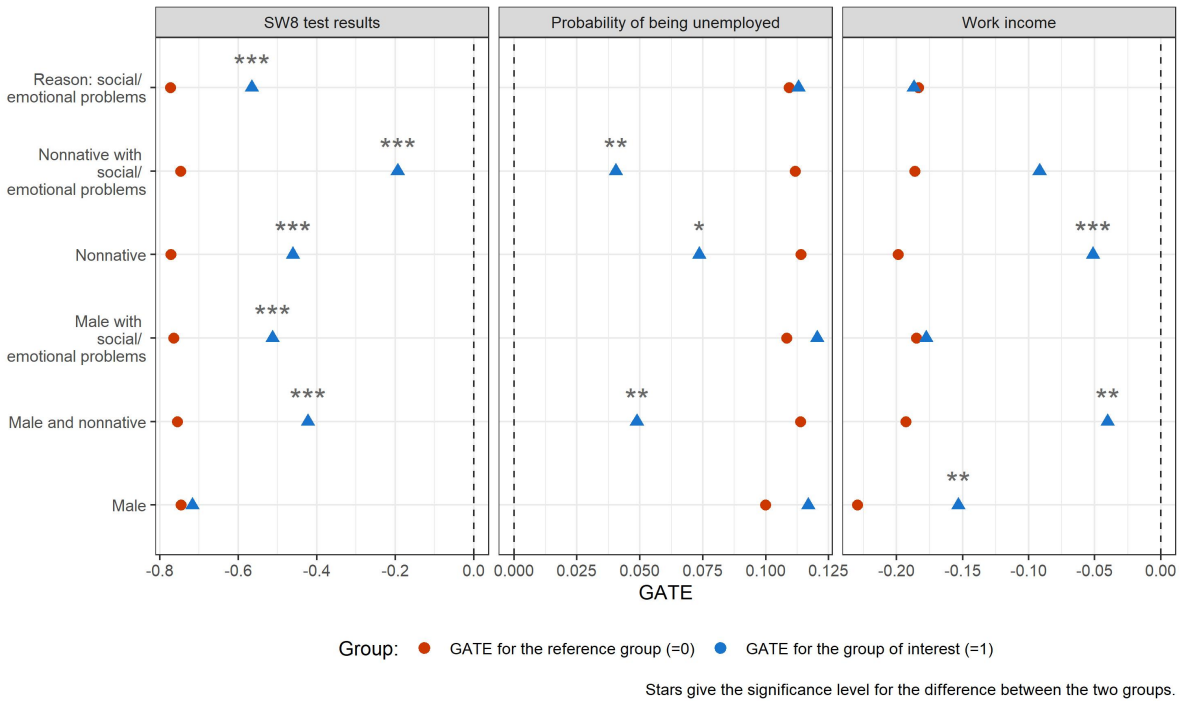}
\caption{GATEs for semi-segregation vs inclusion}
\floatfoot{\emph{Notes:} This graph presents estimated GATEs for the treatment effect of semi-segregation vs. inclusion. The GATEs are represented on the $x$ axis. Red dots indicate the treatment effect for the reference category (those students who do not belong to the category of interest), and blue dots indicate the treatment effect for the category of interest. The stars indicate the statistical significance of the difference between the two groups. For instance, the first row of the first column shows that the treatment effect of semi-segregation vs. inclusion is -0.75 test score standard deviations for students without social or emotional problems. The treatment effect is -0.55 for students with social or emotional problems. The difference in treatment effect between both groups is statistically significant ($p<0.01$). \label{fig:GATE_covariates}}
\end{figure}
\fi

Findings about the ``disruptiveness'' hypothesis are presented in \Cref{fig:GATE_covariates}. Each row of the figure gives the results of a regression where the DR score for the ATE is regressed on the group dummy. The GATEs are represented on the $x$ axis. Red dots indicate the treatment effect for the reference category (those students who do not belong to the category of interest), and blue dots indicate the treatment effect for the category of interest. The stars indicate the statistical significance of the difference between the two groups. For instance, the first row of the first column shows that the treatment effect of semi-segregation vs. inclusion is -0.75 test score standard deviations for students without social or emotional problems. The treatment effect is around -0.55 for students with social or emotional problems. The difference in treatment effect between both groups is statistically significant ($p<0.01$). 

From this analysis, two main conclusions can be drawn. First, heterogeneity in effects along disruptive characteristics are important for school performance, and also for long-term outcomes. Major effect differences between inclusion and semi-segregation can be found for male nonnative students with social or emotional problems. Second, results of the GATE analysis clearly show that ``disruptive'' students tend to benefit more from semi-segregation than non-disruptive students. However, my analysis does not show that ``disruptive'' students would perform better in semi-segregation settings: they would still be better off in inclusive settings. In particular, while semi-segregation negatively impacts SEN students on average, three particular groups of SEN students seem to have systematically higher GATEs: nonnative speakers, students with social and emotional problems, and male students. Any subgroups of students among these three groups exhibit GATEs that are higher than the ATEs. For instance, the effect on test scores of semi-segregation in comparison to inclusion is 0.3 standard deviations higher for nonnative speakers than for native speakers. Nonnative speakers are also less likely to be unemployed when segregated than native speakers, and they expect higher wages. When segregated, they expect a wage premium of .15 wage standard deviations higher than for natives, making semi-segregation as good as inclusion in terms of expected wages. The subgroup that would see the smallest difference between inclusion and semi-segregation are nonnative students with emotional or behavioral problems, who would perform only .19 test score standard deviations less in segregation than in inclusion (and almost 0.6 standard deviations better than other students in semi-segregation). All in all, even though inclusion remains on average better for students with SEN, even for those with ``disruptive'' characteristics, my results show that ``disruptive'' students are the students who benefit the most from semi-segregation.

\subsubsection{Do the effects of semi-segregation vary with school characteristics?}
School characteristics, especially the school socio-economic composition (SES score), could potentially drive the effect of inclusive and semi-segregated programs. To investigate this mechanism, I explore treatment heterogeneity in school characteristics for inclusion and semi-segregation. To estimate CATEs with respect to school characteristics, I regress the continuous school characteristics on the DR score for the ATE following \citet{Zimmert2019} as explained in \Cref{section:estimation}. I present results for the CATE for academic test score and for the probability to be unemployed in \Cref{fig:CATE_SCHOOLS_score} and \Cref{fig:CATE_SCHOOLS_ub}. I do not report per-student spending in my heterogeneity analysis given that it has been measured only once in 2017 and would not reflect schools' change of SpEd policies across the years. 

Results show an interesting tendency in the effects of inclusion in comparison to semi-segregation. Negative effects in school performance get closer to 0 as school size increases, and also as the schools' SES score and the share of nonnative students increase. For instance, the negative effects of semi-segregation become 0.2 standard deviations smaller when the school has a high SES score (meaning that the school's population is of a lower average SES status). The magnitude of this effect is rather substantial, and would indicate that students in schools that have a less homogeneous population (mostly in urban schools) and with more lower SES students are not harmed as much if they are segregated. The same picture is reflected in the effects for the probability to be unemployed: the negative effects of semi-segregation become almost zero in schools with a high number of foreign language speaking students, and with a higher SES score. 

Thus, the negative effects of semi-segregation are shown to slightly fade out for schools that are more socially and economically diverse.  One potential explanation for this phenomenon is that the average level of disruptiveness in schools with more diversity would make semi-segregation comparatively more effective for disruptive students. Another possible explanation is that inclusion works better when teachers are less overwhelmed, i.e. in more homogeneous classrooms. These findings indirectly corroborate teachers' worries that inclusion is difficult to implement when the population of students is already difficult to deal with in the main classroom.\footnote{Such arguments have been expressed by teachers: inclusion is a factor of teachers' overload, incompatibility between non-SEN and included SEN students, and a lack of efficiency for SEN students. Newspaper articles from the \emph{Tagblatt} newspaper in St.\ Gallen regularly refer to the problem. For instance \mbox{\href{https://www.tagblatt.ch/ostschweiz/wenn-die-integration-die-lehrer-zerreisst-ld.1120235}{\emph{``How inclusion divide teachers''}}, \href{https://www.tagblatt.ch/ostschweiz/stgallen-gossau-rorschach/foerderung-braucht-ressourcen-ld.379372}{\emph{``Special education needs resources''}}, \href{https://www.tagblatt.ch/ostschweiz/stgallen-gossau-rorschach/integrieren-statt-separieren-ld.387848}{\emph{``Include instead of segregate''}}}, \mbox{\href{https://www.tagblatt.ch/ostschweiz/zusammen-unterrichten-was-nicht-zusammengehoert-ld.1120241}{\emph{``Teaching the same to students who are not the same''}}, \href{https://www.tagblatt.ch/ostschweiz/das-integrative-schulmodell-bringt-lehrer-an-ihre-grenzen-verbaende-fordern-kleinklassen-bildungsforscher-raten-davon-ab-ld.1122379}{\emph{``The integrative model puts teachers to their limits''}.}}}   

In conclusion, my analysis shows that semi-segregation programs are less effective than inclusion in terms of academic performance and labor market integration for (almost) all students with SEN. However, students who exhibit ``disruptive'' characteristics, i.e. male students, students with social and emotional problems, and nonnative speakers, are the students who benefit most from semi-segregation. Importantly, the success of inclusion or semi-segregation does not depend on IQ, gender or the prevalence of learning disabilities: this is especially true for school performance, but also extends to labor market integration. Finally, the schools' average SES status and population of nonnative students has non-negligible influence on the effect of inclusion and semi-segregation SpEd programs.

\subsection{Further analyses and robustness checks}
I conduct a battery of robustness tests in Appendix \ref{appendix:robustness}.  In Appendix \ref{appendix:robustness.1}, I account for potential overlap and lack of common support problems in the generalized propensity score distribution and implement overlap-weighted average treatment effects \citep{Li2018,Li2019} as well as different trimming schemes \Citep[see][]{Crump2009,Sturmer2010}. I find that point estimates do not vary much when extreme weights are trimmed, and become less precisely estimated. 

In Appendix \ref{appendix:robustness.2}, I investigate how sensitive my estimates are to the inclusion of text covariates in order to see how much confounding my text variables are able to remove. I find that estimates based on both covariates and text information are on average 29\% smaller than estimates that do not leverage the text information. Furthermore, I account for the possibility that my text representations capture the psychologist's biases (towards a certain treatment or a certain writing style) rather than information on the student. I show that my estimates remain consistent with my main findings when I strip out psychologists' effects from the text. 

In Appendix \ref{appendix:robustness.3}, I tackle the problem of potential selective attrition in the measured outcomes. I conduct an attrition analysis by showing the results for the cohorts that are in both the SW8 and the SSA subsamples. Results of this check are in line with main results.

Finally, even though I observe the assignment process in its entirety through written reports, some factors influencing SpEd placements might still remain unobserved. I address this issue by leveraging the within year across school variation in supply of SpEd as an instrument to compare the outcomes of SEN students in inclusive settings with similar peers in semi-segregated settings \citep[in the spirit of ][]{Keslair2012}. In Appendix \ref{appendix:subsection_IV}, I present Local Average Treatment Effects of inclusion on students who would have been segregated, had they lived in a school that implemented semi-segregated SpEd programs. LATE results corroborate my main findings. 

\section{Policy learning: optimal SpEd placement for inclusion and semi-segregation\label{section:policy}}
How would a school psychologist or a policy maker assign SEN students to the program that corresponds the best to their particular characteristics? Getting insights from the literature on statistical treatment rules \citep[e.g.,][]{Kitagawa2018,Manski2004}, I perform optimal treatment allocations simulations based on tree-search based algorithms \Citep{Athey2020,Zhou2018}. I leverage my rich set of covariates as well as the information retrieved from the psychological records to look at whether policy makers might be able to better tailor policies on the basis of observed individual characteristics. I focus on allocation to inclusion and semi-segregation, two programs that are used as quasi-substitutes in St.\ Gallen.

Let $\pi(Z_i)$ be a policy rule that leverages a relatively small number of observed individual characteristics of interest $Z_i$ to assign individuals to a treatment $d$. The number of characteristics in $Z_i$ is small because it contains variables that can be easily interpreted and used by a policy maker. For each policy rule $\pi(Z_i)$ among all candidate policy rules $\Pi$, a policy value summarizes the estimated population potential outcome attained under the policy. The policy value is the average of the individual APOs (defined by \Cref{equation:AIPW_DR}) under the policy rule $\hat{Q}(\pi) = \frac{1}{N}\sum_{d=1}^D\sum_{i = 1}^N \text{\underline{1}}(\pi(Z_i)= d)\hat{\Gamma}_i^d$. The goal of the policy maker is to find the optimal policy, i.e. the particular policy rule among all possible policy rules that maximizes the policy value. This means finding the policy rule $\hat{\pi}^*$ that maximizes the policy value function $\hat{\pi}^* = $ arg max$_{\pi \in \Pi} \hat{Q}(\pi)$ among all candidate policies. Note that the goal of this exercise is not to discriminate students on the basis of  their characteristics, but to guide policy makers on how they could potentially improve policies by leveraging observed characteristics. 
%Let $\pi(Z_i)$ be a policy rule that leverages observed individual characteristics of interest $Z_i$ and assigns individuals to an optimal treatment $d$. The optimal treatment is defined as, for each individual, the treatment that maximizes the APO given $Z_i$, i.e. $E[Y_i^d|Z_i = z]$. For each policy rule $\pi(Z_i)$ among all candidate policy rules $\Pi$, the estimated population potential outcome that could be attained is summarized by the policy value function $\hat{Q}(\pi) = \frac{1}{N}\sum_{d=1}^D\sum_{i = 1}^N \text{\underline{1}}(\pi(Z_i)= d)\hat{\Gamma}_i^d$. This value is computed with the APO under each treatment defined by \Cref{equation:AIPW_DR}. The goal of the policy maker is to find the optimal allocation of SpEd programs such that the average potential outcome for all treated SEN students is maximized, i.e. by finding the policy rule $\pi^*$ that maximizes the policy value function $\hat{\pi}^* = $ arg max$_{\pi \in \Pi} \hat{Q}(\pi)$ among all candidate policies. Note that the goal of this exercise is not to discriminate students on the basis of  their characteristics, but to guide policy makers about how they potentially could improve policies. 

For instance, a policy maker could propose three candidate policy rules for optimal placements: (1) assign all SEN students to inclusion; (2) assign all SEN students to semi-segregation; (3) ``split'' students along their nonnative status, i.e. assign all nonnative speakers to semi-segregation and all other students to inclusion. The policy value of the first rule is the average of all APOs under inclusion, and similarly for the second rule under semi-segregation. For the third rule, the policy value would be the average over the APO under semi-segregation for all nonnative students and over the APO under inclusion for all native students. The policy maker would then choose the policy with the highest overall policy value.

To find the optimal policy, I compute the policy tree algorithm with fixed depth based on double machine learning of \Citet{Zhou2018}.\footnote{The algorithm finds the optimal policy such that it minimizes the regret function, i.e. the difference between the true and the estimated optimal policy value. In general, see algorithm 1 in \Citet{Zhou2018}.} Intuitively, trees (policy rules) split the sample along variables in $Z_i$ many times, and the split that gives the maximal policy value is given as the optimal policy. I experiment with policy trees of depth 2 and 3 (the depth indicates the number of times leaves are split) with two different sets of student attributes $Z$, i.e. the baseline covariates and covariates extracted from the diagnosis (dictionary approach). I compute optimal policy allocations for inclusion or semi-segregation on the subsample of students either sent to inclusion or segregation. As outcomes, I look at test scores and labor-market integration (probability of employment). I subsequently link optimal policy allocations to actual and counterfactual policy costs, which are estimated by taking average costs per student per year of each SpEd placement given by the Canton of St.\ Gallen.\footnote{For costs, I use the following estimates obtained from \emph{SG-Volksschulgesetznachtrag 2013}. A student in mainstreamed environment costs between 15'000 and 20'000 CHF (approximately 16'500 to 22'000 USD) on average per year, depending on the school and the grade. I take the highest estimate, namely 20'000 CHF. A student in semi-segregation costs on average 24'500 CHF (27'000 USD) per year. Individual, hour-long therapy SpEd programs costs on average 5000 CHF (5500 USD) per year. Schooling in full segregation settings costs between 39'000 and 260'000 CHF, on average between 70‘000 to 80'000 CHF (77'000 USD to 88'000 USD) per student per year.} I then compare changes in costs and changes in policy value for each optimal policy. 

\if\figuresintext0
\begin{center}
  [Insert \Cref{table:optimal_policy} here]
\end{center}
\else
\begin{table}[p!]
\footnotesize 
\begin{threeparttable}
\begin{tabular}{lccccc}
\multicolumn{3}{l}{\textbf{Panel A}: Allocation to program in percent and potential cost reduction} \\[1ex]

  \toprule
                                &\% Students    &\% Students        &Policy           &Costs per year      & Percent of   \\ 
                                &sent to        &sent to semi-      &value            &(in mm CHF)        & actual costs   \\ 
                                &inclusion      &segregation        &                 &                    &  \\
  \midrule
\multicolumn{3}{l}{\emph{\textbf{Test scores. N = 2988}}} \\[1ex]
\emph{Actual allocation}        &0.69          &0.31                &-0.46         &63,925           &1       \\
Depth 2 and baseline variables  &0.96          &0.04                &-0.29         &60,271           &0.94    \\
Depth 3 and baseline variables  &0.96          &0.04                &-0.27         &60,235           &0.94   \\
Depth 2 and diagnosis variables &0.96          &0.04                &-0.29         &59,762           &0.93   \\
Depth 3 and diagnosis variables &0.97          &0.03                &-0.27         &60,145           &0.94   \\[4ex]

\multicolumn{3}{l}{\emph{\textbf{Probability of employment. N = 2939}}} \\[1ex]
\emph{Actual allocation}      &0.53          &0.47                &0.67          &64,958           &1       \\
Depth 2 and baseline variables  &0.83          &0.17                &0.79          &61,007           &0.94   \\
Depth 3 and baseline variables  &0.87          &0.13                &0.80          &60,490           &0.93   \\
Depth 2 and diagnosis variables &0.87          &0.13                &0.79          &60,485           &0.93  \\
Depth 3 and diagnosis variables &0.85          &0.15                &0.81          &60,796           &0.94   \\
\bottomrule
\end{tabular}
\begin{tablenotes}[para,flushleft]
\scriptsize
\emph{Notes:} This table shows the treatment assignment from four different policies. The depth indicates the depth of the policy trees. The baseline variables are individual covariates excluding variables from text, and diagnosis variables are individual covariates together with covariates from the diagnoses extracted from the text (dictionary approach). The policy value is the average APO under each policy. Total cost estimates and potential cost reduction from the implemented policies are computed.
\end{tablenotes}
\end{threeparttable}

\vspace{10mm}

\begin{threeparttable}
\begin{tabular}{lccc}
\multicolumn{3}{l}{\textbf{Panel B}: Cross-validated difference between optimal policy value and different policies} \\[1ex]
\toprule
                                  & All inclusion           &All semi-segregation           &Assigned policy \\
\midrule
\multicolumn{3}{l}{\emph{\textbf{Test scores. N = 2988}}} \\[1ex]
Depth 2 and baseline variables    &-0.010$^{**}$             &0.651$^{***}$                 &0.484$^{***}$   \\  
                                  &(0.004)                   &(0.036)                       &(0.029)        \\
Depth 3 and baseline variables    &-0.009$^{*}$              &0.652$^{***}$                 &0.485$^{***}$   \\ 
                                  &(0.004)                   &(0.036)                       &(0.029)        \\
Depth 2 and diagnosis variables   &-0.009$^{**}$             &0.652$^{***}$                 &0.485$^{***}$   \\   
                                  &(0.004)                   &(0.036)                       &(0.029)        \\
Depth 3 and diagnosis variables   &-0.015$^{*}$              &0.646$^{***}$                 &0.478$^{***}$   \\   
                                  &(0.009)                   &(0.035)                       &(0.029)        \\[4ex]

\multicolumn{3}{l}{\emph{\textbf{Probability of no unemployment. N = 2939}}} \\[1ex]
Depth 2 and baseline variables    &-0.027$^{***}$           &0.110$^{***}$         &0.016     \\  
                                  &(0.008)                  &(0.037)               &(0.028)       \\
Depth 3 and baseline variables    &-0.031$^{**}$            &0.106$^{***}$         &0.012  \\ 
                                  &(0.013)                  &(0.035)               &(0.029)        \\
Depth 2 and diagnosis variables   &-0.021$^{**}$            &0.116$^{***}$         &0.022  \\   
                                  &(0.009)                  &(0.037)               &(0.028)        \\
Depth 3 and diagnosis variables   &-0.043$^{***}$           &0.094$^{***}$         &-0.000      \\
                                  &(0.014)                  &(0.035)               &(0.030)        \\

\bottomrule
\end{tabular}

\begin{tablenotes}[para,flushleft]
\scriptsize
\emph{Notes:} This table displays validation tests for policy trees. 10-fold cross-validation is used. Optimal policies are compared to either sending everyone to inclusion, to sending everyone to semi-segregation, or to the (already implemented) observed policy. For each policy, the average difference between the APO under the optimal policy and the APO under one of the three alternative policies is computed.  Inference is done with a one sample $t-$test on the difference. Standard deviations of one-sample $t$-test in parenthesis. $^{***}$: p $<$0.01, $^{**}$: p $<$ 0.05, $^{*}$: p $<$ 0.1.
\end{tablenotes}
\end{threeparttable}

\caption{Optimal policies for inclusion and semi-segregation} \label{table:optimal_policy}
\end{table}
\fi

Panel A of \Cref{table:optimal_policy} compares the actual observed percentage of students assigned to inclusion and to semi-segregation with the allocation of students resulting from optimal reallocation. Around 70\% of the students sent either to inclusion or semi-segregation who took the SW8 test were actually assigned to inclusion, and 53\% of students who were registered in the SSA dataset were assigned to inclusion. For academic performance, the policy value of the actual implemented policy is -0.46, and it is 0.67 for labor market integration (which gives the average employment probability under the implemented policy). I propose four reallocation policies for each outcome. All proposed policies dramatically improve the policy value by reallocating the majority of students to inclusion rather than segregation. All proposed policies would reduce overall costs, but not dramatically so (at around 94\% of actual realized cost). In general, proposed assignment schemes are very similar in terms of improved outcomes and reduced costs.\footnote{One could think that the remaining students sent to segregation are observations with extreme weights in their doubly robust score. I computed the same policy classification after trimming, and the policy rules would consistently send the same amount of students to segregation.} 

What are the rules that allow better allocation of students in terms of better school performance? \Cref{fig:optimalpolicytree_SW8} represents the tree policies for improved test scores. Interestingly, a simple policy tree of depth 2 would automatically assign all nonnative students with emotional or social issues to semi-segregation, as well as all gifted students (with IQ higher than 125) without emotional or social issues to semi-segregation. All other students with SEN would be assigned to inclusive settings. This makes sense since students with social problems are likely to be disruptive and might benefit from segregation, and since gifted students might benefit from ``pull-out'' programs. This policy would outperform the actual assignment by gaining an average of around 0.2 test score standard deviations.\footnote{Note that the variable ``IQ score'' is actually the interaction between the actual score and whether the IQ test has been administered, thus taking 0 when no IQ score exists. I perform a similar analysis with the subsample of students for whom the IQ is observed.} The policy tree of depth 3 with baseline characteristics confirms the importance of issues in emotional or social behaviors, problems with test performance, and IQ score as important variables for optimal policies. Policies using diagnoses are very similar to policies based solely on main covariates, which highlights the fact that diagnoses extracted from text are not very predictive of policy improvement.

Sending more students to inclusive settings has benefits for integration on the labor market. The second part of Panel A in \Cref{table:optimal_policy} shows that by sending less students to semi-segregation, the average employment probability can be increased by around 20 percentage points for quasi-similar policy costs. Interestingly, proposed policies that target better labor-market integration send a larger share of students to semi-segregation. \Cref{fig:optimalpolicytree_UBD} shows that semi-segregation is the most helpful for students with IQ scores lower than average, without performance or learning problems. In general, IQ seems to be the most predictive variable of success of semi-segregation with respect to labor-market integration. Students who were not given an IQ test would also be sent to semi-segregation. From the trees with diagnoses, we learn that children with ADHD would be sent to semi-segregation. The policy tree of depth 3 that leverages the information on ADHD is the most effective policy in terms of labor market integration. 

To test whether the decision trees are stable, I conduct validation tests inspired by \Citet{Zhou2018} and \Citet{Knaus2020}. I test whether the proposed policies perform better than either sending everyone to inclusion, sending everyone to semi-segregation, or implementing the (already implemented) observed policy (the ``Null-hypothesis'' policies). To do this, I use 10-fold cross-validation, i.e. I train the policy tree on the training subsample and use the tree to predict assignment on the left-out fold. I then compute the difference between the APO under the optimal policy and the APO under one of the three alternative policies, and compute a one sample $t-$test to assess whether the difference in means is significantly different than zero. Panel B in \Cref{table:optimal_policy} shows that all optimal reallocations outperform sending all students to semi-segregation for both outcomes. For academic performance, all policies outperform the implemented policies, but perform as well or even slightly worse than sending all students to inclusion. For labor-market integration, sending all students to inclusive settings marginally outperforms the optimal policies; however, all proposed policies fail to significantly improve on realized therapy assignments. This indicates that actual placement by school psychologists already leverages the available information in a relevant way, but that this placement could be improved to increase academic performance.

I implicitly assumed in my optimal allocation exercise that non-SEN students receive a welfare weight of 0. In reality, the reallocation of students with SEN from semi-segregation to the main classroom induce spillover effects that could negatively impact mainstreamed SEN and non-SEN students \citep[as shown by][]{BalestraEtal2020,Rangvid2019}. To measure overall welfare functions, I integrate spillover effects of reassigning students with SEN to the main classroom in my optimal policy rule. I proceed as follows: I merge my dataset with the data containing all students without SEN from the Canton of St. Gallen. I then estimate the peer effect of reassigning students with SEN to mainstream classrooms on the academic performance of their classmates (for classmates with SEN and for classmates without SEN). As students with SEN are randomly allocated to classrooms in St.\ Gallen, I am able to estimate causal spillover functions with flexible estimation procedures \citep[for more information on this quasi-experimental setting, see][]{BalestraEtal2021}.\footnote{I estimate spillovers with machine learning algorithms and an ensemble learner. I use clustered cross-validation to estimate functions at the classroom level.} The classroom spillover function is presented in \Cref{fig:SNvsNonSN} of the Appendix. The function is estimated for students with SEN (in brown) and for students without SEN (in blue). I find that the shape of the spillover function is monotonically decreasing, meaning that including an additional peer with SEN has negative effects on mainstreamed peers with SEN and on mainstreamed peers without SEN. The negative effect worsens with more students with SEN in the classroom. %General equilibrium. Does not take into account the fact that small classes would be smaller, and thus more effective ;-)

Based on my flexible estimation of the spillover function, and with a simplistic back-of-the-envelope utility computation, I compute the average reallocation effects induced by the optimal policy for the whole student population. I base my computations on the following statistics: students with SEN make up 25\% of the mainstreamed students population, mainstream classrooms have on average 19.17 students, and the data contains 2723 classrooms. As presented in Panel A of \Cref{table:optimal_policy}, optimal policies reallocate 807 students from semi-segregation to inclusion for a policy gain of 0.17. The reallocation of 807 students from semi-segregation to inclusion means that there are 0.3 additional students with SEN per classroom, thus increasing the average proportion of students with SEN per classroom from 0.25 to 0.266 (a 0.016 increase). The effect of reallocation for the reallocated students is therefore an average gain of $0.016 *.17 = 0.003$ in policy value per student in the whole population. For the population of mainstreamed students, the spillover functions in \Cref{fig:SNvsNonSN} (and tables available on request) show that the increase in the proportion of peers with SEN in the main classroom from 0.25 to 0.266 generates a loss of expected test score standard deviation of -0.04 for mainstreamed students with SEN, and a loss of -0.03 test score standard deviations for mainstreamed students without SEN. This means that the policy loss is on average around $0.25*(-0.04) + 0.75*(-0.03) =-0.032$ per student. Combining both effects (and ignoring the very small shift in group proportions after reallocation), the average loss in policy value after reallocation is around $-0.03$ standard deviations of test scores per student. 

All in all, this optimal policy exercise exercise delivers valuable insights into improved allocation of SEN students to semi-segregation and inclusion. It strengthens the idea that inclusion works well, as none of my suggested reallocation policies perform better than allocating all students to inclusion. There also seems to be a trade-off between short-term, academic benefits, and longer-term benefits: from the perspective of labor-market integration, it seems that a higher share of semi-segregated students is beneficial, whereas it is not from an academic performance perspective. Finally, it highlights the idea that less segregation is desirable and could be reached without imposing too much harm on students in mainstream classrooms: the individual gains for the newly reallocated students are large and by far offset the individual loss for mainstreamed students, even though the social reallocation effect is negative by $-0.03$ standard deviations of test scores per student.  In light of the results from the heterogeneity analysis and from the reallocation exercise, it might seem wise for policy makers to segregate disruptive students, as these students generate the largest negative spillover effects on their peers, and they are the ones benefiting the most from segregated environments.

\section{Conclusion\label{section:conclusion}}
The present study sheds light on short-term and long-term returns to SpEd programs for students with special needs. Using recent methodological developments in computational text analysis and in causal machine learning, this study leverages psychologists' written reports to address the problem of confounding in the absence of experimental design. My study complements the literature by showing that SpEd should not be considered as a single intervention. Each program differs in its expected returns, and inclusive programs are quite effective at generating academic success and labor market integration. More specifically, I find that returns to SpEd programs in inclusive settings (counseling and individual therapies) are positive for academic performance, and that tutoring has no measurable effect. When compared to receiving no SpEd, all inclusive treatments have zero to positive returns. 

Moreover, this study contributes to our understanding of inclusive and segregated measures for students with SEN. In general, inclusion pays off in terms of academic performance, labor market participation and earnings in comparison to semi-segregation. The results of this study show that inclusion works better for (almost) all students with SEN. However,  semi-segregation is the least detrimental for students with SEN who exhibit disruptive tendencies, or for nonnative students with SEN. Semi-segregation works best in schools with an lower average SES status and more nonnative speakers. Results presented in this study, however, do not extend to the analysis of full segregation, as students placed in fully segregated settings have almost no overlapping characteristics with students in semi-segregated and inclusive settings. Moreover, higher attrition and selection into test participation for students placed in fully segregated school environments make the assessment of academic returns difficult.  

The optimal policy allocations analyses offer further insights into inclusive and segregated SpEd programs for policy makers. With the help of policy trees, I propose placement recommendations to improve aggregate school performance and better labor market integration. These policies are outcome-improving for students with SEN, and cost-reducing. By implementing such optimal policies, a policy maker could significantly increase average school performance and labor market integration for students in semi-segregation by reallocating them to mainstream classrooms. However, this reallocation would incur costs on their peers in mainstream classrooms. In light of the results from the heterogeneity analysis and from the reallocation exercise, it might seem wise for policy makers to segregate disruptive students, as these students generate the largest negative spillover effects on their peers, and they are the ones benefiting the most from segregated environments. However, it is ultimately up to the policy maker and to society at large to decide whether a small decrease in school performance is a cost that should be endured by mainstream students in order to significantly increase the performance of reallocated students. 

This paper invites further research on two fronts. On the one hand, the role of teachers and SpEd teachers in decisions to implement inclusion or segregation must be further investigated. Teachers' input and teachers' productivity in inclusive and segregated settings are important factors of the success of inclusive policies. However, little is known about their value-added for students with SEN. On the other hand, inclusion has benefits that extend beyond measurable academic performance and labor market integration. For instance, inclusion is likely to affect non-cognitive skills such as altruism, self-esteem, and self-image. These aspects need to be investigated to deliver a more complete picture of inclusion. Finally, this paper is an invitation for further methodological research in optimal policy allocation and in the use of text as covariate.

\newpage
{\singlespacing
\bibliographystyle{jpe}
\bibliography{bibliography_1}}

\clearpage
\setcounter{table}{0}
\setcounter{figure}{0}
\setcounter{page}{1}
% !TeX root = Submission_ASallin_v9.tex

% Only if figuresintext is 1
\if\figuresintext0

\section{Figures and tables}

\vfill
  \begin{figure}[h!]
  \centering
    \caption{Share of Special Education placements among students referred to the School Psychological Service over the years \label{fig:kk.isf.year}}
    \begin{minipage}{\linewidth}
        \includegraphics[width=0.9\textwidth]{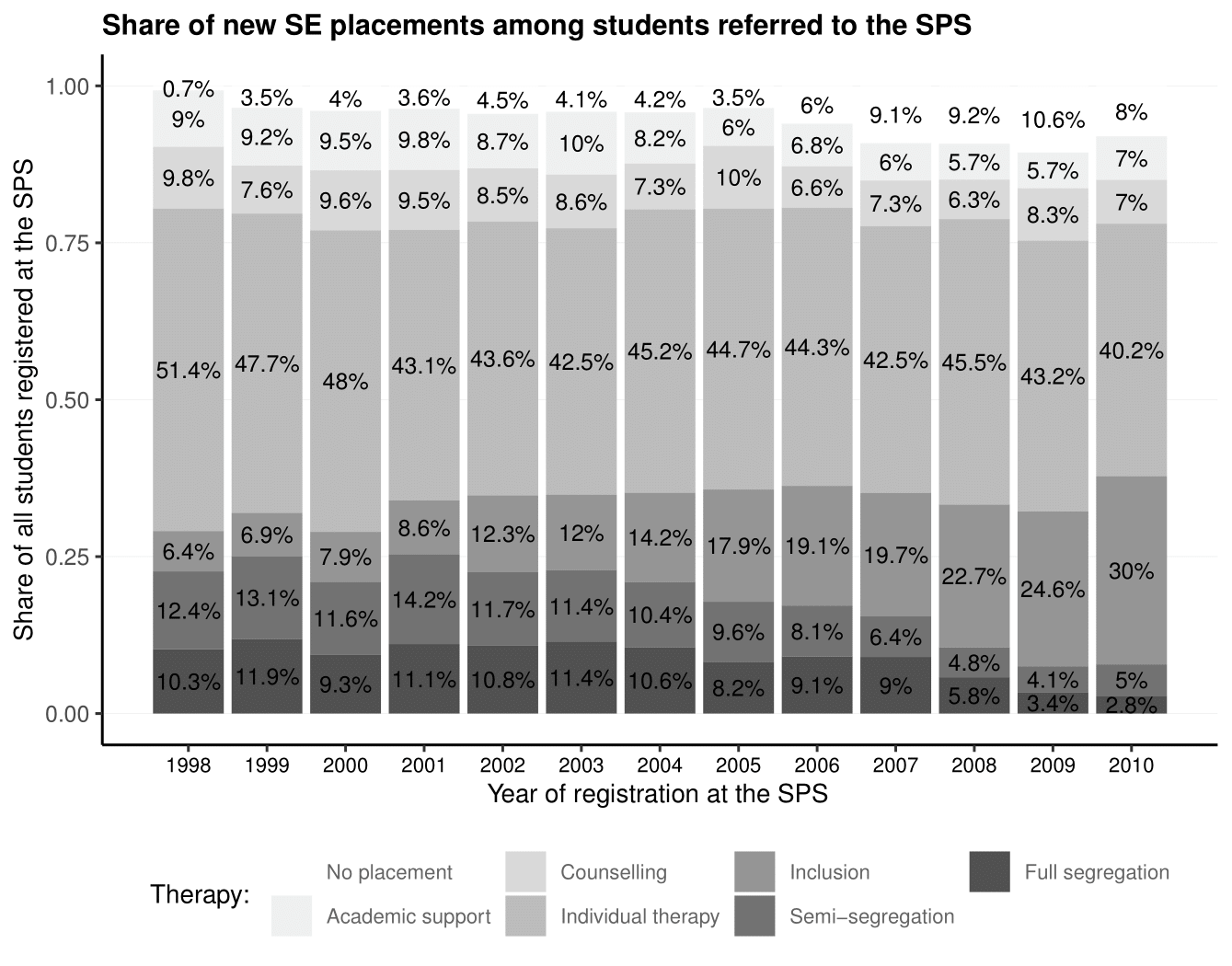}
        
        \footnotesize \emph{Notes:} This figure displays the newly assigned special education interventions per year. It gives the share of students assigned to a particular SpEd program among all students referred to the School Psychological Service over the years. \emph{Source: SPS}.
    \end{minipage}
  \end{figure}
\vfill

\clearpage
\begin{figure}
\centering
\caption{Distribution of school-years deviations in assignment to SpEd from mean year inclusion assignment rate \label{fig:IV_variation_school_year}}
\begin{minipage}{\linewidth}
    \includegraphics[width=0.9\textwidth]{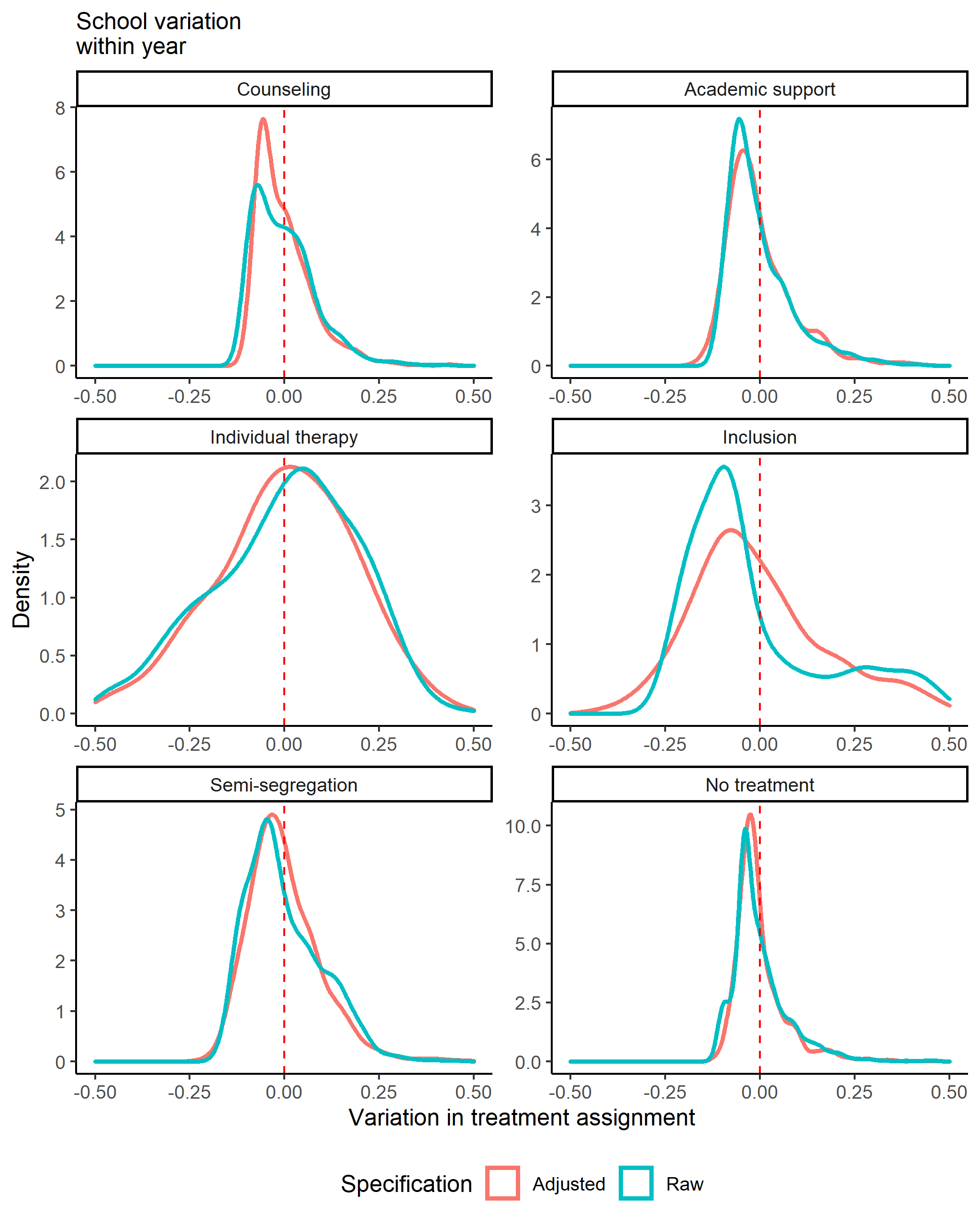}

    \footnotesize \emph{Notes:} This figure shows the distribution of deviations (residuals) in the assignment rate of students assigned to all SpEd programs per school-year from the mean year assignment rate for the population of students with SEN. Both the raw deviation and the regression-adjusted deviation are displayed. The adjusted deviation shows deviation adjusted for student-level and school-level covariates. Student-level covariates include gender, nonnative status, IQ, reason for referral, and who sent the student for referral (Panel A of \Cref{table:summary_stats}). School-level covariates are shown in Panel B of \Cref{table:summary_stats} and include share of nonnative speakers, share of students with SEN, school size, school socio-economic composition, and urban status. Full segregation is not represented. \emph{Source: Pensenpool, SPS}.
\end{minipage}
\end{figure} 

\clearpage
\begin{figure}
  \centering
  \caption{Visualization of the sample structure \label{fig:dataflow}}
  \begin{minipage}{\linewidth}
    \includegraphics[width=\textwidth]{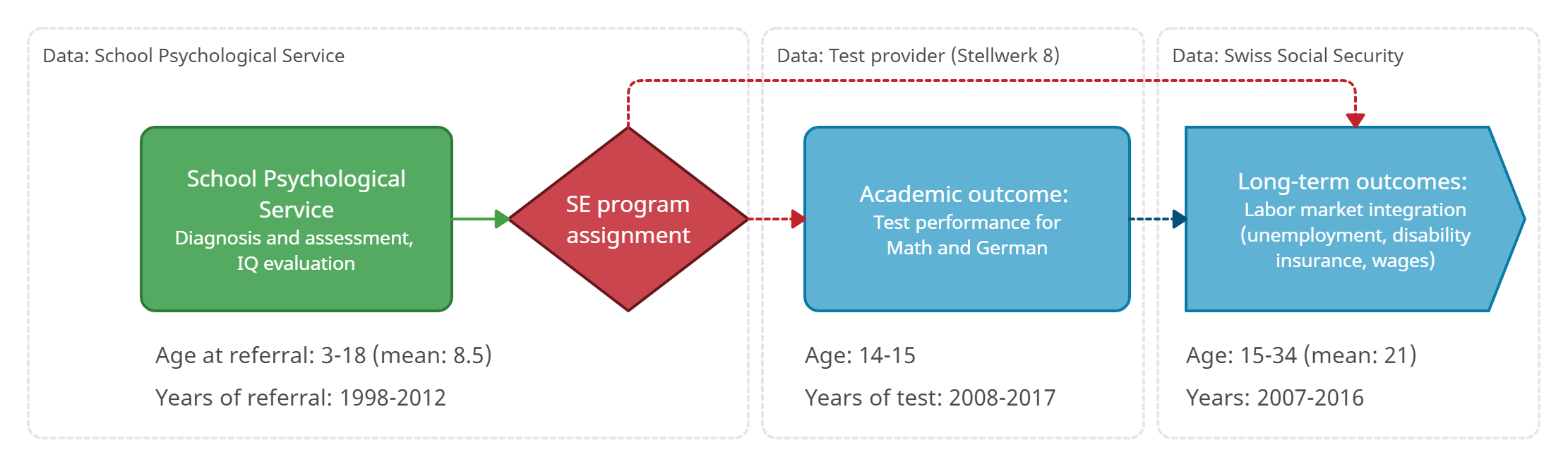} %made with creately.com
    \footnotesize
    \emph{Notes:} This figure presents the sample structure as a timeline. All students referred to the School Psychological Service (SPS) between years 1998 to 2012 are observed and receive a treatment. Students' academic performance is observed in the test data (``Stellwerk8'') for all students reaching the age of 14 or 15 in years 2008 to 2017. Labor market outcomes are observed in the Swiss Social Security Administration (SSA) data for students reaching the labor market in years 2007 to 2016. Because of attrition and the particular data structure, not all students are observed in both the Stellwerk8 and the SSA data (blue arrow).
  \end{minipage}
\end{figure} 

\clearpage
\begin{figure}
  \centering
  \caption{Workflow of Double Machine Learning and text analysis \label{fig:workflow}}
  \begin{minipage}{\linewidth}
    \includegraphics[width=\textwidth]{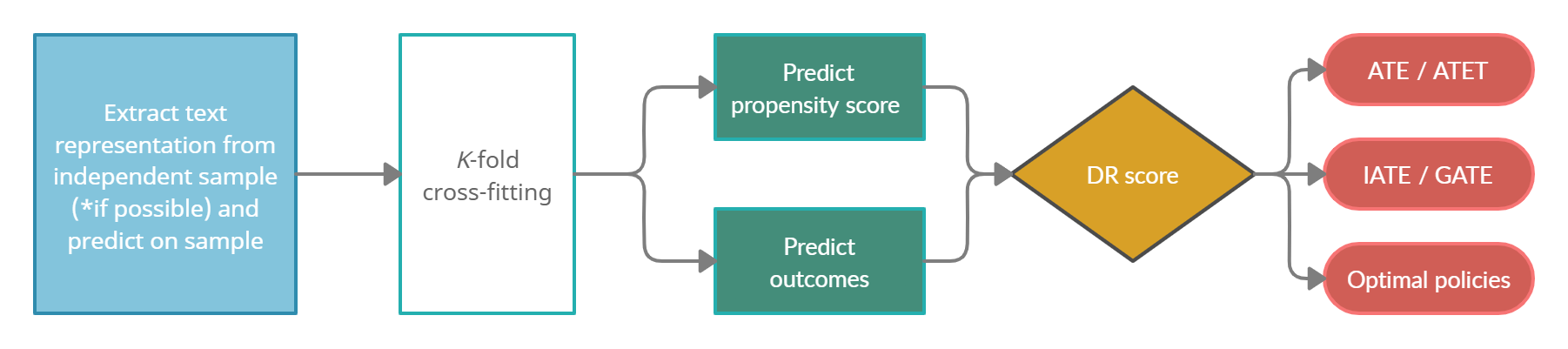}
    \footnotesize
    \emph{Notes:} This figure represents a stylized workflow of the estimation procedure. First information from text is retrieved, then used in $k-$fold cross-fitting to estimate the two nuisance parameters (estimated propensity score and estimated conditional expectation of the outcome). The doubly-robust score is computed and used to estimate the estimands of interest (APO, ATE, ATET, IATE and GATE, optimal policies).
  \end{minipage}
\end{figure}

\clearpage
\begin{figure}[p]
  \caption{Pairwise returns to Special Education programs according to their level of inclusion. \label{fig:results.main}}
  \begin{minipage}{\linewidth}
  \includegraphics[width=\textwidth]{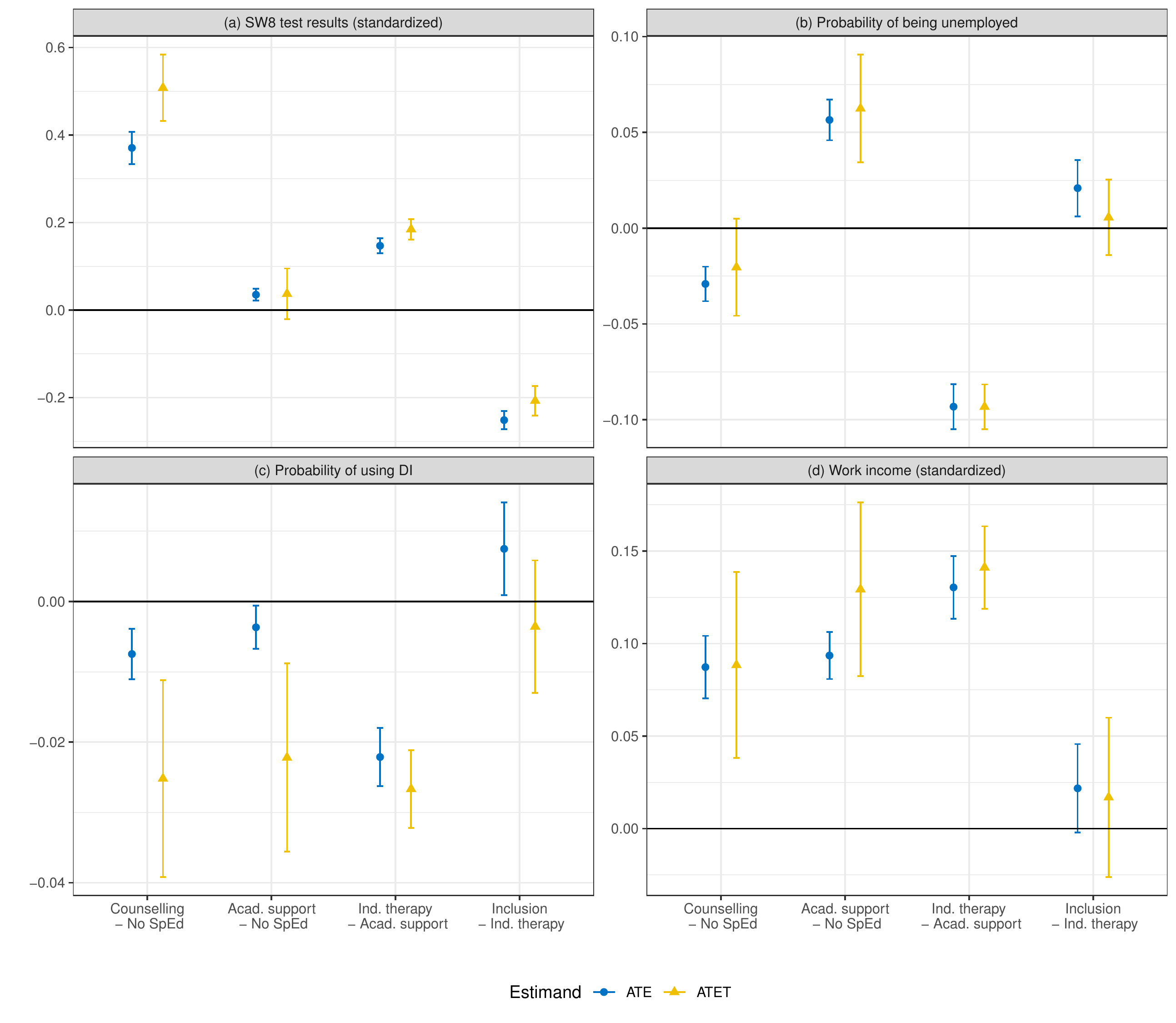}
  \footnotesize
  \emph{Notes:} This figure depicts pairwise treatment effects for Special Education programs in St.\ Gallen. Each pair compares interventions that are the closest in degree of severity and inclusion. Each pairwise treatment effect is the effect of being assigned to the first program in comparison to the second program on one of the four outcomes presented in the panel headers. Both the treatment effect on the whole population (ATE) and on the population of the treated (ATET) are presented. ``Ind. therapy'' is the abbreviation for individual therapies, ``Acad. support'' for academic support, and ``no SpEd'' for receiving no program. Nuisance parameters are estimated using an ensemble learner that includes text representations presented in the ``data'' section. 95\% confidence intervals are represented and are based on one sample $t$-test for the ATE and the ATET. Test results and wages are standardized with mean 0 and standard deviation 1. The baseline  (``No SpEd'') probability of unemployment benefit recipiency is 0.19, and 0.07 for disability insurance recipiency. \emph{Source: SPS}.
  \end{minipage}
\end{figure} 

\clearpage
\begin{figure}[p]
  \caption{Pairwise returns to segregated SpEd programs. \label{fig:results.ISFKKSON}}
  \begin{minipage}{\linewidth}
  \includegraphics[width=\textwidth]{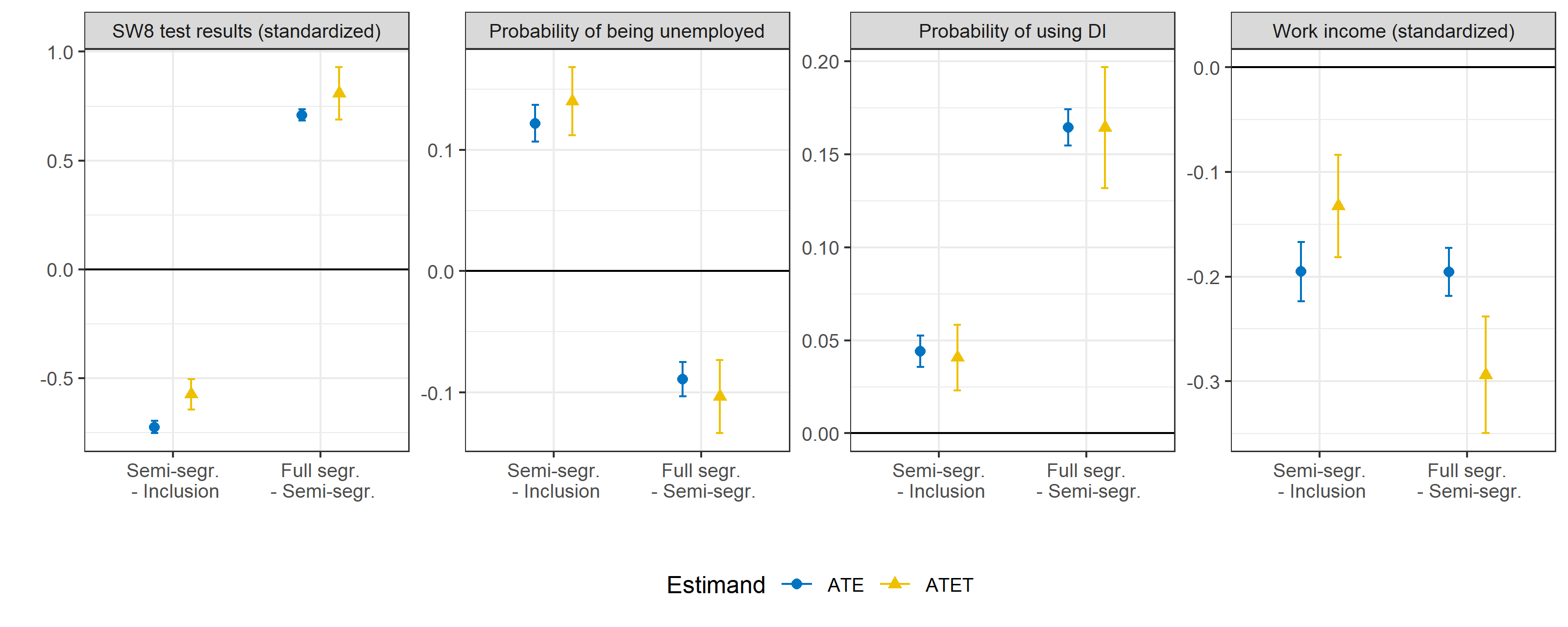}
  \footnotesize
  \emph{Notes:} This figure depicts relevant pairwise treatment effects for Special Education programs in St.\ Gallen. Each pairwise treatment effect is the effect of being assigned to the first program in comparison to the second program on one of the four outcomes presented in the panel headers. Both the treatment effect on the whole population (ATE) and on the population of the treated (ATET) are presented. ``Semi-segr.'' is the abbreviation for semi-segregation (segregation in small classes), and ``Full segr.'' stands for full segregation (in special schools). Nuisance parameters are estimated using an ensemble learner that includes text representations presented in the ``data'' section. 95\% confidence intervals are represented and are based on one sample $t$-test for the ATE and the ATET. Test results and wages are standardized with mean 0 and standard deviation 1. The baseline  (``Inclusion'') probability of unemployment benefit recipiency is 0.19, and 0.04 for disability insurance recipiency. \emph{Source: SPS}.
  \end{minipage}
\end{figure}

\clearpage
\begin{figure}[p]
\caption{Pairwise returns to Special Education vs. No Special Education. \label{fig:results.NO}}
  \begin{minipage}{\linewidth}
  \includegraphics[width=\textwidth]{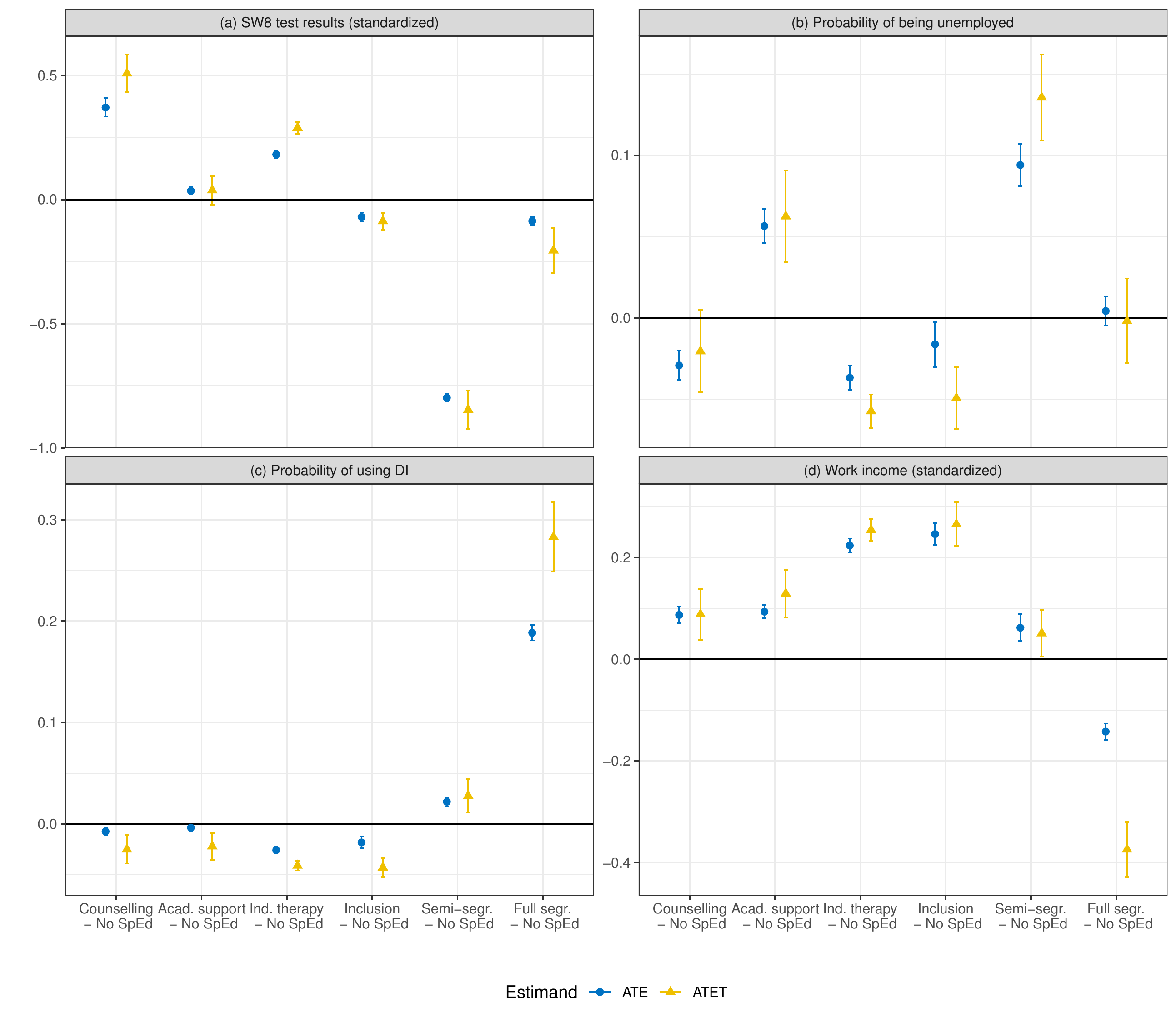}
  \footnotesize
  \emph{Notes:} This figure depicts relevant pairwise treatment effects for Special Education programs in St.\ Gallen. Each pairwise treatment effect is the effect of being assigned to the first program in comparison to being assigned to no program on one of the four outcomes presented in the panel headers. Both the treatment effect on the whole population (ATE) and on the population of the treated (ATET) are presented. ``Ind. therapy'' is the abbreviation for individual therapies, ``Acad. support'' for academic support, ``no SpEd'' for receiving no program, ``Semi-segr.'' for semi-segregation (segregation in small classes), and ``Full segr.'' for full segregation (in special schools). Nuisance parameters are estimated using an ensemble learner that includes text representations presented in the ``data'' section. 95\% confidence intervals are represented and are based on one sample $t$-test for the ATE and the ATET. \emph{Source: SPS}.
\end{minipage} 
\end{figure}

\clearpage
\begin{figure}[p]
\centering
\caption{Distributions of IATEs for segregation vs. inclusion. \label{fig:distr.IATE}}
\begin{minipage}{\linewidth}
  \includegraphics[width=0.9\textwidth]{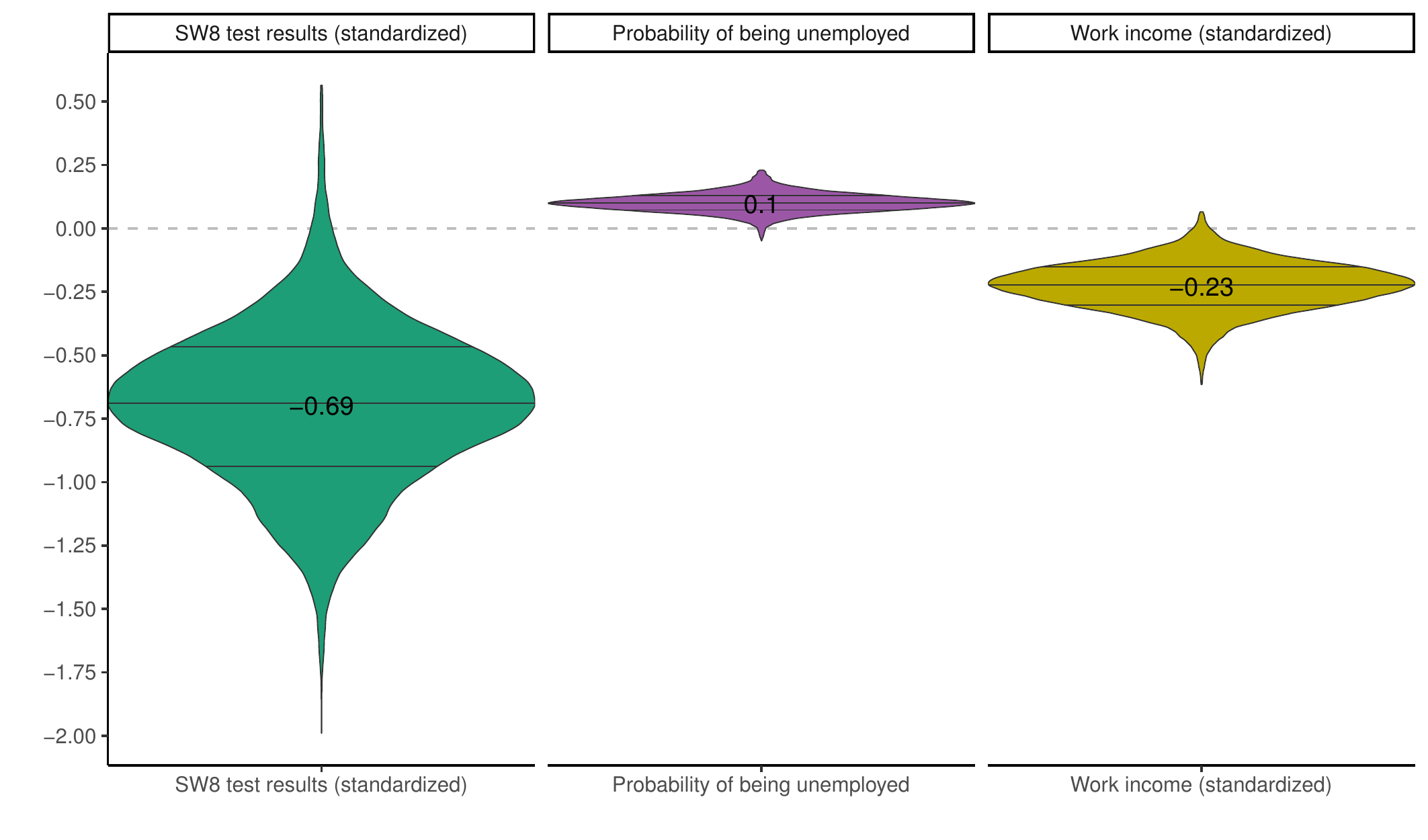}

  \footnotesize
  \emph{Notes:} This graph represents the smoothed distribution of IATEs for the four outcomes of interest. The IATEs were predicted out-of-sample using the DR-learner presented in \Cref{equation:AIPW_DR}. The average effect is reported in the figure. Light quintile bars represent the median and the 1st. and 5th. quintiles of the IATE distribution.
\end{minipage}
\end{figure}

\clearpage
\begin{figure}[p]
\includegraphics[width=1\textwidth]{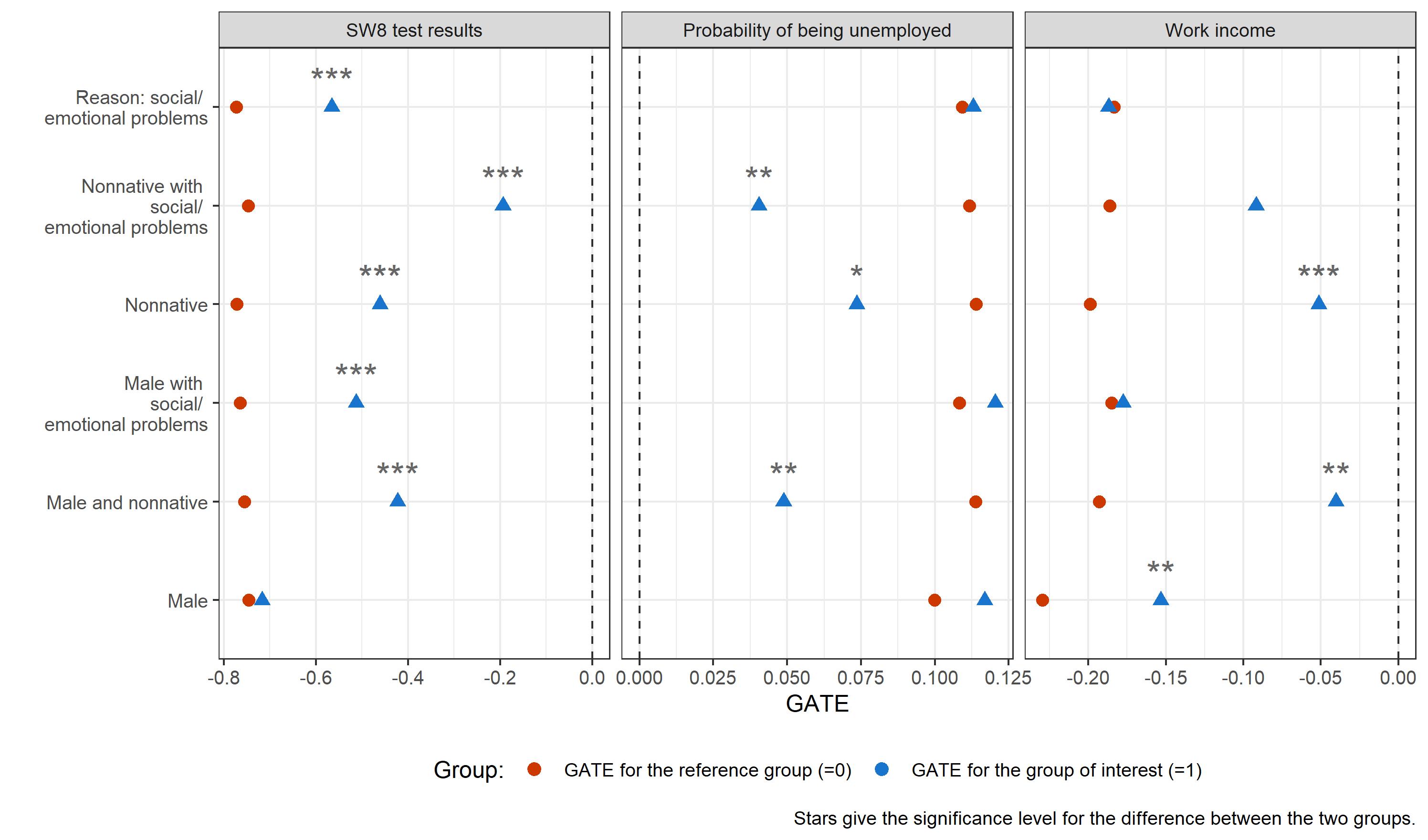}
\caption{GATEs for semi-segregation vs inclusion}
\floatfoot{\emph{Notes:} This graph presents estimated GATEs for the treatment effect of semi-segregation vs. inclusion. The GATEs are represented on the $x$ axis. Red dots indicate the treatment effect for the reference category (those students who do not belong to the category of interest), and blue dots indicate the treatment effect for the category of interest. The stars indicate the statistical significance of the difference between the two groups. For instance, the first row of the first column shows that the treatment effect of semi-segregation vs. inclusion is -0.75 test score standard deviations for students without social or emotional problems. The treatment effect is -0.55 for students with social or emotional problems. The difference in treatment effect between both groups is statistically significant ($p<0.01$). \label{fig:GATE_covariates}}
\end{figure}

\clearpage

\clearpage
\newgeometry{left=1.5cm, right = 1.5cm }

\restoregeometry

%\clearpage

\else

\fi

\clearpage
\doublespacing
\newgeometry{left=2.5cm, bottom = 2.5cm, right = 2.5cm, top = 2.5cm}
\begin{appendices}
\clearpage

\setcounter{table}{0}
\setcounter{figure}{0}
\setcounter{page}{1}
\renewcommand\thetable{A.\arabic{table}}
\renewcommand\thefigure{A.\arabic{figure}}
\renewcommand{\thepage}{\roman{page}}
% !TeX root = Manuscript_ASallin_v9.tex

\section{Appendix: Using text to adjust for confounding \label{appendix:appendix_text}}

The purpose of using computational text analysis methods and Natural Language Processing (NLP) in this paper is for confounding adjustment in the estimation of returns to SE programs. It secondarily serves as an interesting pretreatment variable to explore treatment heterogeneity. 

Extracting information from raw text is difficult for two main reasons: first, text is unstructured and high-dimensional, and, second, it includes latent features. To account for these two difficulties, text must be represented by an unknown $g()$ function that must be discovered in order to make text comparable and interpretable, as well as compress its dimensionality to a lower dimensional space. The main trade-off in discovering $g()$ is to compress the high-dimensionality of text without losing its substantial meaning: marginally extracting more information from the text occurs at the cost of increasing the dimensions of the covariate space, which leads to support and computational problems. Traditionally, $g()$ is discovered by human coders who extract relevant dimensions of the text into a low-dimensional space (e.g., a series of indicators). Recent machine-learning methods discover $g()$ in an unsupervised manner (e.g., \citet{Blei2003}).

%\paragraph{exogeneity of text} With covariates extracted from text, written comments could be related to the treatment the student receives insofar as the caseworker who writes the comment is the one suggesting the therapy. In this sense, the text could be endogenous to treatment if it expresses evaluative, prescriptive or predictive references to treatment. For instance, if the caseworker writes her motivation for assigning, say, speech therapy to a pupil who stutters, then she might write ``because of (1) \emph{stuttering}, I recommend (2) \emph{intense speech therapy} to alleviate (3) \emph{speech impediment} and make the pupil more (4) \emph{self-confident}.'' Using bag-of-words in a propensity score, for instance, would match on the variables `stutter', `intense speech therapy', `speech impediment', `self-confidence'. Whereas (1) and  (3) are likely to be exogenous since they contain description of the student prior to treatment, (4) is an evaluation of the treatment. (2) is a case of prescriptive discourse about the treatment the caseworker is about to give (which is not endogenous, but which would predict the treatment perfectly and violate common support). I remove all occurrences of (2), and I apply models that reduce the risk of (4), such as STM and Word2Vec.

This appendix presents in greater details the way I tackle these two difficulties and how I prepare the written psychological records for the estimation of treatment effects. It provides information on how I preprocessed the text, as well as how I implement the different methods. Summary statistics about the distribution of text statistics across treatment states are also provided.

\subsection{Text preprocessing}
Most text analysis methods implemented in this study require the text to be preprocessed --- with the exception of word embeddings. For preprocessing, I strip the text from words which have low information value (so-called ``stopwords'', numbers, and punctuation signs). Subsequently, I reduce the text to single words (``tokens''), and lemmatize them.\citep[for more details, see][]{Manning2008}. The lemmatization ensures that all tokens are reduced to their stems/roots without inflectional endings. Finally, I represent the text into types, i.e. unique expressions of tokens made out of onegrams (single tokens) and bigrams (two co-occuring tokens). For instance, if the number of tokens in the following document ``I ate an apple apple'' is 5, the number of one-gram types is 4, as the token ``apple'' is repeated. After preprocessing, the psychological records per student contain on average 241 tokens and 137 types.

\subsection{Text representations}
\paragraph{Document term matrix (DTM)}
The simplest way of representing a text is to reduce the text to its components (``tokens'') and, for each token, indicate its frequency of appearance within each document in a ``Document term matrix'' (DTM). Thus, two documents are identical (and comparable) if they use the same tokens with a similar frequency. The limitations of DTMs is that frequency matrices do not account for the context in which tokens appear. Moreover, the dimensionality of DTMs quickly explodes with the number of documents, making comparisons across documents difficult. This requires handling huge sparse matrices, which can be computationally difficult. 

To reduce the dimension of the DTM, I implement two methods: scaling and bounding. For scaling, I weight the DTM either by term frequency \emph{tf}, which simply weights terms according to their number of appearances in a document, or by term-frequency-inverse document frequency (\emph{tf-idf}), which increases with the term frequency of a word in a document and decreases with the number of documents in the corpus that contains the word (thus highlighting words that carry a lot of information about the document). Basically, very often used ``stop words'' have a very low \emph{tf-idf} score, and words that are highly specific to a document have high \emph{tf-idf} scores. 

\begin{figure}[t!]
\centering
\caption{Most common tokens per treatment assignment \label{fig:tf}}
\begin{minipage}{0.75\linewidth}
  \includegraphics[width=\textwidth]{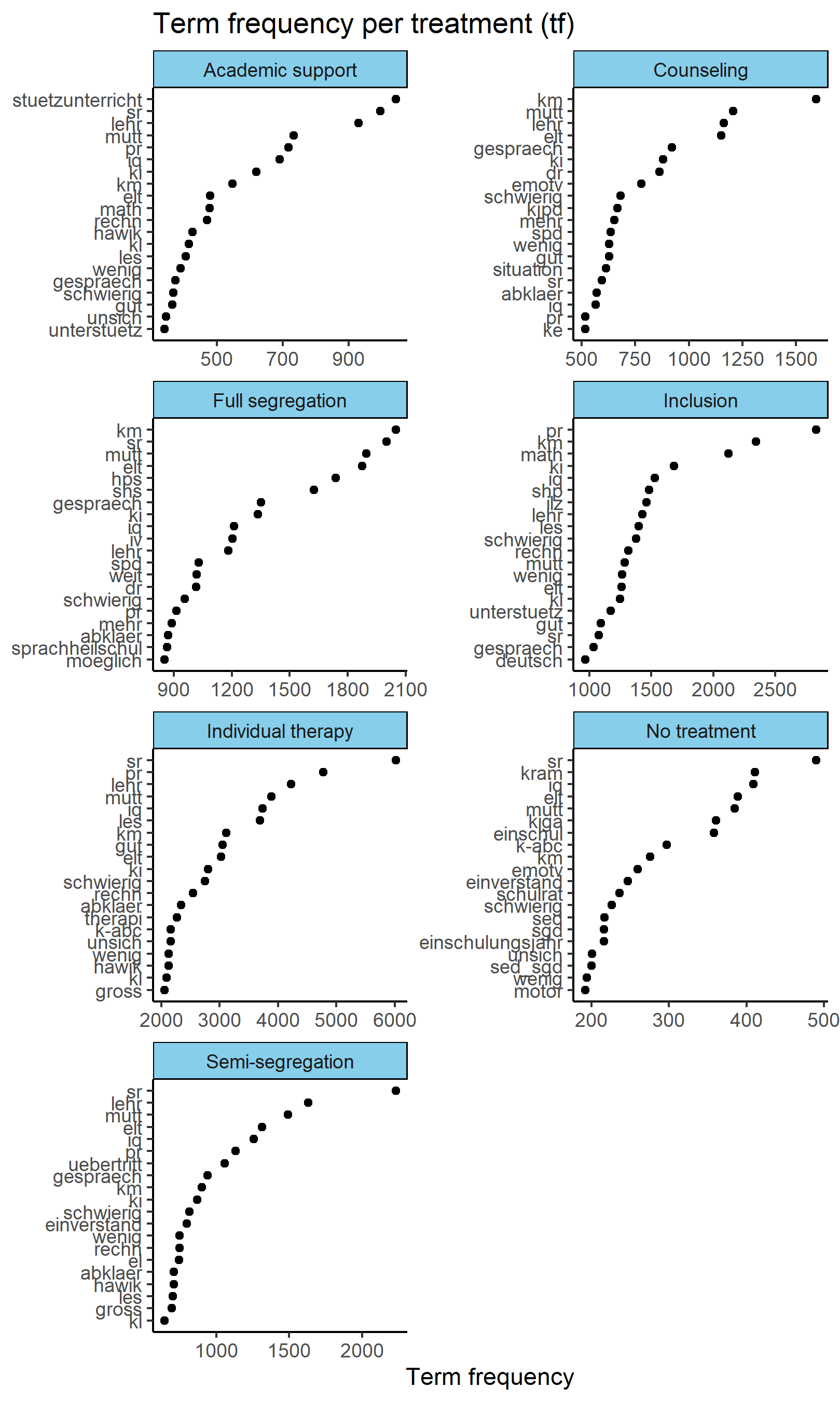}

  \footnotesize\emph{Notes:} This figure represents the 20 most frequent terms (tokens) per treatment assignment. Text is preprocessed via lemmatization. \emph{Source: SPS}.
\end{minipage}
\end{figure}

\begin{figure}[t!]
\centering
\caption{Most frequent weighted tokens per treatment assignment \label{fig:tfidf}}
\begin{minipage}{0.75\linewidth}
  \includegraphics[width=\textwidth]{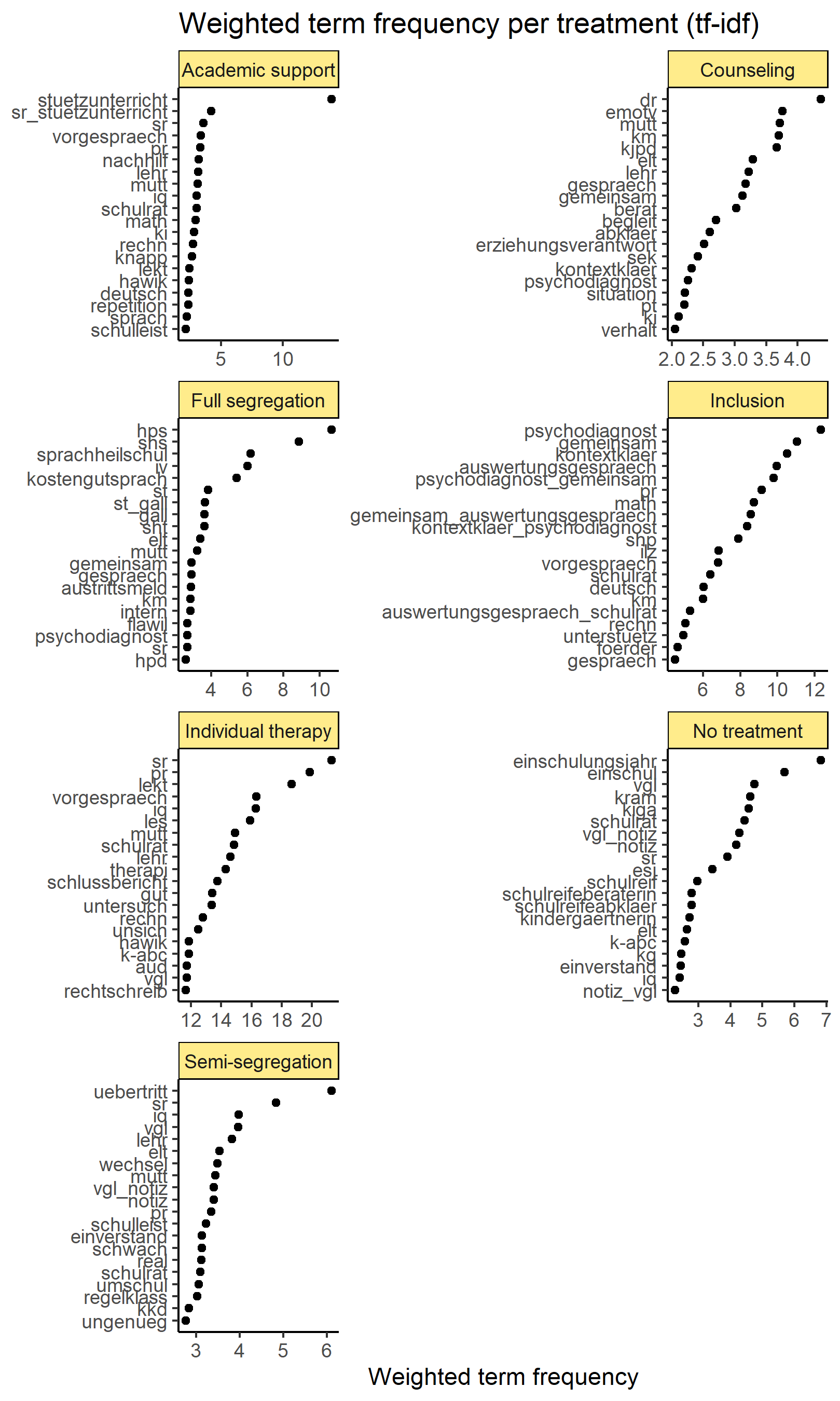}

  \footnotesize\emph{Notes:} This figure represents the 20 most frequent terms (tokens) weighted by \emph{tf-idf} per treatment assignment. Text is preprocessed via lemmatization. \emph{Source: SPS}.
\end{minipage}
\end{figure}

To select terms that are general enough, I bound the number of terms by selecting words that appear at least 350 times (min. term frequence) and in at least 150 documents (min doc. frequency) for \emph{tf}. For \emph{tf-idf}, I select very specific tokens, i.e. \emph{tf-idf} score bounded at the 99.9th percentile of all \emph{tf-idf} scores. I provide a third measure with a mixture of \emph{tf} and \emph{tf-idf} scores: I first select the most frequent tokens, and then weight them by \emph{tf-idf}, which ensures that only frequent words with high significance are selected. As an illustration, \Cref{fig:tf} and \Cref{fig:tfidf} display the 20 most frequent terms per treatment assignment. 

The criterion for the choice of scaling scheme and tuning parameters in the DTM representation depends on the empirical problem at hand: I do not only want tokens that are very predictive of treatment and outcome, but also tokens that are common enough to serve the purpose of comparing documents. Thus, a very high \emph{tf-idf} may suit the purpose of prediction very well; however, scores in the middle range might serve the purpose of comparison better. 

%https://stackoverflow.com/questions/16927494/how-to-select-stop-words-using-tf-idf-non-english-corpus

\paragraph{Structural Topic Modeling (STM) and Topical Inverse Regression Matching (TIRM)}
To mitigate the high-dimensionality and lack of word context in DTM, representations that discover latent features of the text are interesting ways of representing high-dimensional text. I implement Structural Topic Models (STM) as proposed by \citet{Roberts2016} and Topical Inverse Regression Matching (TIRM) \citet{Roberts2020}, which are a variant of the Latent Dirichlet Allocation topic model (LDA, \citet{Blei2003}). Succinctly, LDA and STM first sample a topic from the distribution of tokens in a given document, and then sample each observed token from the distribution of words given each topic. Unlike LDA, STM allows for covariates to affect the proportion of a document attributed to a topic (``topical prevalence'') and the distribution of tokens within a topic (``topical content''). STM assumes that covariates influence the way tokens are distributed: it includes covariates in the prior distribution of topics per documents (logistic normal distribution) and the distribution of tokens per topic (multinomial logistic regression). In STM, the number of topics is chosen \emph{ad-hoc}, and topics are not directly interpretable. 

The advantage of STM is that it reduces the text dimensionality into a finite number of topics, and offers a good comparison of documents, as documents that cover the same topics at the same rates are represented as similar documents. In this application, I estimate STM and define the number of topics $k$ to either 10 or 80. I then use the vector of $k$ topic proportions ($k\times N$) directly in the propensity score. Other variants are possible, such as taking the $k-x$ most important topics or the topics that explain topical content the most \citep[see, for instance, ][]{Mozer2020}.

Topical Inverse Regression Matching (TIRM) is an interesting method that allows to directly model the treatment assignment process from text. It builds on STM by estimating an additional reduction, i.e. a document-level propensity score based on STM and the treatment status as a content covariate. This additional reduction is used together with the STM topics to perform traditional matching. This method ensures that matched documents are similar in their topics and within-topic treatment propensity. In this study, I predict the TIRM sufficient prediction score for each treatment status (similarly to the propensity score estimation). I then use the score as additional covariate for the estimation of the nuisance parameter for the propensity score.  

\begin{figure}[t!]
\caption{Prevalence of STM topics per treatment assignment \label{fig:topic.distr.10}}
\centering
\begin{minipage}{0.9\textwidth}
  \includegraphics[width=\textwidth]{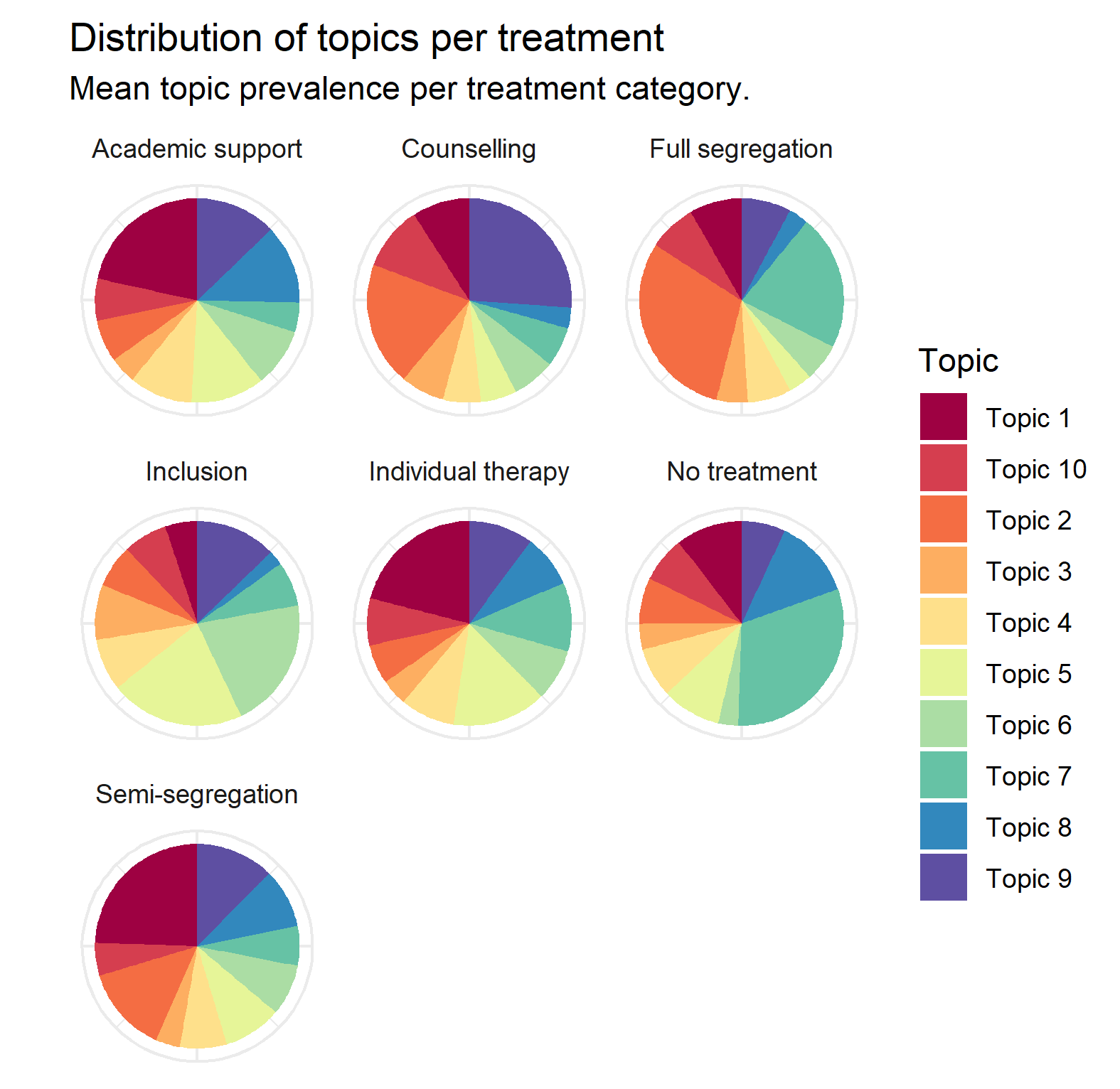}

  \footnotesize\emph{Notes:} The prevalence of each STM topic per treatment assignment is represented. I compute the mean prevalence of each of the 10 topics per treatment category. Topic prevalence in STM is given by the main covariates. \emph{Source: SPS}.
\end{minipage}
\end{figure}

% latex table generated in R 4.0.2 by xtable 1.8-4 package
% Mon Jun 14 16:32:45 2021
\begin{table}[t!]
\centering
\footnotesize
\caption{Topics from a STM on main sample} \label{table:topics10}

\begin{threeparttable}
\begin{tabularx}{\textwidth}{cXX}
  \toprule
Topic & Highest probability & Most frequent and exclusive \\ 
  \midrule
1 & sr, mutt, km, schulrat, les, pr, kiga, vgl, lehr, pr & sr\_lekt, vb, jug, u'ergebnis, iq\_sed, gemeinsam\_auswertungsgespraech, trog-d, vgl\_notiz, proz, sht \\ 
  2 & iq, elt, ki, besprech, rechn, hawik, einschul, notiz, ke, kl & untersuch\_hawik, herrn, inform, besprech\_u'ergebnis, auditiv\_merkfaeh, psychodiagnost\_gemeinsam, wwt, notiz\_vgl, sek, leistungs-\_lernverhalt \\ 
  3 & lekt, gespraech, kv, elt, k-abc, iq, shs, mutt, mutt, elt & untersuch\_k-abc, bad\_sond, beob, leg.-th, dyskalkulie-therapi, wld\_agd, einschulungsjahr, vgl, ke, pl \\ 
  4 & hawik, dr, math, abklaer, sed, sv, motor, ki, schwierig, iq & lekt\_vorlaeuf, beistand, z.h, abklaer\_wunsch, forst, agd\_vg, hpd, antragsschreib, li, lernverhalt\_aktuell \\ 
  5 & untersuch, spd, jedoch, mutt, sgd, vg, auffaell, vgl\_notiz, mehr, gut & testsitz, km, semesterbericht, macht\_mueh, agd, shs, notiz, befind, lernverhalt \\ 
  6 & lehr, kjpd, sp, lehrerin, sed\_sgd, wld, abklaer, lehr, bess, cpm & rav\_pr, moegl, dc-ther, les\_langsam, kontextklaer, hs\_ds, vg\_mutt, familia, cpm\_rav \\ 
  7 & schulleist, weit, kl, vorgespraech, unsich, agd, kg, abkl, wenig, kram & kkd, langhald, zz, wunsch\_lehrerin, k-abc\_sed, interview, hs, legasthenie-schlussbericht, diagnost\_termin, wn \\ 
  8 & uebertritt, situation, noetig, therapi, mueh, sv\_wld, einverstand, austausch, termin, pl & schlussgespraech\_sr, sond, thera, schulpsycholog\_abklaer, lektion\_woechent, psychodiagnost, vg\_abkl, th, leistungs- \\ 
  9 & kram, sr, ilz, wunsch, sif, motti, fortschritt, vg, gespraech, evtl & intelligenzstatus, bad, kle, kit, woechent, sv\_wld, ej, mutt\_abkl, ngste, audi \\ 
  10 & elt, info, spd, problem, schwierig, wld\_agd, sed, gut, srp & sr\_schlussbericht, time-out, z.h\_sr, lrs-ther, textverstaendnis, rt, spezial, antragsschreib\_schulrat, thema, slp \\ 
   \bottomrule
\end{tabularx}

\begin{tablenotes}[para,flushleft]
\footnotesize \emph{Notes:} This table represents the tokens occurring with the highest probability in each of the 10 topics, as well as the tokens that are the most frequent and most exclusive in each topic. The method of topic extraction is STM, and topics are not necessarily interpretable.  
\end{tablenotes}
\end{threeparttable}
\end{table}

To give an example on how topic modeling can be used to remove confounding, I extract an STM on 10 topics with the covariates presented in \Cref{table:summary_stats} as topic prevalence covariates. The topic content is presented in \Cref{table:topics10} (in German) and the topic distribution across treatment states is shown in \Cref{fig:topic.distr.10}.\footnote{Note that this topic distribution is presented as an example. In the main analysis, it will slightly differ because of the cross-fitting strategy. However, if data are randomly cross-validated, there must be no big discrepancy between \Cref{table:topics10} and the topics extracted in each fold (even though stm can slightly vary with changing samples).} Even though STM topics are not always directly interpretable, \Cref{table:topics10} shows interesting patterns. Topic 2, for instance, seems to relate to the evaluation of learning disabilities with an IQ test and the discussion of the results with parents, whereas topic 5 reports behavioral issues and the discussion of school reports. Interestingly, topics highlight the fact that parents are active in the process (see the occurrence of the tokens ``mutt'' and ``elt''). This is valuable information, as it allows me to account for parental influence in my estimates.

When looking at whether topics are discriminatory in terms of treatment assignment, all topics are represented in all treatments, but some seem to be more prevalent in some treatments. For instance, topics 1 and 2 are more represented in children who have been segregated in special classrooms, while topics 5 and 6 are more represented in students sent to inclusion. Even though it could be interesting to find meaning in topics, the \emph{true} number of topics is never known, which makes topics sensitive to the ad-hoc choice of $k$ \citep{Roberts2016}. For this reason, I estimate STM models with different $k$.

\paragraph{Word embedding with Word2Vec}
The word embedding representation estimates the semantic proximity of words. The algorithm I use in this study is the Word2vec of \citet{Mikolov2013}, who propose a neural network architecture to represent words in vectors as a function of their use and the words that most commonly co-occur with them. I use the \emph{Word2Vec} embedding with word vectors of length $K$=50, 100 (I also tried with 200 and 500) based on text which is not preprocessed. For each embedding, I compute a document-level vector of length $K$ by taking the average of the numeric vectors of all the words within a document (similar to \citet{Mozer2020}). The result is an $N \times K$ matrix. Documents that are similar to each other in terms of words and their context will therefore be similar on the $K$ dimensions. Word embeddings are famous for the word associations they can produce, and I give some illustrative examples in \Cref{table:word2vec.simil}. For instance, the words that are the closest to ``ADHD'' and ``dyslexia'' (literally, ``spelling disorder'') are close in meaning to the two expressions. This method is therefore a good way to capture conceptual proximity between words, and accounts for the context of words. 

\begin{table}[t!]
\centering
\footnotesize
\begin{tabular}{llr}
  \hline
Word & Most similar & Similarity \\ 
  \hline
  %fremdsprache & subtr & 0.66 \\ 
  %fremdsprache & groessen & 0.64 \\ 
  %fremdsprache & rechn & 0.64 \\ 
  %fremdsprache & einmaleins & 0.64 \\ 
  %fremdsprache & textaufgaben & 0.64 \\ 
  %fremdsprache & zahlenstrahl & 0.63 \\ 
  %fremdsprache & brueche & 0.63 \\ 
  %fremdsprache & subtraktionen & 0.63 \\ 
  %fremdsprache & bruchrechnen & 0.62 \\ 
  %fremdsprache & zr & 0.62 \\ 
  adhs & ads & 0.84 \\ 
  adhs & adhd & 0.76 \\ 
  adhs & neuropsycholog & 0.74 \\ 
  adhs & pos & 0.73 \\ 
  adhs & autismus & 0.73 \\ 
  adhs & medizinische & 0.71 \\ 
  adhs & erhaertet & 0.70 \\ 
  adhs & mediz & 0.70 \\ 
  adhs & neurologische & 0.70 \\ 
  adhs & asperger & 0.70 \\[1em]
  rechtschreibstoerung & erschwerten & 0.77 \\ 
  rechtschreibstoerung & rezeptive & 0.76 \\ 
  rechtschreibstoerung & beeintraechtigung & 0.75 \\ 
  rechtschreibstoerung & auditiver & 0.74 \\ 
  rechtschreibstoerung & sprachstoerung & 0.73 \\ 
  rechtschreibstoerung & rechtschreibschwaeche & 0.72 \\ 
  rechtschreibstoerung & spracherwerbsstoerung & 0.72 \\ 
  rechtschreibstoerung & teilleistungsstaerung & 0.71 \\ 
  rechtschreibstoerung & rechtschreibschwierigkeiten & 0.71 \\ 
  rechtschreibstoerung & ausgepraegte & 0.71 \\ 
   \hline
\end{tabular}
\label{table:word2vec.simil}
\caption{Word2Vec word similarities} 
\floatfoot{\emph{Notes:} Illustrative examples of word associations produced by word embeddings and \emph{Word2Vec} in German. ``ADHS'' is the German abbreviation for ``ADHD'', and ``rechtschreibstoerung'' is the German translation of ``spelling disorder''.}
\end{table}

\paragraph{Mapping mental diagnoses with \emph{ad-hoc} dictionary approach}
The text representation that is the most directly interpretable and perhaps the most convincing in this context is a dictionary approach that leverages independent mental health diagnoses. I use an independent sample on children from the City of St.\ Gallen, which contains an additional observed diagnosis variable given by the caseworker. I build a lexicon that, for each of the 16 formal diagnoses assigned in the City sample, identifies the tokens that are the most ``key'' in each diagnosis (namely, the ``keywords'' that are the most exclusive, or predictive, of each diagnosis). There are many ways to define ``keyness'', and I propose a combination of different measures that are common in computational linguistics.\footnote{\emph{Keyness} is the importance of a keyword within its context. To compute keyness, the frequency of a keyword in a target category for the observed frequency of a word (one particular diagnosis) is compared with its frequency in a reference category (the expected frequency, in all the other diagnoses).} Namely, I first take the 40 most frequent tokens per diagnosis (based on count frequency), then the 60 most frequent tokens weighted by \emph{tf-idf}, the 40 tokens with the highest chi-squared keyness value, the 40 tokens with the highest likelihood ratio G2 statistics and the 40 tokens with the highest pointwise mutual information statistics. I then take the union of all tokens provided by each measure.\footnote{Alternatively, this purely frequency analysis could be done by training a classifier on text tokens.} 

To classify documents into diagnoses, I first map each document to the diagnoses using the key tokens of each diagnosis extracted in the St.\ Gallen sample. I then compute the share of diagnoses per document by dividing the frequency of tokens assigned to a given diagnosis in a document by the total number of tokens per document. I weight diagnoses such that the sum of each document's diagnosis frequency sums up to 1. As a result, I obtain a vector of length 16 with the predicted proportion of diagnoses per document. In a second measure, I ensure that the most prominent diagnosis assigned to a document is discriminatory enough. I hot-encode the diagnosis if the proportion of the given diagnosis is 1.5 standard deviations above the mean of diagnosis frequency within the document.

The dictionary/keyword approach maps documents into clinically meaningful concepts by using natural language which is almost exactly similar to the language used by therapists in the Canton of St.\ Gallen. Therefore, it is the closest to what a human coder would do if she had to classify the caseworkers' comments. Moreover, contrary to STM and word embeddings, it is supervised. Since therapists from the City and therapists from the Canton work in the same environment, under the same rules, and use the same lexicon, texts are very similar. 

\begin{figure}[t!]
\caption{Average diagnosis per treatment assignment \label{fig:diagnosisdistr}}
\centering
\begin{minipage}{0.9\textwidth}
  \includegraphics[width=\textwidth]{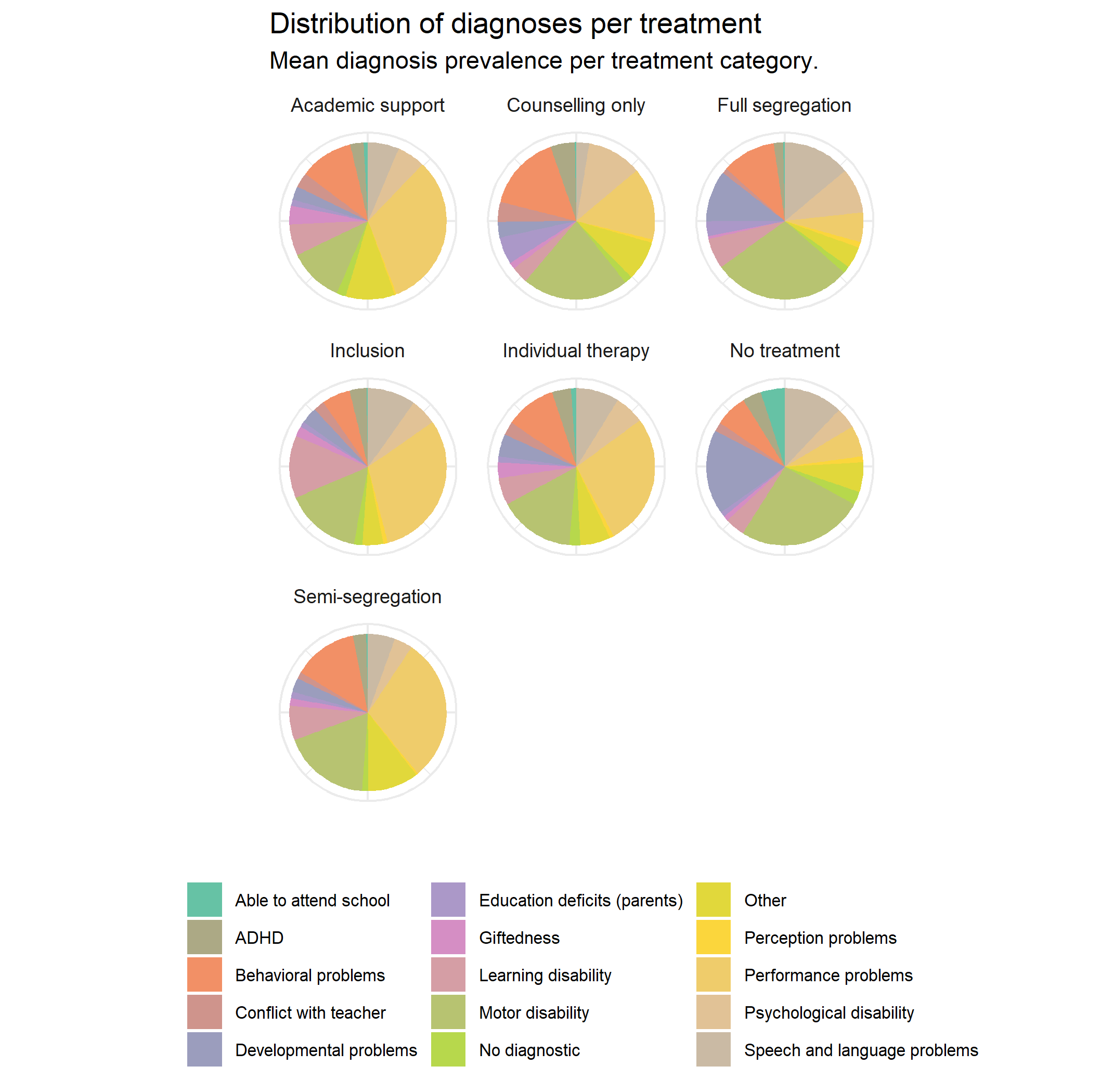}

  \footnotesize \emph{Notes:} The main 15 diagnoses extracted from an independent dataset are represented (the diagnosis ``foreign language'' is not displayed here). Diagnosis are extracted from the data of the City of St.\ Gallen \emph{Source: SPS}.
\end{minipage}
\end{figure}

Distribution of diagnoses across treatment status is presented in \Cref{fig:diagnosisdistr}. It is interesting to notice that individual therapies, inclusion and semi-segregation share a roughly similar population of students, namely students diagnosed with problems related to school performance. Inclusion has a relatively higher share of students with learning disabilities, while semi-segregation has a higher share of students with behavioral problems. In contrast, hard segregation is particularly targeted to students with more severe disabilities, such as motor problems, and students with parental educational deficit. As I can expect, students not given any treatment display a more equal share of all diagnoses.\footnote{A similar descriptive picture holds when I use 1 standard deviations above the mean to classify diagnoses.} 
\clearpage

\setcounter{table}{0}
\setcounter{figure}{0}
\setcounter{page}{1}
\renewcommand\thetable{B.\arabic{table}}
\renewcommand\thefigure{B.\arabic{figure}}
\renewcommand{\thepage}{\roman{page}}
% !TeX root = Submission_ASallin_v9.tex

\clearpage

\section{Appendix: Robustness and sensitivity checks \label{appendix:robustness}}

\subsection{Overlap: ATO and alternative trimming schemes \label{appendix:robustness.1}} 
The problem of overlap is especially exacerbated in settings with many treatments. It becomes even more important with a high-dimensional, highly predictive, covariate space \citep[see][]{Damour2020}. In such settings, overlap is difficult to obtain, which induces bias and extreme variability of ATEs estimates. To ensure overlap and remove extreme weights, I check the robustness of my results by estimating the Average Treatment Effect on Overlap (ATO), and by using different trimming rules. 

\paragraph{Average treatment effect on the population of Overlap (ATO)}
If students with SEN differ greatly across programs and overlap is poor, it might be relevant from a policy perspective to look at pairwise treatment effects for the population which is the most similar in terms of covariates across multiple treatments. I estimate overlap-weighted average treatment effects \citep{Li2018,Li2019}, i.e. average treatment effects on the population of propensity score overlap $\text{ATO}_{d,d'} = E_{\text{overlap}}[Y_i^d - Y_i^{d'}]$. The ATO estimand is the following:
\begin{alignat}{3}
&\text{ATO}_{d,d'}      &&= E_{\text{overlap}}[\Gamma^{h}(d,X_i)  - \Gamma^{h}(d',X_i) ],       &&\quad h(x)= \sum_{k=1}^D (\frac{1}{p_k(x)})^{-1} 
\end{alignat}
The ATO score gives the most relative weight to the covariate regions in which none of the propensities are close to zero. It is the product of the IPW and the harmonic mean of the generalized propensity scores. Beyond focusing on an interesting population, the advantage of using the ATO is that it mitigates problems of extreme propensity scores in the ATE computation. The ATO score is not doubly-robust, and is sensitive to potential misspecifications in the propensity score.\footnote{Since the weight is a function of the propensity score, it is not consistent if the propensity score is misspecified. However, as pointed out by \citet{Li2019}, outcome regression may still increase the efficiency of the weighting estimator. I estimated the variance of the ATO the same way I estimated the variance of the ATE and the ATET. For more information, see \citet{Li2019} and the \texttt{R}-Package \texttt{psweight}.}

\begin{figure}[t!]
\centering
\caption{Pairwise treatment effects with ATO estimates \label{fig:results_ATO}}
	\begin{minipage}{\linewidth}
	\includegraphics[width=\textwidth]{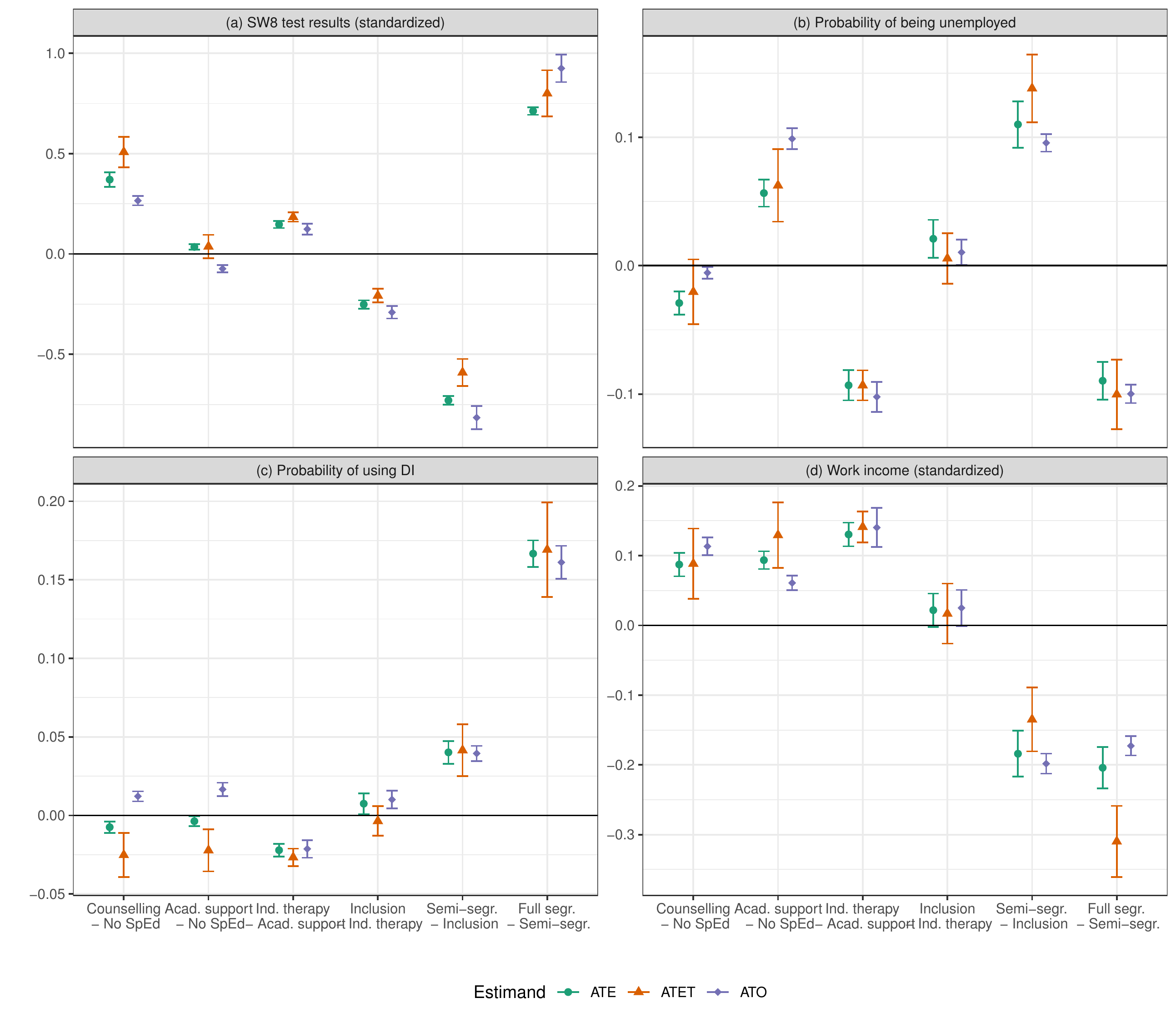}

	\footnotesize\emph{Notes:} This figure depicts pairwise treatment effects for Special Education programs in St.\ Gallen. Each pair compares interventions that are the closest in degree of severity and inclusion. Each pairwise treatment effect is the effect of being assigned to the first program in comparison to the second program on one of the four outcomes presented in the panel headers. The treatment effect on the whole population (ATE), on the population of the treated (ATET), and on the population of overlap (ATO) \citep[see][]{Li2019} are presented. ``Ind. therapy'' is the abbreviation for individual therapies, ``Acad. support'' for academic support, ``segr'' for segregation, and ``no SpEd'' for receiving no program. Nuisance parameters are estimated using an ensemble learner that includes text representations presented in the ``data'' section. 95\% confidence intervals are represented and are based on one sample $t-test$ for the ATE and the ATET. Test results and wages are standardized with mean 0 and standard deviation 1. \emph{Source: SPS}.
	\end{minipage}
\end{figure}

Results for the ATO are presented for each pairwise treatment effect in \Cref{fig:results_ATO}, and they are compared to the ATE and the ATET. The point estimates of the ATO are, in most cases, quite similar to the ATE point estimates, which indicates that there is good overlap in the overall population. This suggests that my ATE results do not suffer from lack of overlap.

\paragraph{Trimming}
The idea of trimming is to remove observations with weights $h(x)$ in \Cref{equation:weighting} below a certain threshold, such that $h(x) = \text{\underline{1}}(x \in C)$ with $C$ denoting the target subpopulation defined by a threshold of the propensity score distribution $\alpha$. Each trimming rule slightly shifts the population of interest $C$. In other words, the more conservative the trimming rule, the more homogeneous the population across treatment states. 

\begin{figure}[t!]
\centering
\caption{Pairwise treatment effects with different trimming rules \label{fig:results.robustness.trimmed}}
\begin{minipage}{\linewidth}
	\includegraphics[width=\textwidth]{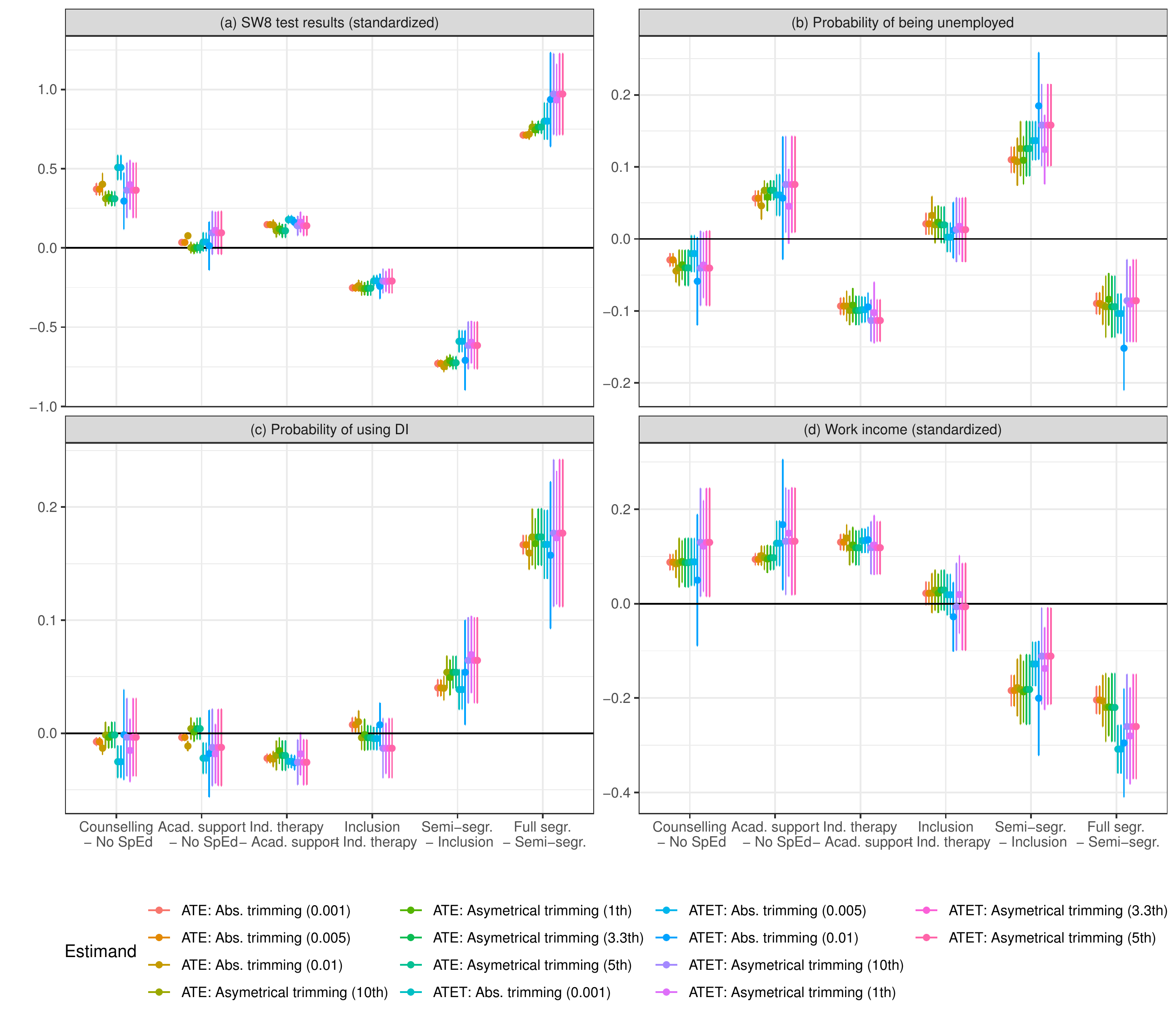}
	
	\footnotesize\emph{Notes: }This figure depicts relevant pairwise treatment effects for Special Education programs in St.\ Gallen at different trimming levels. Absolute trimming of \citet{Crump2009} and asymmetrical trimming of \citet{Sturmer2010} are represented at different trimming thresholds. Each pairwise treatment effect is the effect of being assigned to the first program in comparison to the second program on one of the four outcomes presented in the column headers. Both the treatment effect on the whole population (ATE) and on the population of the treated (ATET) are presented. ``Ind. therapy'' is the abbreviation for individual therapies, ``Acad. support'' for academic support, ``segr.'' for segregation and ``no SpEd'' for receiving no program. Nuisance parameters are estimated using an ensemble learner that includes text representations presented in the ``data'' section. 95\% confidence intervals are represented and are based on one sample $t-test$ for the ATE and the ATET. \emph{Source: SPS}.
\end{minipage}
\end{figure}

To define the trimming schemes and trimming thresholds $\alpha$, I explore different options. Following the adaptation of trimming schemes for the multi-treatment case in \citet{Yoshida2019}, I trim, in a first setting, observations with all propensity scores below a certain threshold of the propensity score \Citep[see][]{Crump2009}. This referenced under ``absolute trimming'' with $\alpha = \{0.001, 0.005, 0.01\}$.  In a second setting, the ``asymmetrical trimming'', I trim treated observations with their corresponding propensity score within a chosen quantile of each propensity score (``asymmetrical trimming'', see \Citet{Sturmer2010}). I use $\alpha = \{0.01, 0.033, 0.05, 0.1\}$ for the 1st., 3.3th, 5th and 10th quantiles. %I also implement the optimal trimming rule of \citet{Yang2016} (following the trimming rule for binary treatment of \citet{Crump2009}), which finds a trimming threshold such that the variance of the estimated ATE is minimized. However, this trimming rule assumes homoscedasticity and constant treatment effects. 
Results are presented in \Cref{fig:results.robustness.trimmed}: the majority of results discussed in \Cref{section:results} persist across different trimming schemes. The interpretation of effect variation under different trimming rules must however be done with caution, as each trimming rule ``shifts'' the population of interest.

\subsection{Handling text as covariate\label{appendix:robustness.2}}

In this section, I check that my results do not depend heavily on the way I retrieve information from the text. First, I investigate the sensitivity of my estimates to the inclusion of text covariates. Second, I explore the problem of ``text-induced endogeneity'', which might arise if the text representation captures the psychologist's biases (towards a certain treatment or a certain writing style) rather than information on the student.

\begin{figure}[t!]
\centering
\caption{Effect comparison with specification without text covariates \label{fig:noTEXTvariation}}

\begin{minipage}{\linewidth}
	\includegraphics[width=\textwidth]{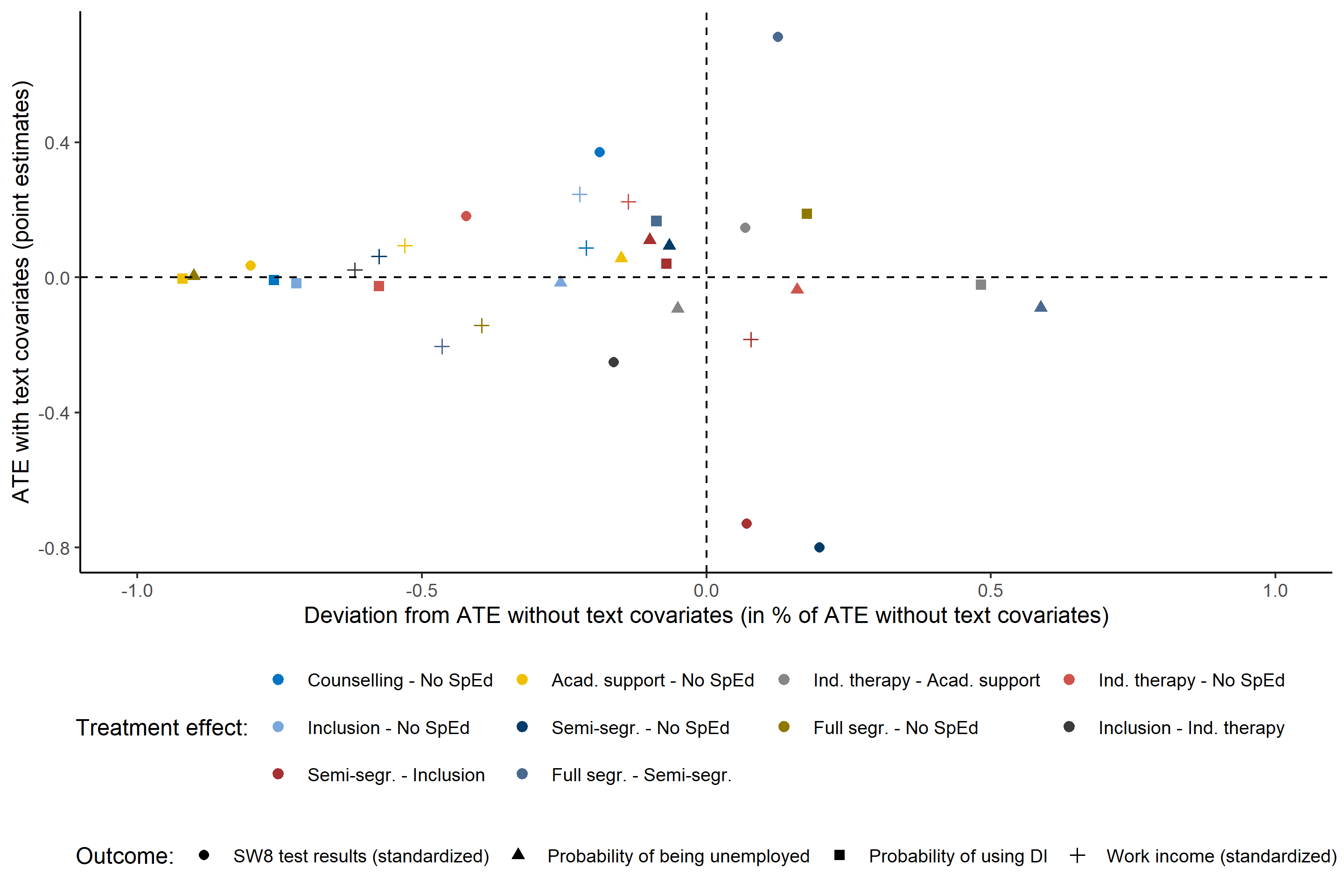}
	
	\footnotesize \emph{Notes: } This figures gives the effect variation between the specification with text covariates and the specification without text. On the $y$ axis, the ATE with text covariates is represented for each outcome and pairwise treatment effect investigated in this paper. The $x$ axis gives the difference between the ATE estimated with text covariates and the ATE estimated without text covariates in percent of the ATE estimated without text covariates.
\end{minipage}
\end{figure}

\paragraph{Sensitivity to text covariates} To investigate the sensitivity of my results to the inclusion of text covariates, I compute all my results using only nontext covariates. I then show how much the estimates based on text vary with respect to estimates obtained without controlling for text. This exercise gives me an approximate indication on how much confounding I can remove by using the text. 

\Cref{fig:noTEXTvariation} gives the variation in effect between the specification with text covariates and the specification without text. On the $y$ axis, the ATE with text covariates is represented for each outcome and pairwise treatment effect investigated in this paper. The $x$ axis gives the difference between the ATE estimated with text covariates and the ATE estimated without text covariates in percent of the ATE estimated without text covariates. On average, I find that estimates based on both covariates and text information are 29\% smaller than estimates that do not leverage the text information. Interestingly, differences between estimates with text and estimates without text are particularly pronounced for comparisons with the ``No SpEd'' intervention, which suggests that there is valuable confounding information contained in the text for this particular population of students.

%\Cref{fig:results.robustness.notext} compares main results and results without text confounding on the same figure. Estimates that do not leverage information contained in text sometimes lead to effects of different magnitudes, or uncovers effects that were nonexistent when adding text covariates. Interestingly, differences between text and non text point estimates are particularly pronounced for comparisons with the ``No SE'' group, which suggests that there is valuable confounding information contained in the text for this particular population of students. 

\paragraph{Text-induced endogeneity} It is inherent to the discovery of latent features in text data that some dimensions (latent or not) of text might be exogenous to the child's characteristics but influencing treatment assignment. An example is if a given psychologist is biased towards a particular treatment, and that this psychologist always writes using a similar set of words-tokens, then the information retrieved from test will capture the psychologist's biases together with the information on the student.

To tackle this problem, I conduct two analyses. In the first analysis, I explore whether the text reflects the psychologist's writing style. I program a classifier to predict from the text the psychologist who wrote the report. If the classifier is not able to capture the psychologists' writing styles, the risk of including unwanted exogenous variation in my estimates is minimized. Running a random forest classifier to predict the caseworker, my best measure of text (frequency weighted term frequency matrix) has a prediction error of 40\%, meaning that 40\% of all psychologists are misclassified. However, this measure goes as low as 59\% for word embeddings with 100 features and 73\% for stm with 80 topics. In conclude that the psychologists do not influence the distribution of text covariates in a systematic manner. 

In the second analysis, I systematically remove, for each psychologist, the most frequently and uniquely used tokens (I compute the "keyness" score across psychologists using different measures such as the chi-squared measure, likelihood ratio and point-wise mutual information). Thus, words that are too predictive of a given psychologists do not influence the estimates. I then compute my main text measures on this reduced set of features. The intuition of this procedure is somewhat similar in spirit to including psychologists' fixed effects in the regression. Results are similar to the main results presented (figures available on request).

\subsection{Selective Attrition \label{appendix:robustness.3}}

\begin{figure}[t!]
\includegraphics[width=\textwidth]{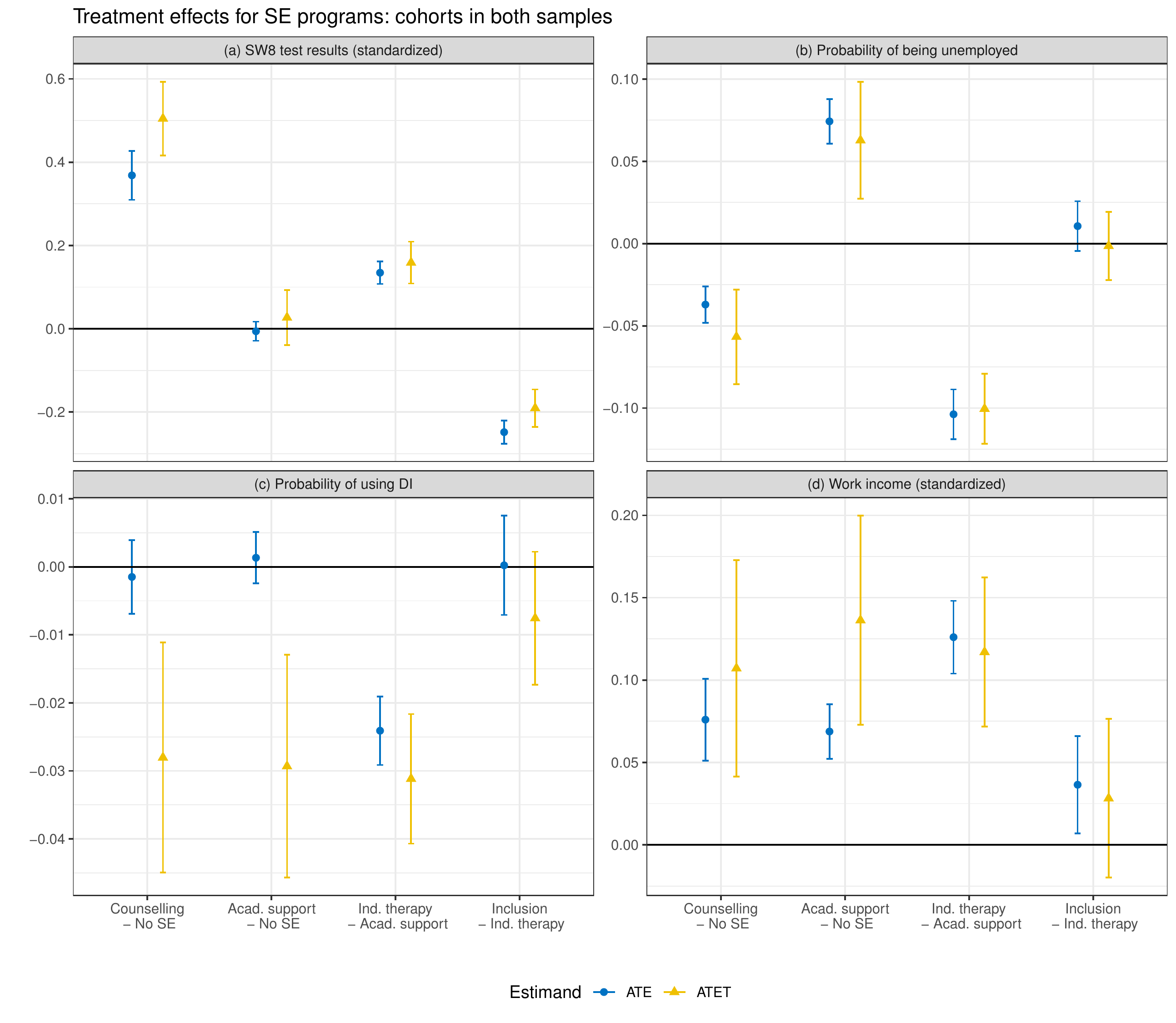}
\caption{Pairwise treatment effects for cohorts observed in all outcomes \label{fig:results.main.cohort}}
\floatfoot{This figure depicts relevant pairwise treatment effects for Special Education programs in St.\ Gallen. Each pairwise treatment effect is the effect of being assigned to the first program in comparison to the second program on one of the four outcomes presented in the column headers. Both the treatment effect on the whole population (ATE) and on the population of the treated (ATET) are presented. ``Ind. therapy'' is the abbreviation for individual therapies, ``Acad. support'' for academic support, ``segr.'' for segregation and ``no SpEd'' for receiving no program. Nuisance parameters are estimated using an ensemble learner that includes text representations presented in the ``data'' section. 95\% confidence intervals are represented and are based on one sample $t-test$ for the ATE and the ATET. \emph{Source: SPS}.\xspace}
\end{figure}

Potential selective attrition in the measured outcomes could undermine the validity of the results insofar as outcomes are not observed for all individuals. For test scores, I explicitly modeled selection into test taking and presented results above. To tackle the problem of selective attrition, I further narrow the sample to individuals for which I observe all outcomes ($N = 8993$). Estimates are presented in \Cref{fig:results.main.cohort}. Results are in line with main results.

\subsection{Exogenous placement into inclusive vs. semi-segregated Special Education programs \label{appendix:subsection_IV}}

I turn my attention to remaining potentially unaddressed selection in the estimation of the causal effect of inclusion vs. semi-segregation. I leverage residual unexplained variation in the probability to assign students to inclusion using instrumental variables in the spirit of ``judge-design'' studies \citep[see, e.g.,][]{Maestas2013}. The main source of exogenous variation in the assignment of students to SpEd education programs and to inclusion instead of semi-segregation is the variation in the probability to assign students with SEN to inclusion \emph{across} schools \emph{within} years. This variation is shown in \Cref{fig:IV_variation_school_year}. Some variation remains even when years fixed effects are introduced, and more importantly, when school-year characteristics are accounted for. Characteristics include reasons for referral, main characteristics, IQ, and whether parents decided to send the child to the SPS. School characteristics include the percentage of nonnative students, the percentage of students with SEN, the socio-economic index, the school size, and the expenditures per student.

I measure variations in the probability to assign students with SEN to inclusion \emph{across} schools \emph{within} years as the school-year deviations in assignment to inclusion from the mean inclusion assignment rate at the year level. This variation gives how much individual schools differ in the probability to assign inclusion from all the other schools within a year. I exploit treatment assignment deviations from mean rates of assignment to inclusive treatment as a instruments for assignment to inclusion. Preferences at the school level for inclusive treatment increase the likelihood that students with SEN will be sent to inclusive SpEd, especially for students with moderate difficulties. The key assumption that underlies my approach is that the assignment of students with SEN to a school-year is uncorrelated with unobserved characteristics (such as SEN severity) conditional on observed characteristics. As pointed out by \citet{Maestas2013}, this amounts to an assumption of conditional random assignment to school-year within a year. 

One potential threat to this assumption is that students could select into schools. However, my rich information on school characteristics and the referral process allows me to control for this. Moreover, a careful reading of ministry reports in St. Gallen (for instance, the \emph{Nachtrag zum Volkschulgesetz 2013}) and school documents show that schools' preferences in terms of inclusion and segregation do not, for the most case, follow a predictable pattern. Moreover,  the diagnosis and treatment assignment are done in a centralized manner by independent psychologists, and they are not influenced by financial constraints at the school level. Finally, it is documented that student mobility between school is rare in the canton of St.\ Gallen. Families must move to another municipality if they want to change school (or enroll their students in private schools) \citep{BalestraEtal2020,Balestra2020a,BalestraEtal2021}.

In \Cref{table:iv_first_stage}, I present first-stage estimates and add covariates sequentially to the regression in order to indirectly test for random assignment within year on the basis of observable characteristics. The idea is that only covariates that are correlated with the deviation will affect the estimated coefficient on the deviation when included. The coefficient on the variation is not significantly affected by the addition of variables, not even by the addition of school-year characteristics such as socio-economic status of the school or by the IQ score. This is the case both when the raw and the adjusted variation is considered. Thus, these results show that the measured variations are not correlated with student or school characteristics, and thus the preference for inclusion depends on schools' preferences. Note that, since, in a multiple treatment setting, a sound IV approach requires one instrument per treatment \citep[see ][]{Kirkeboen2016}, I focus only on SN students having received either inclusion (excluding individual therapies) or semi-segregation and restrict my dataset accordingly. This also ensures that I measure the effect for a homogeneous population of students with SN (students who have issues such that they would either assigned to inclusion or semi-segregation). However, the first stage estimates hold also when the whole sample is considered. 

\begin{table}[t!]
\centering
\footnotesize
\caption{First-Stage Regressions: Effect of preference for inclusion on assignment to inclusion} 
\label{table:iv_first_stage}
\begin{threeparttable}
\begin{tabular}{lccccc}
  
  \toprule
                                &\multicolumn{5}{c}{Probability to be assigned to inclusion} \\
                                \cmidrule(lr){2-6} 
                                &(I)          &(II)         &(III)        &(IV)         &(V)   \\
  \midrule

  Coefficient on raw variation      &0.89$^{***}$ &0.87$^{***}$ &0.82$^{***}$ &0.83$^{***}$ &0.78$^{***}$ \\
  (within year across schools)  \\  
  Coefficient on raw variation      &0.76$^{***}$ &0.76$^{***}$ &0.76$^{***}$ &0.76$^{***}$ &0.75$^{***}$ \\
  (within year across schools)  \\[1em]

  Coefficient on raw variation      &0.44$^{***}$ &0.48$^{***}$ &0.45$^{***}$ &0.44$^{***}$ &0.45$^{***}$ \\
  (within school across years)  \\  
  Coefficient on adjusted variation &0.37$^{***}$ &0.38$^{***}$ &0.39$^{***}$ &0.39$^{***}$ &0.40$^{***}$ \\
  (within school across years)  \\[2em]
  
  \textbf{Control variables included} \\
  Students' characteristics         &             &X            &X            &X            &X \\  
  IQ score                          &             &             &X            &X            &X \\  
  Reasons for referral              &             &             &             &X            &X \\  
  School characteristics per year   &             &             &             &             &X \\    
  \bottomrule

\end{tabular}
\begin{tablenotes}[para,flushleft]
\footnotesize
\emph{Notes:} First stage regression of the schools preferences for inclusion on the assignment to inclusion. All regressions include schools fixed effects for the regressions within schools, and year fixed effects for the regressions within years. Number of observations: 2932 for raw variation and 2497 for adjusted variation. $^{*}p<0.1$; $^{**}p<0.05$; $^{***}p<0.001$. \emph{Source: SPS}
\end{tablenotes}
\end{threeparttable}
\end{table}

\begin{figure}[t]
\centering
\caption{LATE for inclusion vs. semi-segregation \label{fig:IV_results}}
\begin{minipage}{0.8\linewidth}
    \includegraphics[width=1\textwidth]{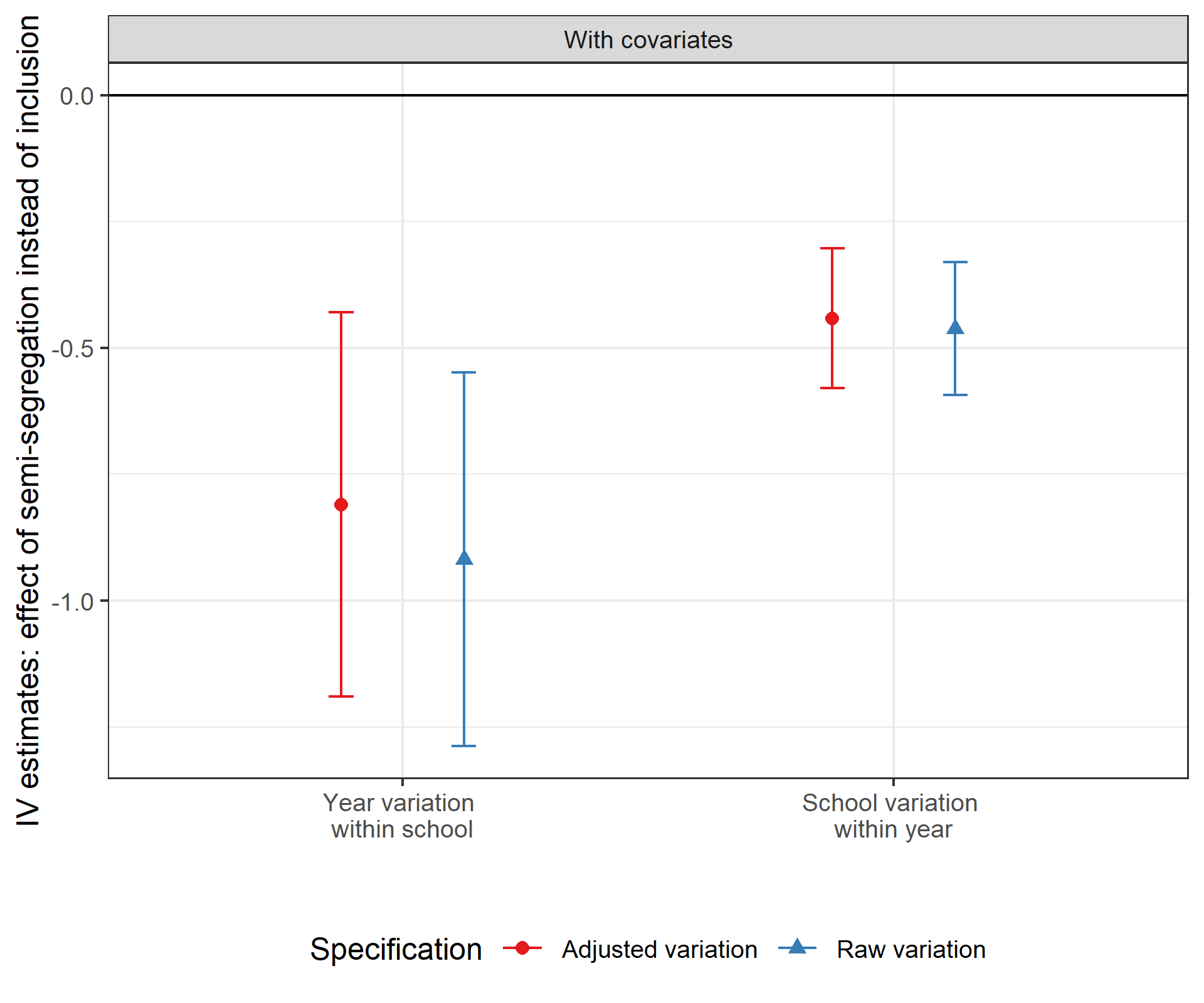}

    \footnotesize
    \emph{Notes:} This figure shows LATE estimates of the effect of inclusion for academic performance and unemployment probability for students either in inclusive or segregated settings. The estimates are found with 2SLS, and covariates include student and school characteristics. 95\% confidence intervals are represented. Both the effect for the adjusted and the raw deviations in the first stage are represented. 
\end{minipage}
\end{figure}

I estimate (Conditional) Local Average Treatment Effects of being assigned to inclusion rather than segregation on test scores using unexplained variation in the decision to implement inclusion by school within years or by year within school. Similar to \citet{Maestas2013}, I estimate 2SLS with and without covariates, both using the raw and the adjusted variation. I do not use text information in this case. Results are presented in \Cref{fig:IV_results}. LATE estimates show that semi-segregation has a negative impact on students' academic performance. The impact of semi-segregation is around -0.45 standard deviations of test score in comparison to inclusion, and it is of 10 percentage points on the probability to be unemployed. The fact that the test score estimates are substantially lower than the ATE estimates is mostly due to the fact that the populations of interest are not exactly the same: ATEs measure the effect for the whole population of students with SEN, whereas the LATE measures the effect on compliers, i.e. students who would have been assigned to semi-segregation if the school did not have a ``preference'' for inclusion. 

In conclusion, this IV exercise allowed me to exploit some existing exogenous variation in treatment assignment across schools within years to estimate treatment effects. It is, however, not the panacea. First, potential difficulties with the monotonicity might persist when more than two treatments exist. Even though I restricted my analysis to the subsample of students observed in inclusion and semi-segregation, the instrument could still move some students to other available treatments. Second, even though I observe many characteristics about schools, the (almost unsolvable) issue of not being able to observe the actual teaching behaviors and teaching styles of both the main teachers and the SpEd teachers at the school level still leaves the black box open. 

\clearpage

\setcounter{table}{0}
\setcounter{figure}{0}
\setcounter{page}{1}
\renewcommand\thetable{C.\arabic{table}}
\renewcommand\thefigure{C.\arabic{figure}}
\renewcommand{\thepage}{\roman{page}}
% !TeX root = Submission_ASallin_v9.tex

\section{Appendix: Supplementary Material}

\vspace{4cm}

\begin{table}[h]
\scriptsize
\centering
\begin{tabular}{lll}
  \toprule
Therapy       & German       & Treatment group                                  \\ 
  \midrule
Inclusive special education (ISF)           & Integrierte Schülerförderung (ISF),                            & Inclusion\\                      
                                            & Schulische Heilpädagogik                                       & \\
Special (small) classes                     & Kleinklasse                                                    & Semi-segregation \\ 
%\midrule
%Introductory classes                        & Einführungsklasse                                              & Intro. classes \textbf{(only before primary school)}\\ 
\midrule
Speech therapy                              & Logotherapie                                                   & Individual therapy\\ 
Psychomotor therapy                         & Psychomotoriktherapie                                          & Individual therapy\\ 
Dyslexia therapy                            & Legasthenie (Lese- und Rechtschreibstörung)                    & Individual therapy\\ 
Dyscalculia therapy                         & Dyskalkulietherapie                                            & Individual therapy\\
Rhythm therapy (Dalcroze eurhythmics)       & Rhythmik                                                       & Individual therapy\\ 
\midrule
Tutoring, language tutoring                 & Hilfe                                                          & Academic support \\ 
%Individual learning goals                   & (reinforcement of ISF)                                         & ILZ \\ 
\midrule
Counseling                                  & Psychologische Hilfe, Beratung                                 & Counseling \\
   \bottomrule
\end{tabular}
\caption{List of available therapies from the Cantonal offer} 
\label{table:list_therapies}
\floatfoot{\emph{Source}: ``Sonderpädagogikkonzept für die Regelschule'' from the Ministry of Education, 18.3.2015, retrieved on the official website of the Ministry of Education of St. Gallen, https://www.sg.ch/bildung-sport/volksschule/rahmenbedingungen/rechtliche-grundlagen/konzepte.html.}
\end{table}

\clearpage
\begin{table}
\centering
\footnotesize
\begin{tabular}{lr}
\toprule
\emph{Data restrictions }                                           & Number of observations \\
\midrule
Full sample of students in contact with SPS                 &28584 \\
%- Visits to the SPS prior to 1998 and after 2012            &-8941 \\
%- City of St. Gallen                                        &-5691 \\
- Trim cohorts (1982 to 2003) and missing birthdates        &-972 \\
- Native language non-imputable                             &-1288 \\
- Treatment not in cantonal offer                           &-8612 \\
%- Remove observations with no pschologist data              &-654 \\
- Treatment not identified                                  &-478 \\
\midrule
\textbf{TOTAL}                                              &\textbf{17822} \\
IQ not computed                                             &4801 \\
\bottomrule
\end{tabular}
\caption{Attrition analysis \label{table:attrition}} 
\floatfoot{}
\end{table}

\clearpage
\newgeometry{left=0.75cm, bottom = 1.5cm, top = 1.5cm }
\begin{landscape}
%\begin{sidewaystable}
\begin{table}[h]
\tiny
\begin{tabular}{lccccccc}
  \toprule
    & Counselling     & Academic support    & Individual therapy    & Inclusive special ed.  & Semi-segregation & Full segregation & No therapy  \\ 
    &                 &                     &                       & (ISF)                  &                  &                  & (but sent to SPS) \\
  \midrule
  $N$ ($N = 17,822$)   & 1,450            & 1,381           & 7,997           & 2,705           & 1,690          & 1,603             & 996 \\ 
  
  \addlinespace[1em]
  \multicolumn{5}{l}{\textbf{A: Individual characteristics}}\\
  Female                                 & 0.34             & 0.50            & 0.40            & 0.46           & 0.43           & 0.30            & 0.38  \\ 
  Foreign language                       & 0.08             & 0.15            & 0.09            & 0.11           & 0.28           & 0.14            & 0.20  \\ 
  IQ                                     & 101.47 (13.08)   & 94.24 (10.54)   & 98.26 (10.58)   & 92.94 (9.41)   & 86.17 (8.96)   & 87.01 (14.99)   & 93.63 (11.25) \\ 
  IQ measured                            & 0.62             & 0.73            & 0.74            & 0.80           & 0.78           & 0.67            & 0.61 \\ 
  Birth year                             & 1994.43 (4.53)   & 1993.86 (4.25)  & 1995.14 (4.33)  & 1996.68 (3.73) & 1993.61 (3.64) & 1995.80 (4.31)  & 1999.18 (3.41) \\ 
  Age at first interview                 & 9.12 (285)       & 9.47 (2.29)     & 8.69 (1.96)     & 8.75 (2.03)    & 9.11 (2.50)    & 7.19 (2.63)     & 6.24 (0.60) \\ 
  Had bridge year                        & 0.08             & 0.08            & 0.08            & 0.07           & 0.13           & 0.08            & 1.00 \\ 
  Reasons: other                         & 0.04             & 0.03            & 0.03            & 0.02           & 0.06           & 0.10            & 0.10 \\ 
  Reasons: social and emotional problems & 0.48             & 0.19            & 0.15            & 0.18           & 0.22           & 0.28            & 0.24 \\ 
  Reasons: performance and learning problems   & 0.68       & 0.91            & 0.92            & 0.94           & 0.88           & 0.78            & 0.87 \\ 
  Reasons: problems with teachers or school    & 0.06       & 0.02            & 0.01            & 0.05           & 0.02           & 0.04            & 0.03 \\ 
  Reasons: not specified                 & 0.03             & 0.01            & 0.01            & 0.00           & 0.01           & 0.02            & 0.00 \\ 
  Referred by: Caseworker                & 0.00             & 0.01            & 0.04            & 0.02           & 0.01           & 0.05            & 0.01 \\ 
  Referred by: Other                     & 0.04             & 0.02            & 0.02            & 0.01           & 0.03           & 0.07            & 0.02 \\ 
  Referred by: Parents                   & 0.15             & 0.06            & 0.05            & 0.04           & 0.02           & 0.06            & 0.03 \\ 
  Referred by: Parents and teacher       & 0.62             & 0.66            & 0.71            & 0.64           & 0.61           & 0.53            & 0.58 \\ 
  Referred by: Teacher                   & 0.19             & 0.25            & 0.18            & 0.28           & 0.33           & 0.29            & 0.36 \\ 
  Total number of SPS visits             & 10.69 (8.08)     & 8.77 (5.94)     & 9.10 (6.18)     & 10.27 (7.72)   & 13.58 (9.25)   & 19.62 (14.79)   & 6.14 (3.93) \\ 
  Regional office: Ro                    & 0.12             & 0.22            & 0.15            & 0.04           & 0.15           & 0.14            & 0.15  \\ 
  Regional office: Go                    & 0.09             & 0.07            & 0.13            & 0.02           & 0.12           & 0.12            & 0.04  \\ 
  Regional office: Wi                    & 0.15             & 0.12            & 0.18            & 0.12           & 0.25           & 0.13            & 0.12  \\ 
  Regional office: Wa                    & 0.22             & 0.10            & 0.12            & 0.17           & 0.07           & 0.19            & 0.11  \\ 
  Regional office: Ra                    & 0.12             & 0.09            & 0.07            & 0.38           & 0.06           & 0.20            & 0.13  \\ 
  Regional office: Sa                    & 0.23             & 0.16            & 0.18            & 0.20           & 0.18           & 0.15            & 0.20  \\ 
  Regional office: Re                    & 0.07             & 0.24            & 0.17            & 0.06           & 0.18           & 0.08            & 0.25  \\ 

  School: percent nonnatives             & 0.22 (0.10)      &0.21 (0.10)      & 0.22 (0.10)     & 0.20 (0.09)    & 0.26 (0.10)    & 0.22 (0.11)     & 0.24 (0.10)   \\
  School: percent SEN students           & 0.18 (0.10)      &0.18 (0.09)      & 0.18 (0.10)     & 0.18 (0.11)    & 0.18 (0.10)    & 0.18 (0.11)     & 0.19 (0.11)       \\
  School: social index                   & 0.98 (0.07)      &0.98 (0.07)      & 0.98 (0.07)     & 0.95 (0.06)    & 1.01 (0.07)    & 0.98 (0.07)     & 0.99 (0.07)  \\
  School: total number of students       & 176.37 (158.92)  &151.55 (123.93)  & 161.58 (133.71) & 174.97 (194.00)& 211.08 (161.88)& 173.50 (150.32) & 165.58 (139.30)    \\
  School: expenditures per student (2017)& -0.01 (0.95)     &0.06 (1.05)      &-0.05 (0.97)     & 0.28 (1.03)    & -0.23 (0.88)   & 0.03 (1.10)     & 0.01 (1.05)    \\
  School: urban                          & 0.48 (0.50)      &0.43 (0.50)      &0.45 (0.50)      & 0.37 (0.48)    & 0.60 (0.49)    & 0.47 (0.50)     & 0.44 (0.50)     \\

  \addlinespace[1em]
  \multicolumn{5}{l}{\textbf{B: Sample attrition}}  \\
  In a SW8 cohort                        & 0.72             & 0.67            & 0.75            & 0.89           & 0.69           & 0.82            & 0.99  \\ 
  In AHV data                            & 0.73             & 0.76            & 0.70            & 0.58           & 0.81           & 0.64            & 0.33  \\ 
  In both SW8 and AHV data               & 0.47             & 0.46            & 0.46            & 0.47           & 0.53           & 0.47            & 0.32  \\ 
  
  \addlinespace[1em]
  \multicolumn{5}{l}{\textbf{C: Outcomes}}  \\
  SW8 Test taken (in SW8 cohort)         & 0.74             & 0.81            & 0.82            & 0.86            & 0.80          & 0.41            & 0.63  \\ 
  SW8 Math and German (std)              & 0.57 (1.05)      & -0.09 (0.86)    & 0.24 (0.88)     & -0.24 (0.85)    & -1.12 (0.92)  & -0.20 (1.12)    & 0.17 (0.95) \\ 
  %Chose VET                              & 0.54  & 0.64  & 0.61  & 0.54  & 0.65  & 0.36  & 0.30  \\ 
  %Chose no further education             & 0.37  & 0.33  & 0.35  & 0.44  & 0.35  & 0.62  & 0.66  \\ 
  Used disability insurance              & 0.06  & 0.05  & 0.03  & 0.04  & 0.11  & 0.38  & 0.07  \\ 
  Used unemployment insurance            & 0.24  & 0.32  & 0.19  & 0.19  & 0.39  & 0.25  & 0.19  \\ 
  Last registered yearly wage (std., SSA cohort) &   -0.09 (1.03) &   -0.06 (0.96) &    0.15 (0.97) &    0.12 (0.98) &   -0.16 (1.00) &   -0.57 (0.95) &   -0.15 (0.95) \\ 
   \bottomrule
\end{tabular}
\caption{Summary statistics per treatment group} 
\floatfoot{\scriptsize Summary statistics for the population of students referred to the SPS in the Canton of St.\ Gallen. The names of Regional offices are abbreviated for confidentiality purposes. The sample is composed of SN students from the Canton of St.\ Gallen having visited the SPS between 1998 and 2010 and being born between 1982 and 2003. Mean per treatment groups are reported, and standard deviations are reported in parentheses for continuous variables. \emph{Source: SPS}}
\label{table:summary_stats_treatment}
\end{table}
%\end{sidewaystable}
\end{landscape}
\restoregeometry

\begin{landscape}
\begin{table}[h]
\centering 
 \tiny
\begin{tabular}{lcccccccccccccccccccccc}
  \toprule
  & \rot{Average } & \rot{Counseling vs Acad. support } & \rot{Counseling vs Ind. ther. } & \rot{Counseling vs ISF } & \rot{Counseling vs Semi-segr. } & \rot{Counseling vs Full segr. } & \rot{Counseling vs No treatment } & \rot{Acad. support vs Ind. ther. } & \rot{Acad. support vs ISF } & \rot{Acad. support vs Semi-segr. } & \rot{Acad. support vs Full segr. } & \rot{Acad. support vs No treatment } & \rot{Ind. ther. vs ISF } & \rot{Ind. ther. vs Semi-segr. } & \rot{Ind. ther. vs Full segr. } & \rot{Ind. ther. vs No treatment } & \rot{ISF vs Semi-segr. } & \rot{ISF vs Full segr. } & \rot{ISF vs No treatment } & \rot{Semi-segr. vs Full segr. } & \rot{Semi-segr. vs No treatment } & \rot{Full segr. vs No treatment} \\ 
 %& Average & Couns. vs & Couns. vs & Couns. vs & Couns. vs & Couns. vs & Couns. vs & Tutor. vs & Tutor. vs & Tutor. vs & Tutor. vs & Tutor. vs & Ind. ther. vs & Ind. ther. vs & Ind. ther. vs & Ind. ther. vs & ISF vs & ISF vs & ISF vs & Semi-segr. vs & Semi-segr. vs & Full segr. vs \\ 
 %&      & Tutor.   & Ind.ther.   & ISF   & Semi-segr.    & Full segr.    & No    & Ind. ther.  & ISF  & Semi-segr.  & Full segr. & No & ISF & Semi-segr. & Full segr. & No & Semi-segr. & Full segr. & No & Full segr. & No & No. \\
  \midrule
Female & 0.17 & 0.33 & 0.13 & 0.26 & 0.19 & 0.08 & 0.09 & 0.20 & 0.08 & 0.14 & 0.41 & 0.24 & 0.12 & 0.05 & 0.21 & 0.04 & 0.07 & 0.34 & 0.16 & 0.27 & 0.09 & 0.17 \\ 
  Foreign language & 0.23 & 0.23 & 0.04 & 0.11 & 0.53 & 0.19 & 0.37 & 0.19 & 0.12 & 0.30 & 0.04 & 0.14 & 0.07 & 0.49 & 0.15 & 0.33 & 0.42 & 0.07 & 0.25 & 0.35 & 0.17 & 0.18 \\ 
  IQ & 0.58 & 0.61 & 0.27 & 0.75 & 1.37 & 1.03 & 0.64 & 0.38 & 0.13 & 0.82 & 0.56 & 0.06 & 0.53 & 1.23 & 0.87 & 0.42 & 0.74 & 0.47 & 0.07 & 0.07 & 0.73 & 0.50 \\ 
  IQ measured & 0.20 & 0.24 & 0.27 & 0.40 & 0.36 & 0.10 & 0.01 & 0.03 & 0.16 & 0.12 & 0.14 & 0.25 & 0.12 & 0.09 & 0.17 & 0.29 & 0.03 & 0.30 & 0.41 & 0.26 & 0.38 & 0.11 \\ 
  Birth year & 0.58 & 0.13 & 0.16 & 0.54 & 0.20 & 0.31 & 1.19 & 0.30 & 0.71 & 0.06 & 0.45 & 1.38 & 0.38 & 0.38 & 0.15 & 1.04 & 0.83 & 0.22 & 0.70 & 0.55 & 1.58 & 0.87 \\ 
  Age at first interview & 0.67 & 0.13 & 0.18 & 0.15 & 0.01 & 0.71 & 1.40 & 0.36 & 0.33 & 0.15 & 0.92 & 1.92 & 0.03 & 0.19 & 0.65 & 1.69 & 0.16 & 0.66 & 1.67 & 0.75 & 1.57 & 0.49 \\ 
  Had bridge year & 1.36 & 0.02 & 0.01 & 0.05 & 0.16 & 0.00 & 4.64 & 0.01 & 0.04 & 0.18 & 0.02 & 4.79 & 0.04 & 0.17 & 0.01 & 4.71 & 0.21 & 0.06 & 5.11 & 0.16 & 3.59 & 4.61 \\ 
  Reasons: other & 0.16 & 0.07 & 0.08 & 0.12 & 0.06 & 0.22 & 0.20 & 0.01 & 0.05 & 0.13 & 0.29 & 0.27 & 0.04 & 0.13 & 0.29 & 0.27 & 0.17 & 0.33 & 0.31 & 0.17 & 0.15 & 0.02 \\ 
  Reasons: social and emotional problems & 0.27 & 0.64 & 0.75 & 0.66 & 0.57 & 0.43 & 0.51 & 0.11 & 0.02 & 0.06 & 0.20 & 0.12 & 0.09 & 0.17 & 0.31 & 0.23 & 0.09 & 0.22 & 0.15 & 0.13 & 0.06 & 0.07 \\ 
  Reasons: performance and learning problems & 0.30 & 0.59 & 0.64 & 0.71 & 0.50 & 0.22 & 0.46 & 0.04 & 0.13 & 0.09 & 0.37 & 0.14 & 0.08 & 0.14 & 0.41 & 0.18 & 0.22 & 0.49 & 0.26 & 0.27 & 0.04 & 0.23 \\ 
  Reasons: problems with teachers or school & 0.11 & 0.20 & 0.22 & 0.01 & 0.17 & 0.07 & 0.15 & 0.03 & 0.19 & 0.03 & 0.13 & 0.05 & 0.22 & 0.06 & 0.16 & 0.08 & 0.17 & 0.06 & 0.15 & 0.11 & 0.02 & 0.09 \\ 
  Reasons: not specified & 0.11 & 0.10 & 0.11 & 0.22 & 0.11 & 0.07 & 0.23 & 0.01 & 0.14 & 0.01 & 0.03 & 0.16 & 0.14 & 0.00 & 0.04 & 0.15 & 0.14 & 0.17 & 0.04 & 0.04 & 0.15 & 0.18 \\ 
  Referred by: Caseworker & 0.14 & 0.06 & 0.25 & 0.16 & 0.06 & 0.29 & 0.03 & 0.20 & 0.11 & 0.01 & 0.25 & 0.03 & 0.10 & 0.20 & 0.06 & 0.23 & 0.11 & 0.15 & 0.14 & 0.26 & 0.03 & 0.28 \\ 
  Referred by: Other & 0.09 & 0.09 & 0.07 & 0.14 & 0.11 & 0.09 & 0.10 & 0.02 & 0.05 & 0.03 & 0.17 & 0.01 & 0.08 & 0.05 & 0.15 & 0.03 & 0.03 & 0.22 & 0.04 & 0.20 & 0.01 & 0.18 \\ 
  Referred by: Parents & 0.17 & 0.32 & 0.37 & 0.40 & 0.46 & 0.29 & 0.42 & 0.05 & 0.09 & 0.15 & 0.03 & 0.10 & 0.04 & 0.11 & 0.08 & 0.06 & 0.07 & 0.12 & 0.02 & 0.19 & 0.05 & 0.14 \\ 
  Referred by: Parents and teacher & 0.13 & 0.08 & 0.18 & 0.04 & 0.01 & 0.16 & 0.08 & 0.10 & 0.04 & 0.08 & 0.24 & 0.16 & 0.14 & 0.19 & 0.35 & 0.26 & 0.05 & 0.20 & 0.12 & 0.16 & 0.07 & 0.08 \\ 
  Referred by: Teacher & 0.18 & 0.16 & 0.01 & 0.22 & 0.33 & 0.25 & 0.38 & 0.18 & 0.06 & 0.16 & 0.09 & 0.22 & 0.24 & 0.34 & 0.26 & 0.40 & 0.10 & 0.02 & 0.16 & 0.08 & 0.06 & 0.13 \\ 
  Referred by: NA & 0.10 & 0.02 & 0.11 & 0.11 & 0.04 & 0.12 & 0.04 & 0.09 & 0.09 & 0.06 & 0.14 & 0.02 & 0.00 & 0.14 & 0.21 & 0.07 & 0.14 & 0.22 & 0.07 & 0.09 & 0.08 & 0.16 \\ 
  Total number of SPS visits & 0.55 & 0.27 & 0.22 & 0.05 & 0.33 & 0.75 & 0.72 & 0.06 & 0.22 & 0.62 & 0.96 & 0.52 & 0.17 & 0.57 & 0.93 & 0.57 & 0.39 & 0.79 & 0.68 & 0.49 & 1.05 & 1.25 \\ 
  Regional office: Ro & 0.17 & 0.26 & 0.07 & 0.30 & 0.07 & 0.06 & 0.07 & 0.19 & 0.55 & 0.19 & 0.20 & 0.19 & 0.37 & 0.00 & 0.01 & 0.00 & 0.37 & 0.36 & 0.37 & 0.01 & 0.01 & 0.02 \\ 
  Regional office: Go & 0.19 & 0.05 & 0.13 & 0.28 & 0.11 & 0.09 & 0.19 & 0.19 & 0.23 & 0.16 & 0.14 & 0.13 & 0.40 & 0.02 & 0.04 & 0.31 & 0.38 & 0.36 & 0.10 & 0.02 & 0.29 & 0.27 \\ 
  Regional office: Wi & 0.14 & 0.09 & 0.09 & 0.08 & 0.25 & 0.06 & 0.07 & 0.19 & 0.01 & 0.34 & 0.04 & 0.02 & 0.17 & 0.16 & 0.15 & 0.16 & 0.33 & 0.02 & 0.01 & 0.30 & 0.32 & 0.01 \\ 
  Regional office: Wa & 0.19 & 0.34 & 0.26 & 0.12 & 0.43 & 0.09 & 0.30 & 0.09 & 0.22 & 0.09 & 0.26 & 0.04 & 0.14 & 0.18 & 0.17 & 0.05 & 0.31 & 0.03 & 0.18 & 0.35 & 0.13 & 0.22 \\ 
  Regional office: Ra & 0.32 & 0.07 & 0.14 & 0.65 & 0.19 & 0.23 & 0.05 & 0.07 & 0.72 & 0.12 & 0.30 & 0.13 & 0.79 & 0.05 & 0.37 & 0.20 & 0.84 & 0.42 & 0.60 & 0.42 & 0.24 & 0.18 \\ 
  Regional office: Sa & 0.08 & 0.17 & 0.13 & 0.09 & 0.14 & 0.21 & 0.08 & 0.04 & 0.09 & 0.03 & 0.04 & 0.09 & 0.05 & 0.00 & 0.08 & 0.05 & 0.05 & 0.13 & 0.00 & 0.07 & 0.06 & 0.13 \\ 
  Regional office: Re & 0.27 & 0.46 & 0.29 & 0.04 & 0.32 & 0.03 & 0.49 & 0.18 & 0.50 & 0.15 & 0.44 & 0.02 & 0.33 & 0.03 & 0.26 & 0.20 & 0.36 & 0.07 & 0.53 & 0.29 & 0.17 & 0.46 \\ 
  In a SW8 cohort & 0.39 & 0.11 & 0.07 & 0.43 & 0.07 & 0.24 & 0.83 & 0.18 & 0.54 & 0.04 & 0.35 & 0.94 & 0.35 & 0.14 & 0.17 & 0.76 & 0.49 & 0.18 & 0.44 & 0.31 & 0.89 & 0.60 \\ 
  In AHV data & 0.40 & 0.06 & 0.08 & 0.33 & 0.19 & 0.21 & 0.89 & 0.14 & 0.39 & 0.13 & 0.27 & 0.97 & 0.25 & 0.27 & 0.13 & 0.80 & 0.53 & 0.12 & 0.53 & 0.40 & 1.13 & 0.65 \\ 
  In both SW8 and AHV data & 0.13 & 0.03 & 0.02 & 0.00 & 0.12 & 0.01 & 0.32 & 0.01 & 0.03 & 0.14 & 0.02 & 0.30 & 0.02 & 0.14 & 0.01 & 0.30 & 0.12 & 0.01 & 0.32 & 0.13 & 0.45 & 0.32 \\ 
  SW8 Test taken (in SW8 cohort) & 0.39 & 0.17 & 0.19 & 0.30 & 0.14 & 0.70 & 0.22 & 0.02 & 0.13 & 0.03 & 0.89 & 0.39 & 0.11 & 0.05 & 0.92 & 0.42 & 0.16 & 1.05 & 0.53 & 0.86 & 0.36 & 0.46 \\ 
  SW8 Math and German (std) & 0.64 & 0.68 & 0.34 & 0.85 & 1.71 & 0.71 & 0.40 & 0.38 & 0.18 & 1.15 & 0.11 & 0.28 & 0.56 & 1.52 & 0.44 & 0.08 & 0.99 & 0.04 & 0.46 & 0.90 & 1.38 & 0.35 \\ 
  %Chose VET & 0.33 & 0.19 & 0.15 & 0.01 & 0.22 & 0.37 & 0.50 & 0.05 & 0.19 & 0.03 & 0.57 & 0.71 & 0.14 & 0.07 & 0.52 & 0.66 & 0.21 & 0.37 & 0.51 & 0.60 & 0.74 & 0.13 \\ 
  %Chose no further education & 0.32 & 0.08 & 0.05 & 0.14 & 0.05 & 0.52 & 0.59 & 0.04 & 0.22 & 0.04 & 0.61 & 0.68 & 0.18 & 0.00 & 0.57 & 0.64 & 0.18 & 0.38 & 0.45 & 0.57 & 0.64 & 0.07 \\ 
  Used disability insurance & 0.34 & 0.02 & 0.12 & 0.10 & 0.19 & 0.84 & 0.05 & 0.10 & 0.08 & 0.21 & 0.86 & 0.07 & 0.02 & 0.31 & 0.94 & 0.17 & 0.29 & 0.93 & 0.15 & 0.65 & 0.15 & 0.80 \\ 
  Used unemployment insurance & 0.20 & 0.18 & 0.12 & 0.13 & 0.33 & 0.01 & 0.12 & 0.30 & 0.31 & 0.15 & 0.17 & 0.30 & 0.01 & 0.45 & 0.13 & 0.00 & 0.46 & 0.14 & 0.01 & 0.32 & 0.45 & 0.13 \\ 
  Income (std.) & 0.16 & 0.01 & 0.05 & 0.07 & 0.10 & 0.32 & 0.09 & 0.07 & 0.08 & 0.10 & 0.34 & 0.09 & 0.02 & 0.17 & 0.41 & 0.16 & 0.18 & 0.40 & 0.17 & 0.26 & 0.00 & 0.24 \\ 
   \bottomrule
\end{tabular}
\caption{Summary statistics: SMD comparison} 
\floatfoot{Standardized mean differences (SMD) across SE programs. For two SE programs $w$ and $w'$, SMDs are computed as $\frac{\bar{x}_w - \bar{x}_{w'}}{\frac{\sqrt{s^{2}_{w} + s^{2}_{w'}}}{2}}$, where $\bar{x}_w$ is the mean of the covariate in treatment group $w$ and $s^2_w$ is the sample variance of covariate in treatment group $w$. A SMD above 0.2 is considered as an important difference across groups. ``Acad. support'' is academic support, ISF is inclusion, Semi-segr. is semi-segregation. \emph{Source: SPS}}
\label{table:summary_stats_SMD}
\end{table}
\end{landscape}

%\newgeometry{left=2.5cm, bottom = 1.5cm, top = 1.5cm }
\begin{landscape}
\begin{figure}[t!]
\centering
\includegraphics[width = \textwidth]{"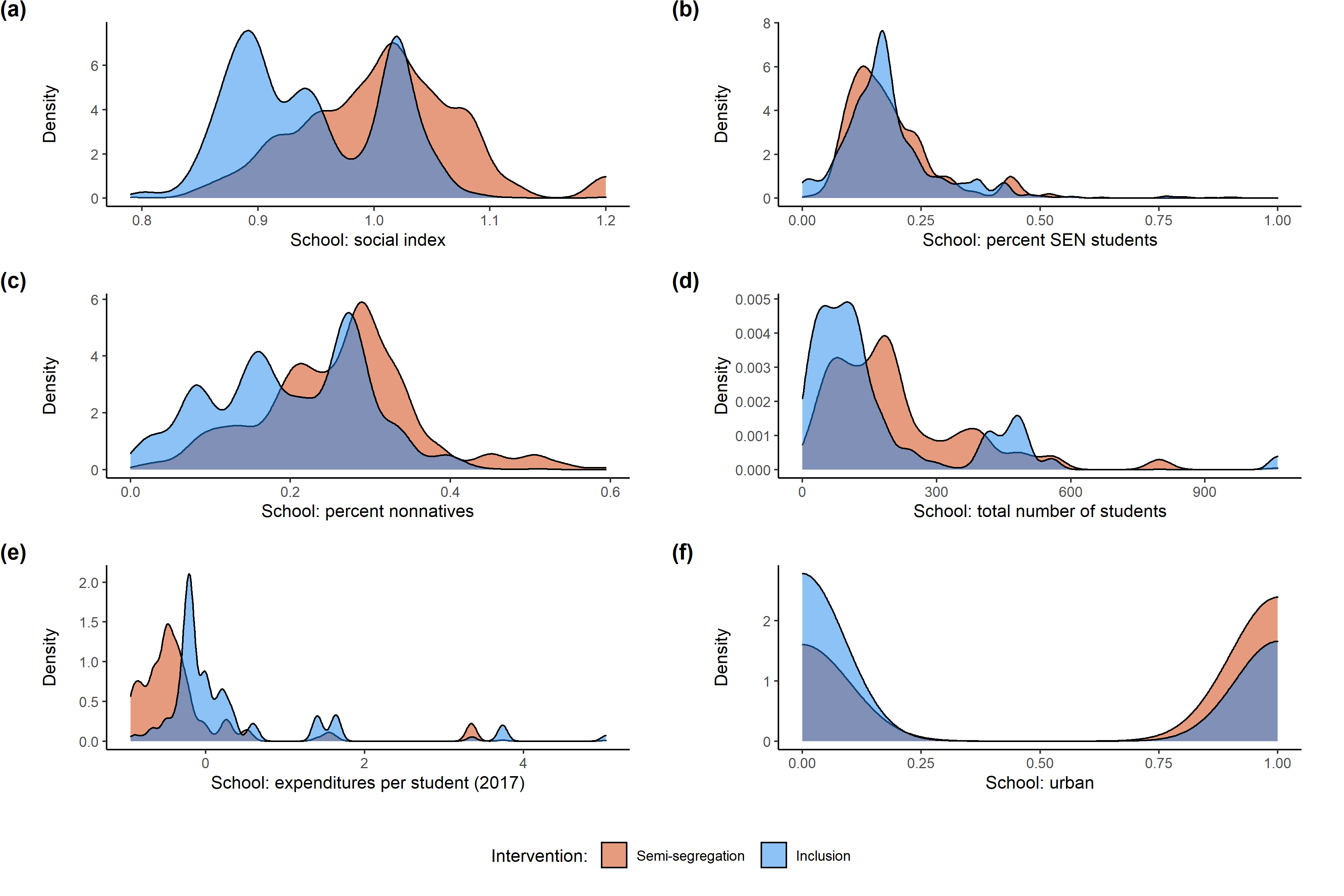"}
\caption{Overlap of school characteristics for schools with inclusion and for schools with semi-segregation \label{fig:KKvsISF_SCHOOLS}}
\floatfoot{\emph{Notes:} This figure represents the distribution of schools with inclusion and of schools with semi-segregation along main school characteristics.}
\end{figure}
\end{landscape}
%\restoregeometry

%\clearpage
%\input{"Tables/optimal.policy.COMPOSITE.tex"}

\begin{comment}
\clearpage
\begin{figure}[t!]
\includegraphics[width=\textwidth]{Graphs/IDENTIFICATION_prediction_covariates.png} 
\caption{Prediction of main covariates from text \label{fig:identification_prediction_covariates}}
\floatfoot{\emph{Notes:} This figure represents the fraction of missclassified variables for the main covariates. For each covariate, I train a classifier on my text measures to predict the covariate status. For the IQ, I run a regression and compute the Mean Absolute Error to assess the accuracy of my predictors. Classification and regression algorithms are the methods I use in my main specification (random forest, lasso and elastic net). See \Cref{table:listmeth} and \Cref{appendix:appendix_text} for more details on the text retrieval methods used in this figure.}
\end{figure}
%\clearpage
%\floatfoot{\setstretch{1.0}First stage estimates of whether a given school has implemented inclusive SpEdprograms on the probability for children to be assigned to inclusive programs. Sample includes all SN students but the ones sent to full segregation. \emph{Source: SPS, Pensenpool}.}
%\input{"Tables/IV_FULL.tex"}
\end{comment}

\begin{figure}[p]
\centering
\caption{Distribution of age and grade at registration \label{fig:registration}}
\begin{minipage}{\linewidth}
    \includegraphics[width=1\textwidth]{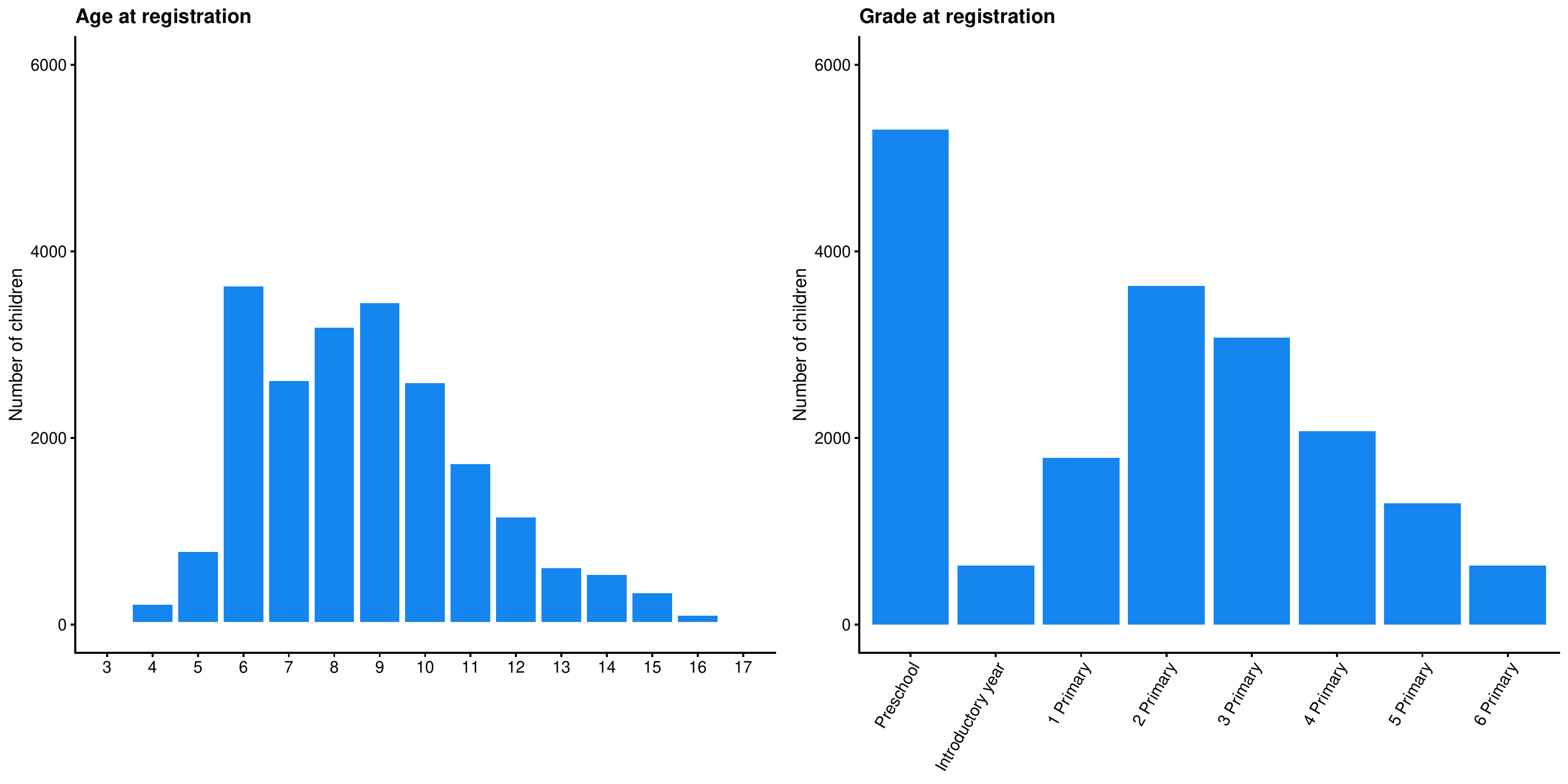}

    \footnotesize
    \emph{Notes:} This figure shows the distribution of age and grade at registration to the School Psychological Service. The ``Preschool'' category groups three years of Kindergarten, which explains why it appears to be the largest category. The number of children who have been registered at the SPS (assigned and non-assigned to a particular therapy) is represented. \emph{Source: SPS}.
\end{minipage}
\end{figure}

\clearpage
\begin{figure}
\centering
\caption{Pairwise treatment effects for probability of taking the SW8 test \label{fig:results.SW8.nontrimmed}}
\begin{minipage}{\linewidth}
    \includegraphics[width=\textwidth]{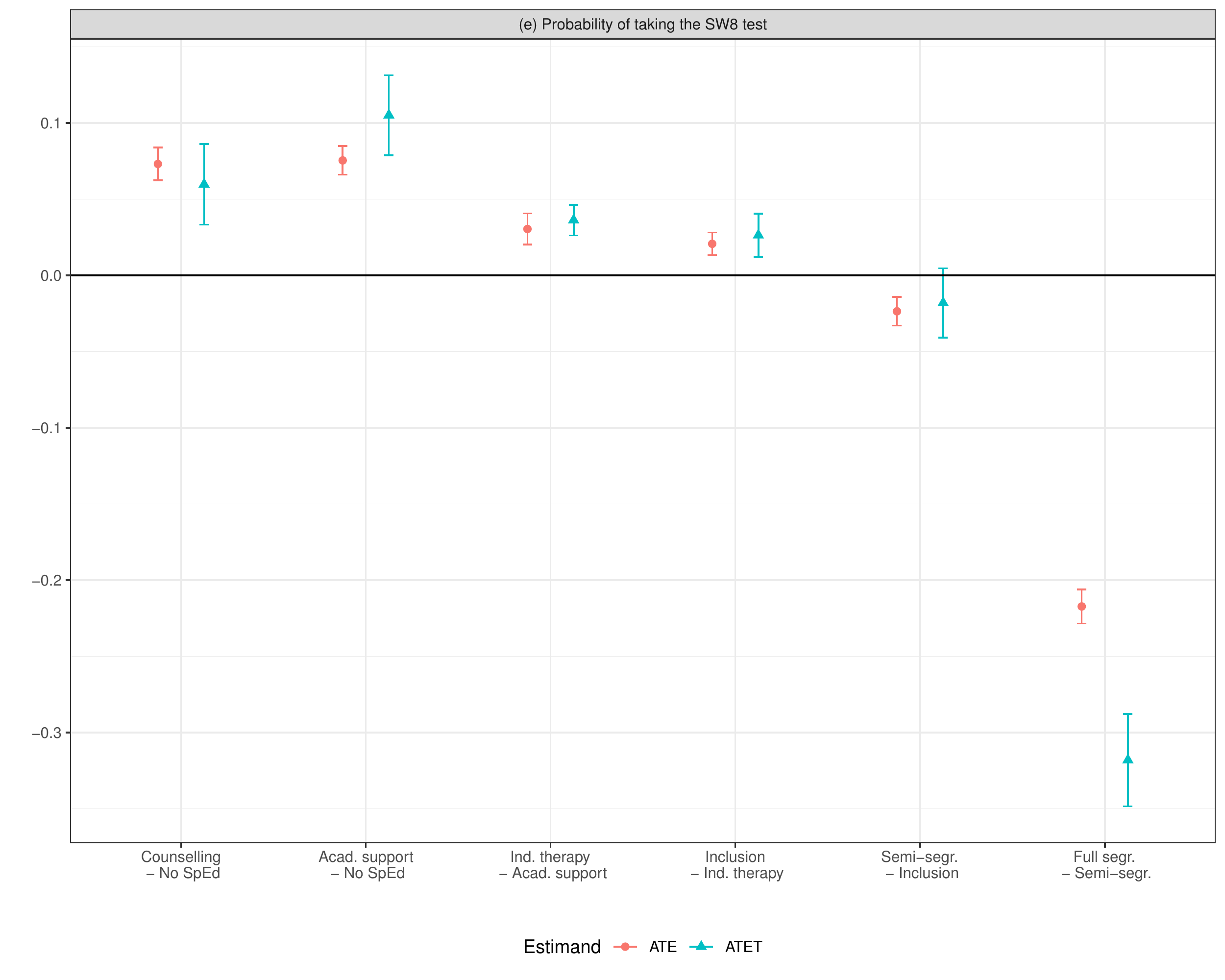}
    
    \footnotesize \emph{Notes:} 
    This figure depicts relevant pairwise treatment effects for Special Education programs in St.\ Gallen. Each pairwise treatment effect is the effect of being assigned to the first program in comparison to the second program on one of the four outcomes presented in the panel headers. Both the treatment effect on the whole population (ATE) and on the population of the treated (ATET) are presented. ``Ind. therapy'' is the abbreviation for individual therapies, ``Acad. support'' for academic support, and ``no SpEd'' for receiving no program, ``Semi-segr.'' for semi-segregation (segregation in small classes), and ``Full segr.'' for full segregation (in special schools). Nuisance parameters are estimated using an ensemble learner that includes text representations presented in the ``data'' section. 95\% confidence intervals are represented and are based on one sample $t-test$ for the ATE and the ATET. \emph{Source: SPS}.
\end{minipage}
\end{figure}

\newgeometry{left=1.3cm, right = 1.3cm}
\begin{figure}[p]
\centering
\caption[short]{Academic performance: continuous CATE for school characteristics \label{fig:CATE_SCHOOLS_score}}
\begin{minipage}{\linewidth}
\begin{subfigure}{.5\textwidth}
    \centering
    \includegraphics[width=0.95\textwidth]{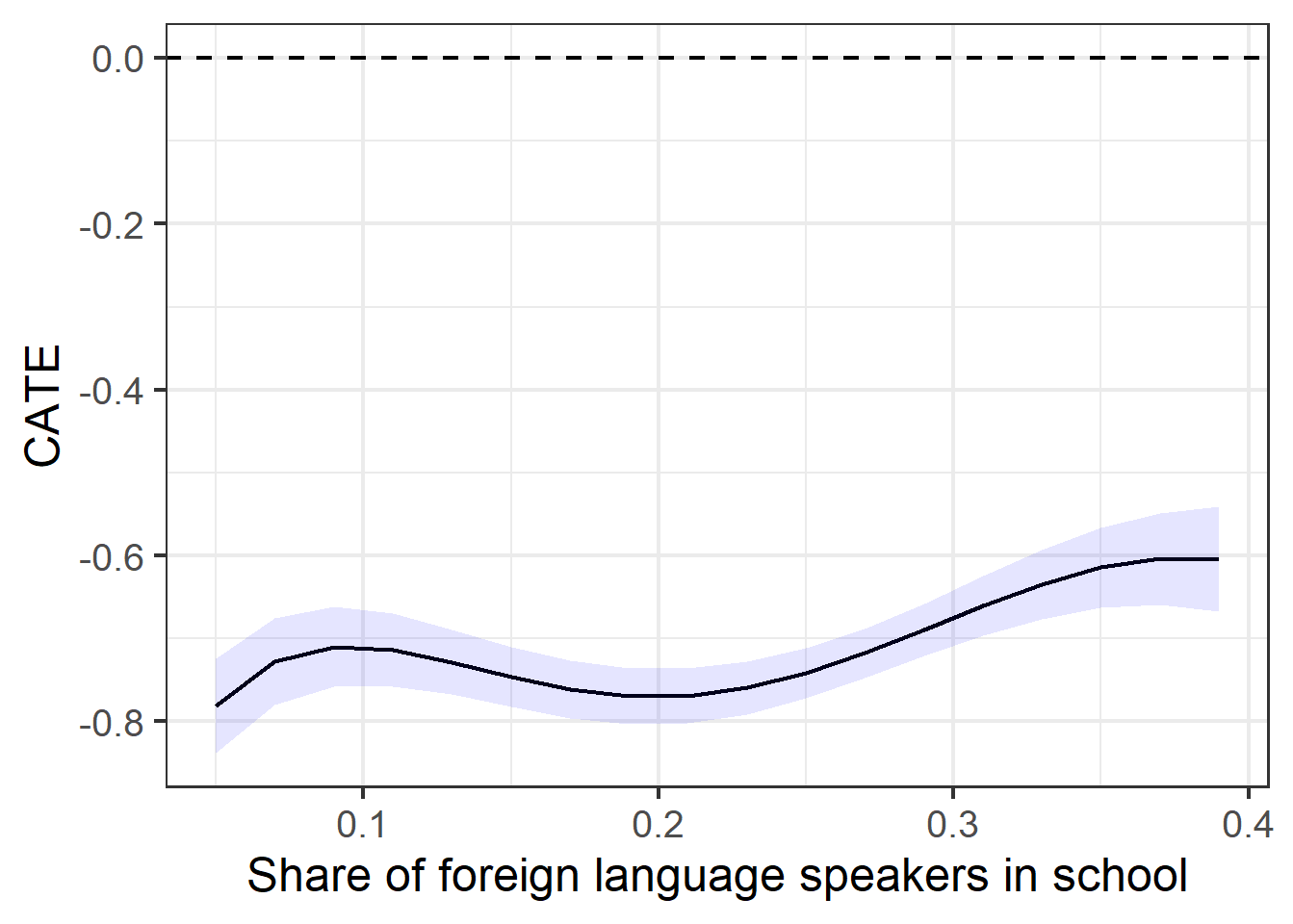}
    \caption{Schools: share of foreign students}
\end{subfigure}%
\begin{subfigure}{.5\textwidth}
    \centering
    \includegraphics[width=0.95\textwidth]{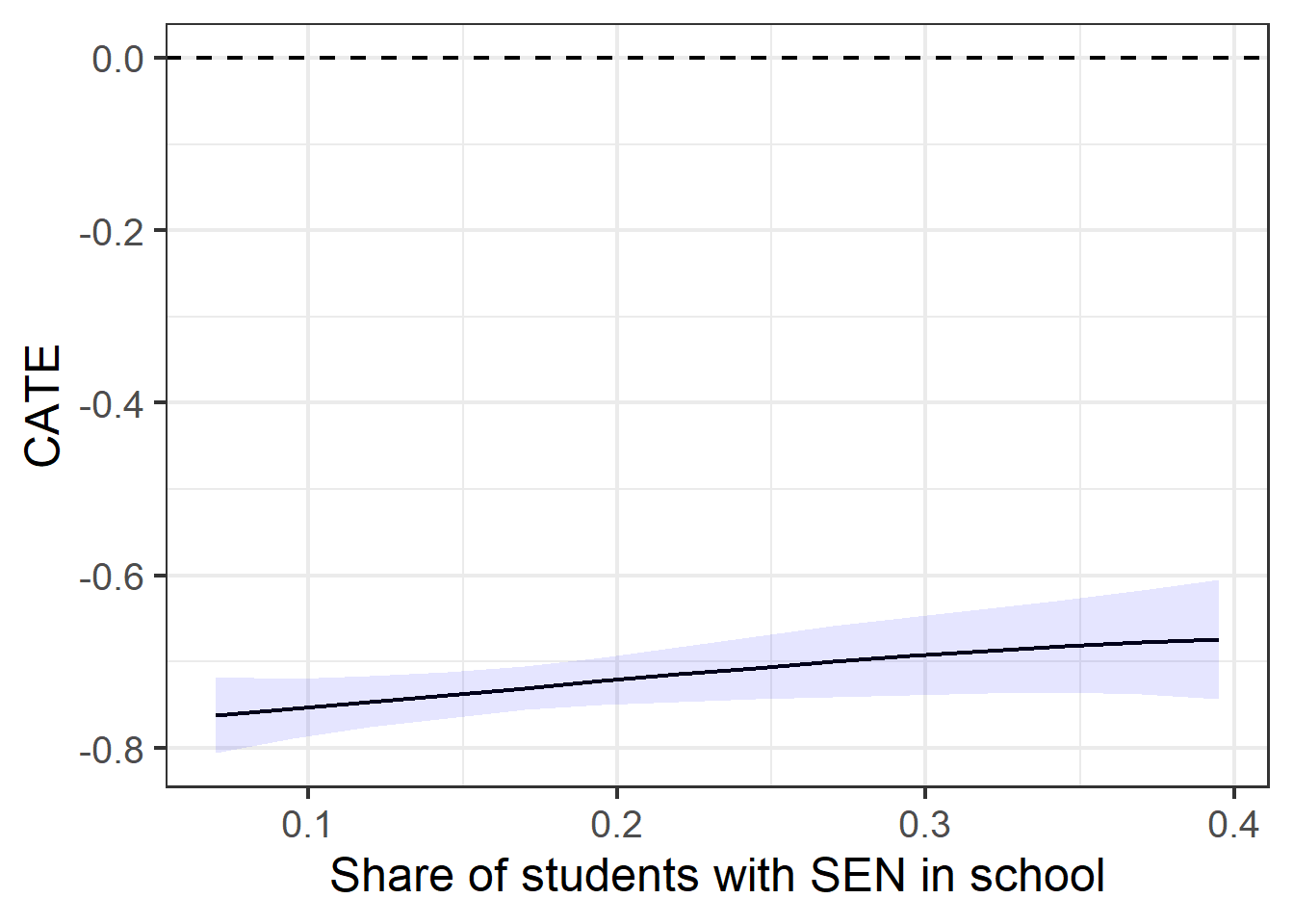}
    \caption{Schools: share of SEN students}
\end{subfigure}
\vspace{10mm}
\begin{subfigure}{.5\textwidth}
    \centering
    \includegraphics[width=0.95\textwidth]{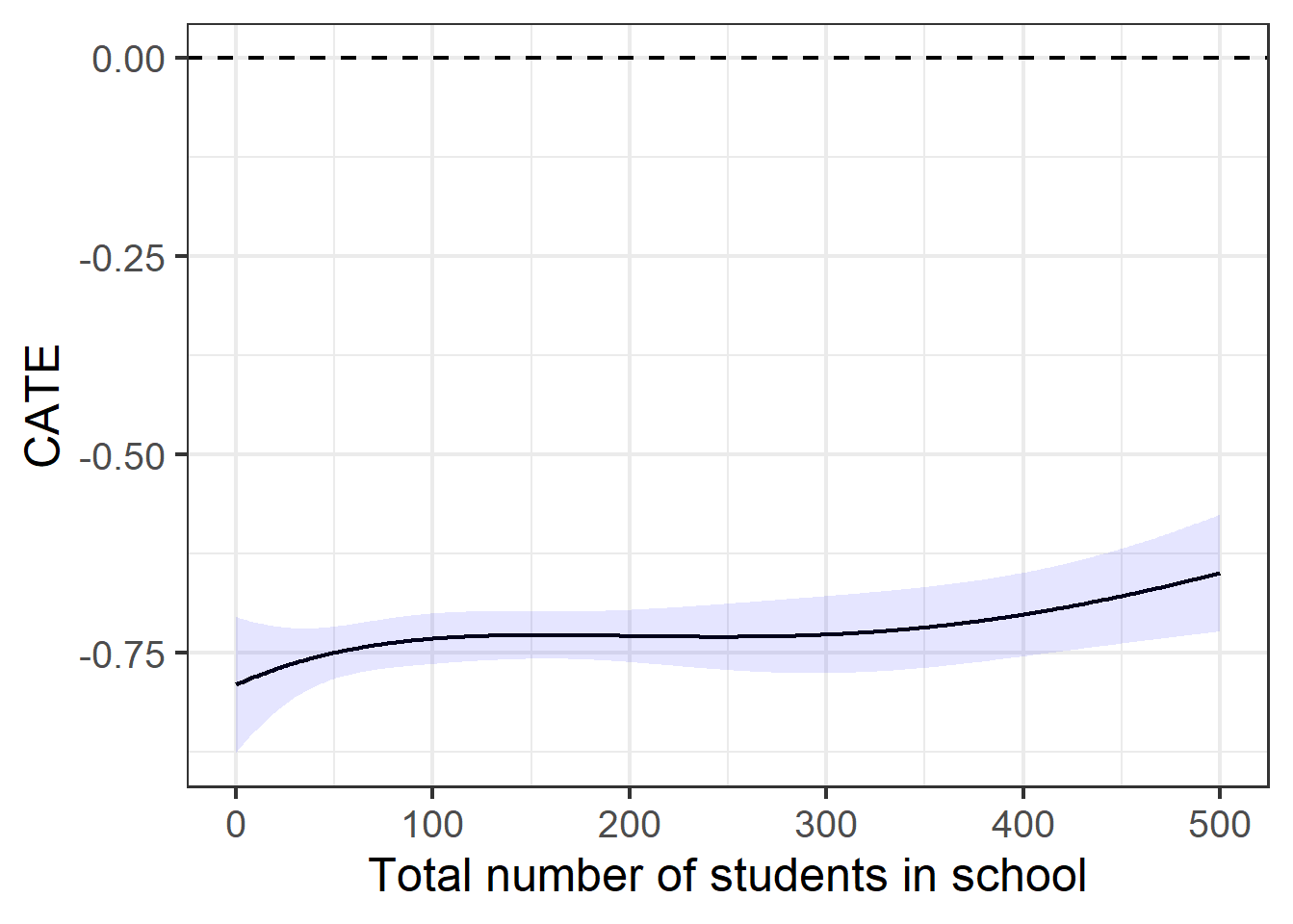}
    \caption{Schools: total number of students}
\end{subfigure}%
\begin{subfigure}{.5\textwidth}
    \centering
    \includegraphics[width=0.95\textwidth]{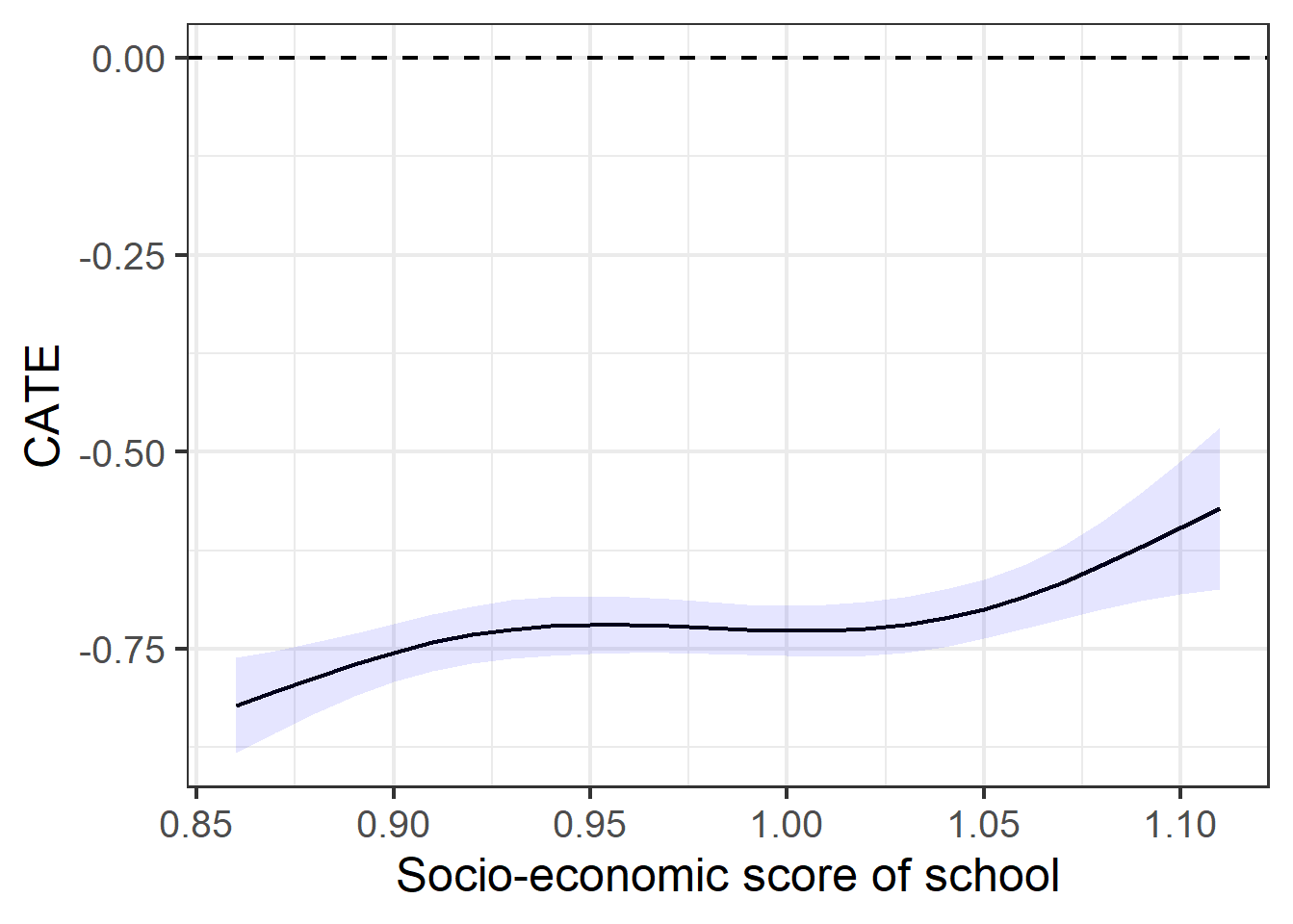}
    \caption{Schools: ``SES score''}
\end{subfigure}
%\begin{subfigure}{.5\textwidth}
%    \centering
%    \includegraphics[width=0.95\textwidth]{Graphs/GATEs/CATE_score_Costs_per_student_std.png}
%    \caption{Schools: expenditures per student (2017, std.) }
%\end{subfigure}
\vspace{5mm}

\footnotesize
\emph{Notes:} Continuous Conditional Average Treatment Effects in academic performance for inclusion vs. semi-segregation are depicted. The CATEs show how much the ATE would change at different levels of school characteristics (on the $x$ axis).
\end{minipage}
\end{figure}

\begin{figure}[p]
\centering
\caption[short]{Probability of unemployment: continuous CATE for school characteristics \label{fig:CATE_SCHOOLS_ub}}
\begin{minipage}{\linewidth}
\begin{subfigure}{.5\textwidth}
    \centering
    \includegraphics[width=0.95\textwidth]{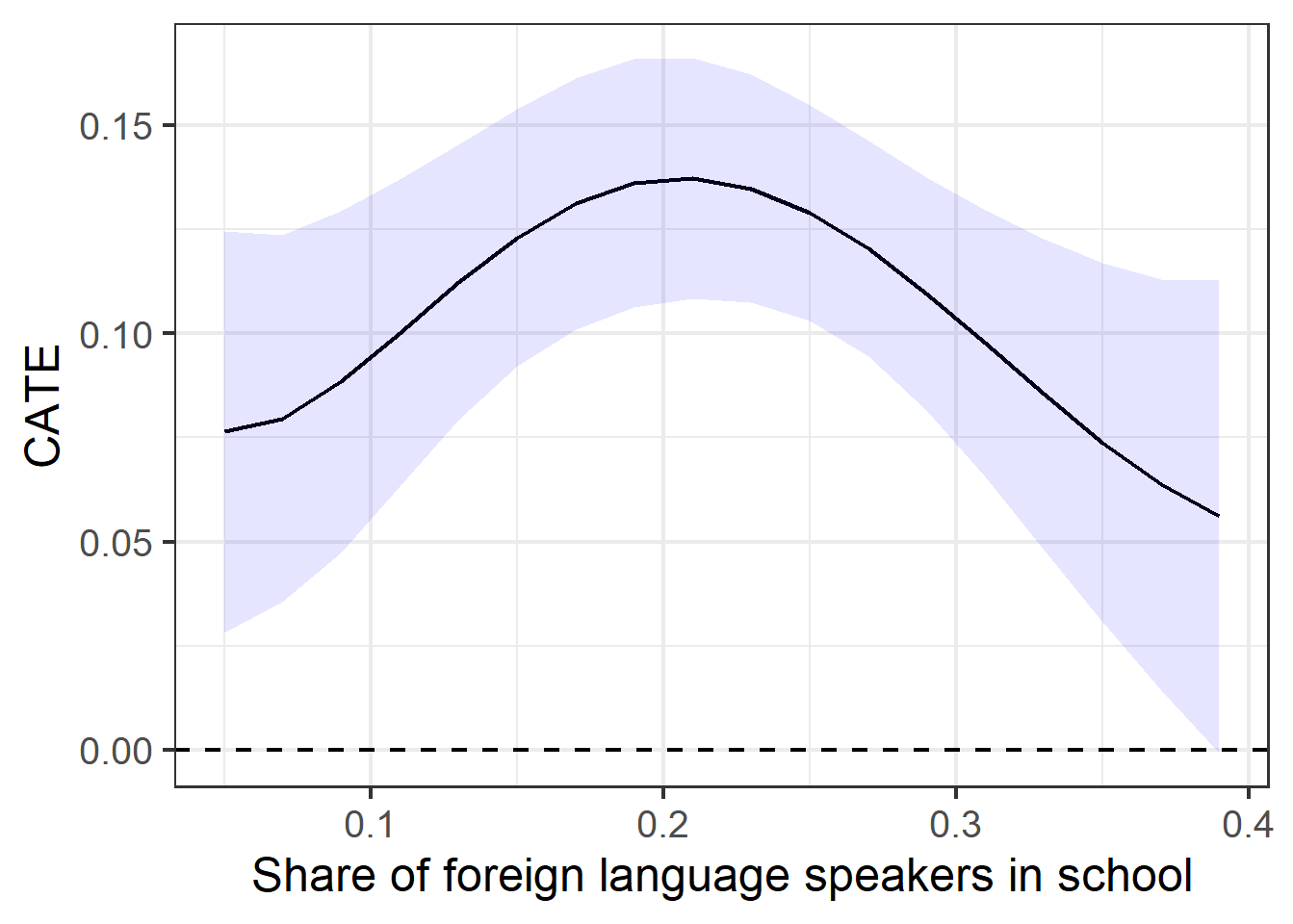}
    \caption{Schools: share of foreign students}
\end{subfigure}%
\begin{subfigure}{.5\textwidth}
    \centering
    \includegraphics[width=0.95\textwidth]{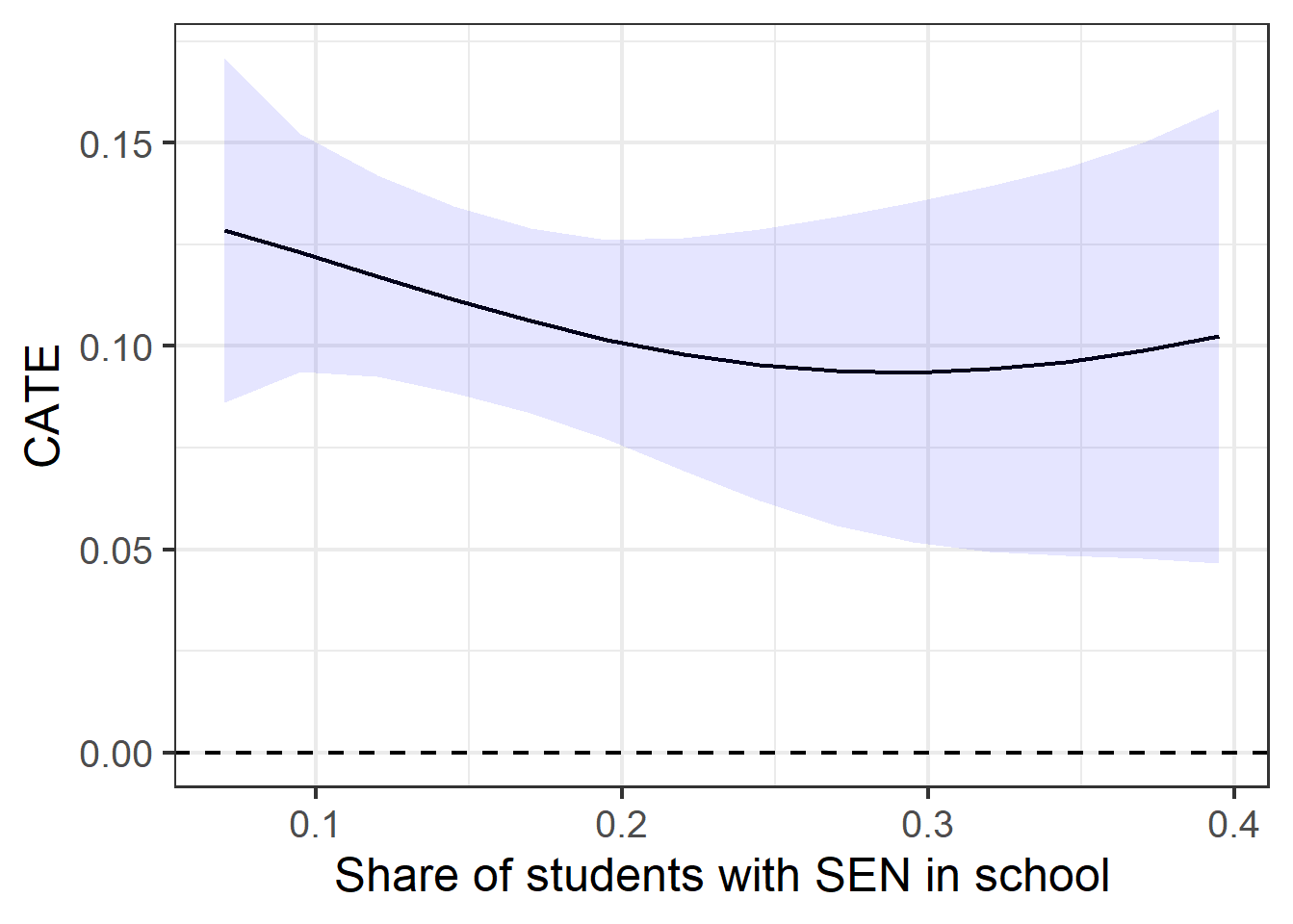}
    \caption{Schools: share of SEN students}
\end{subfigure}
\vspace{10mm}
\begin{subfigure}{.5\textwidth}
    \centering
    \includegraphics[width=0.95\textwidth]{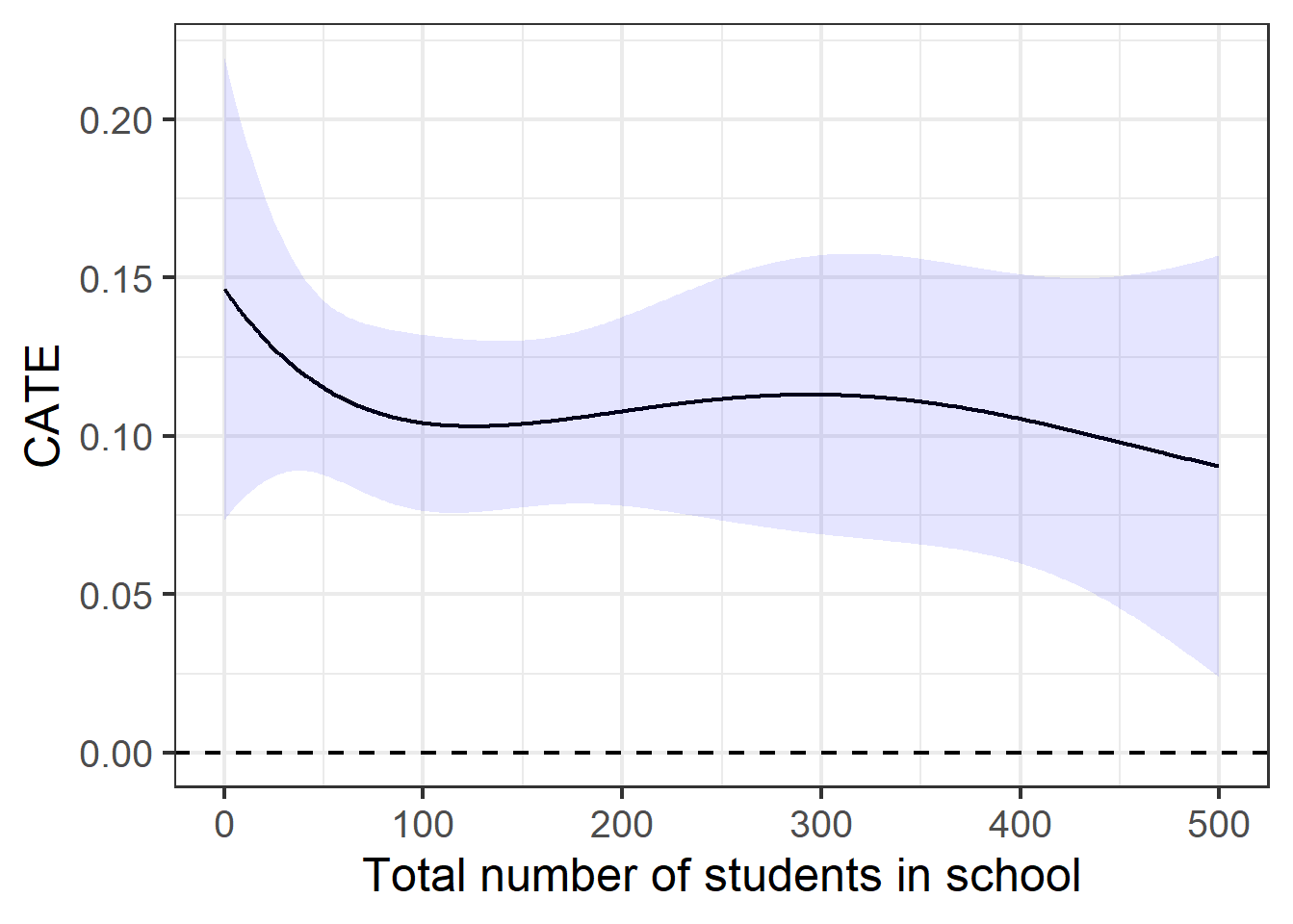}
    \caption{Schools: total number of students}
\end{subfigure}%
\begin{subfigure}{.5\textwidth}
    \centering
    \includegraphics[width=0.95\textwidth]{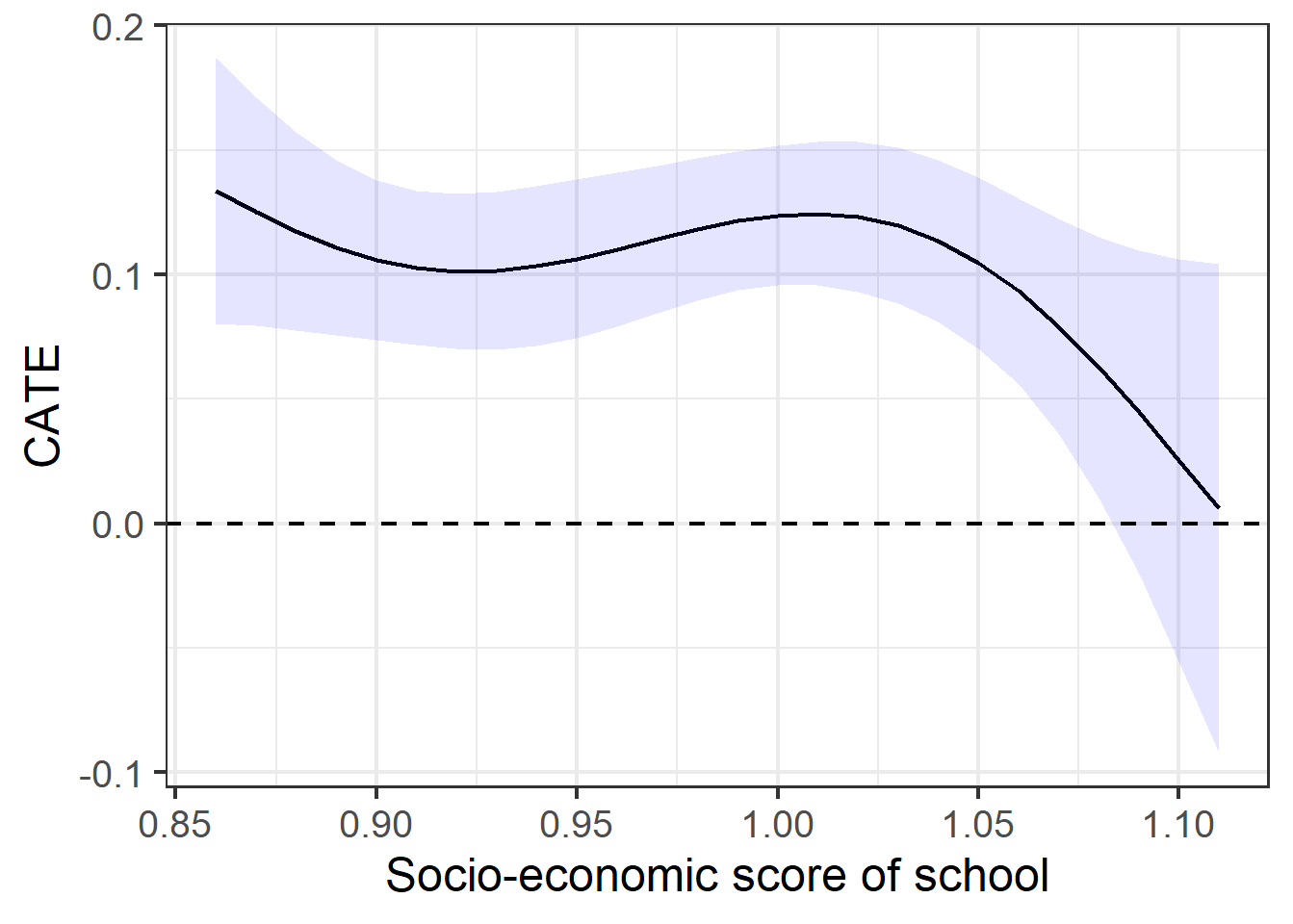}
    \caption{Schools: ``SES score''}
\end{subfigure}
%\begin{subfigure}{.5\textwidth}
%    \centering
%    \includegraphics[width=0.95\textwidth]{Graphs/GATEs/CATE_ub_Costs_per_student_std.png}
%    \caption{Schools: expenditures per student (2017, std.) }
%\end{subfigure}
\vspace{5mm}

\footnotesize
\emph{Notes:} Continuous Conditional Average Treatment Effects in the probability to be unemployed for inclusion vs. semi-segregation are depicted. The CATEs show how much the ATE would change at different levels of school characteristics (on the $x$ axis).
\end{minipage}
\end{figure}
\restoregeometry

\begin{figure}[p]
\centering
\caption[short]{Optimal policy trees for test scores \label{fig:optimalpolicytree_SW8}}

\begin{minipage}{\linewidth}
\begin{subfigure}{.375\textwidth}
    \centering
    \includegraphics[width=0.95\textwidth]{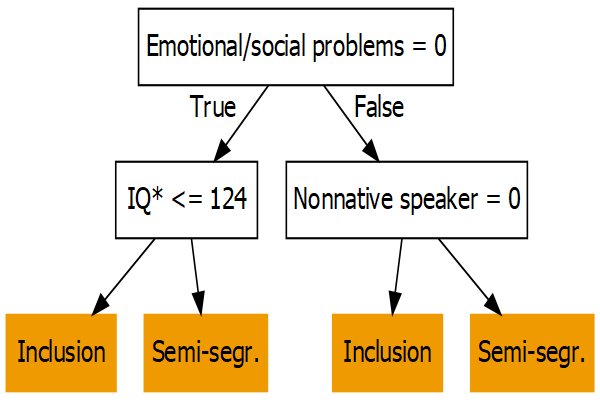}
    \caption{Depth 2, baseline}
\end{subfigure}%
\begin{subfigure}{.625\textwidth}
    \centering
    \includegraphics[width=0.95\textwidth]{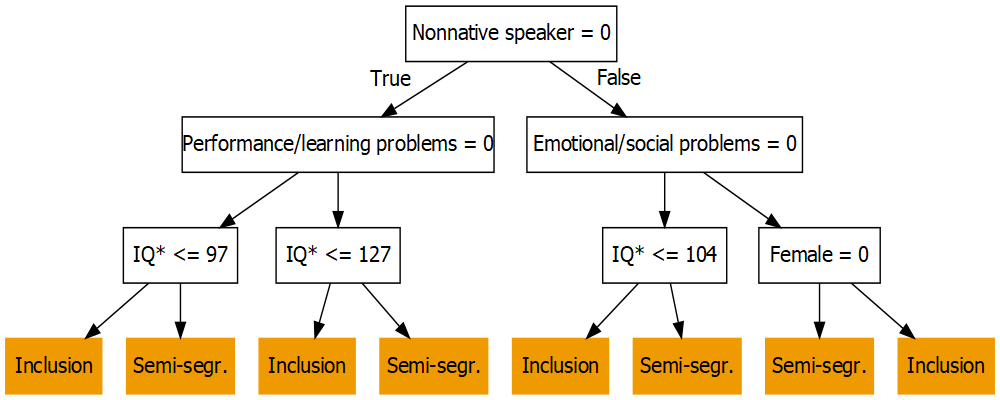}
    \caption{Depth 3, baseline}
\end{subfigure}
\vspace{10mm}
\begin{subfigure}{.375\textwidth}
    \centering
    \includegraphics[width=0.95\textwidth]{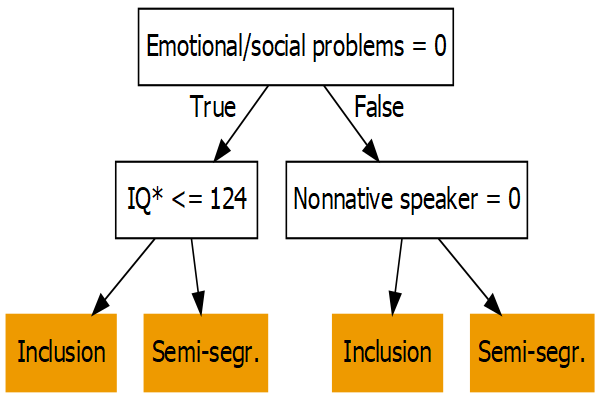}
    \caption{Depth 2, baseline $+$ diagnosis}
\end{subfigure}%
\begin{subfigure}{.625\textwidth}
    \centering
    \includegraphics[width=0.95\textwidth]{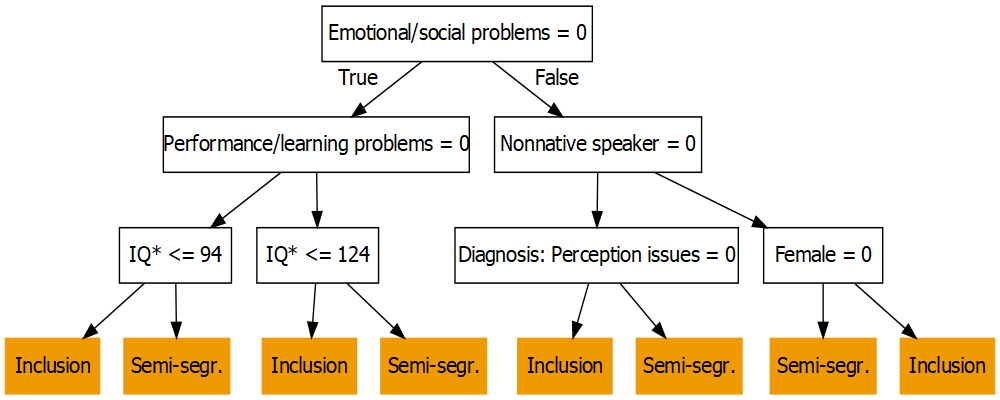}
    \caption{Depth 3, baseline $+$ diagnosis}
\end{subfigure}
\vspace{5mm}

\footnotesize
\emph{Notes:} Decision trees for optimal policy allocations of depth 2 and 3 are depicted. ``Baseline'' means that only baseline students' characteristics are included, and ``baseline $+$ diagnosis'' includes both baseline characteristics and diagnosis characteristics extracted from psychological records with the dictionary approach. The IQ variable IQ* is the interaction between the IQ score and the indicator whether an IQ score has been taken. 
\end{minipage}
\end{figure}

\begin{figure}[p]
\centering
\caption[short]{Optimal policy trees for probability to be employed \label{fig:optimalpolicytree_UBD}}
\begin{minipage}{\linewidth}
\begin{subfigure}{.375\textwidth}
    \centering
    \includegraphics[width=0.95\textwidth]{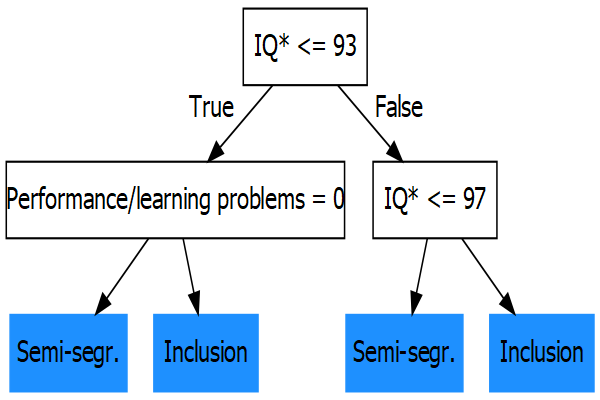}
    \caption{Depth 2, baseline}
\end{subfigure}%
\begin{subfigure}{.625\textwidth}
    \centering
    \includegraphics[width=0.95\textwidth]{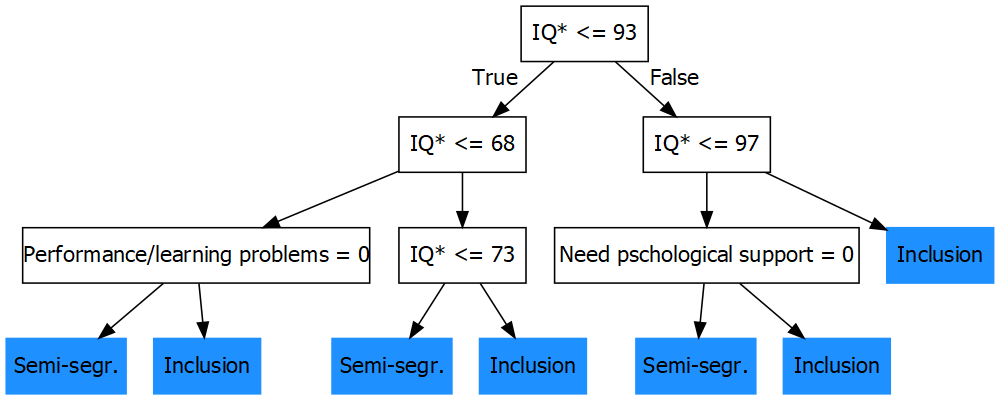}
    \caption{Depth 3, baseline}
\end{subfigure}

\vspace{10mm}

\begin{subfigure}{.375\textwidth}
    \centering
    \includegraphics[width=0.95\textwidth]{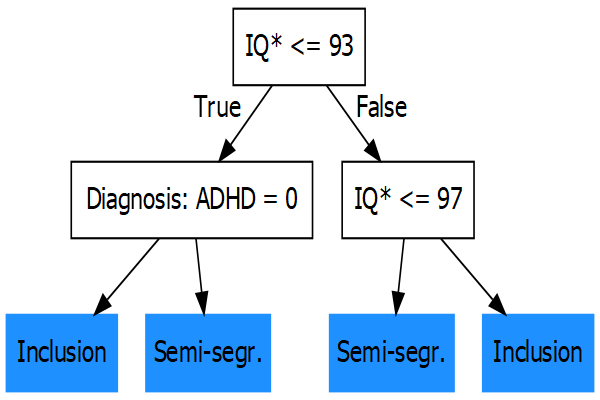}
    \caption{Depth 2, baseline $+$ diagnosis}
\end{subfigure}%
\begin{subfigure}{.625\textwidth}
    \centering
    \includegraphics[width=0.95\textwidth]{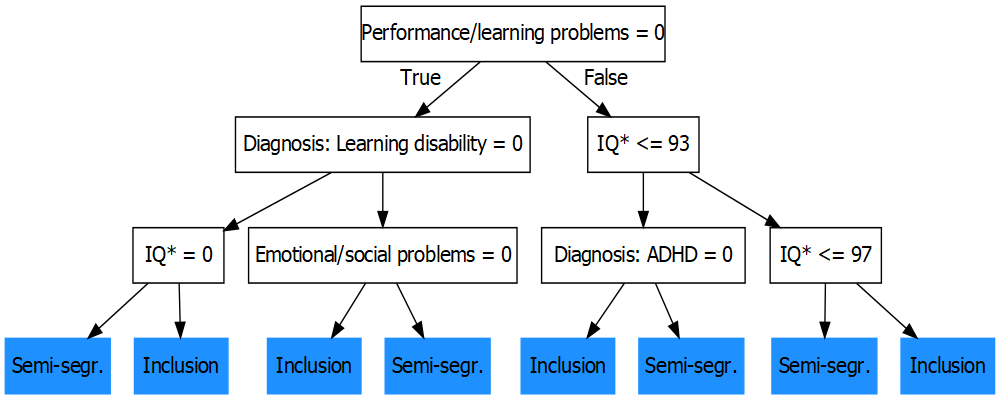}
    \caption{Depth 3, baseline $+$ diagnosis}
\end{subfigure}

\vspace{5mm}

\footnotesize \emph{Notes:} Decision trees for optimal policy allocations of depth 2 and 3 are depicted. ``Baseline'' means that only baseline students' characteristics are included, and ``baseline $+$ diagnosis'' includes both baseline characteristics and diagnosis characteristics extracted from psychological records with the dictionary approach. The IQ variable is the interaction between the IQ score and the indicator whether an IQ score has been taken. 
\end{minipage}
\end{figure}

\clearpage
\begin{figure}[p]
\centering
\caption{Classroom spillover effects of students with SEN on their peers without SEN. \label{fig:SNvsNonSN}}

\begin{minipage}{\linewidth}
\includegraphics[width=0.8\textwidth]{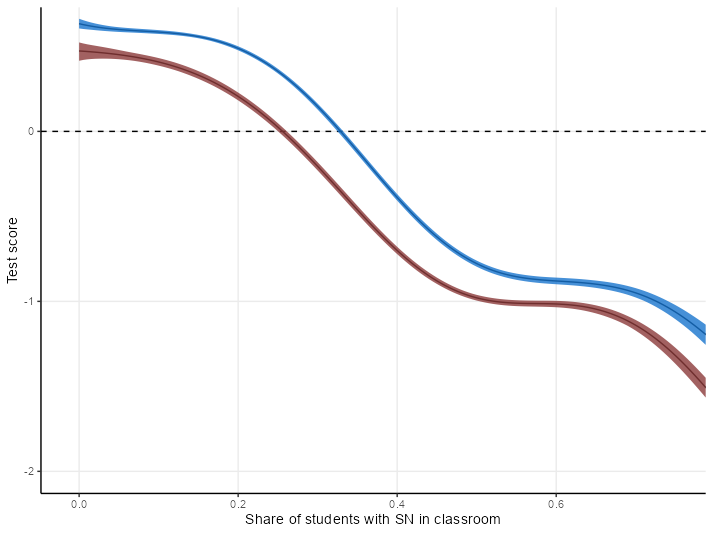}

\footnotesize \emph{Notes:} This graph depicts the spillover functions for the effect of the presence of students with SEN on the test scores of their peers with and without SEN in inclusive classrooms. All effects follow the identification strategy and results by \citet{BalestraEtal2020}. Flexible spillover functions are estimated with an ensemble learner similar to the estimation procedure used in this paper. Clustered cross-validation procedures at the classroom level are implemented. 95\% confidence intervals are represented.
\end{minipage}
\end{figure}

\restoregeometry
\end{appendices}

\end{document}